\documentclass[12pt]{article}
\usepackage{eurosym}
\usepackage{amssymb}
\usepackage{amsmath}
\usepackage[authoryear]{natbib}
\usepackage{geometry}
\usepackage{graphicx}
\usepackage{xcolor}
\usepackage{hyperref}
\usepackage[title]{appendix}
\usepackage{multirow}
\usepackage{dsfont}

\usepackage[multiple]{footmisc}

\emergencystretch=\maxdimen
\definecolor{CustomColor}{rgb}{0,0.04,0.04}
\definecolor{blue}{rgb}{0,0.44,0.74}
\usepackage{comment}
\hypersetup{linkcolor=blue, citecolor=blue, urlcolor=blue, colorlinks=true}

\usepackage{setspace}

\usepackage[labelfont=bf]{caption}

\usepackage[belowskip=-3pt,aboveskip=0pt]{caption}

\begin{document}
\onehalfspacing

\title{The Inflation Attention Threshold and Inflation Surges\thanks{\protect\linespread{1}\protect\selectfont
The University of Texas at Austin, Department of Economics, 2225 Speedway, 78712 Austin TX, \href{pfaeuti.oliver@gmail.com}{%
pfaeuti.oliver@gmail.com}.
I thank Hassan Afrouzi, Rudi Bachmann, Luca Benati, Jonathan Benchimol, Pierpaolo Benigno, Jaroslav Borovicka, Oli Coibion, Leland Farmer, Francesco Furlanetto, Jose-Elias Gallegos, Laura G\'{a}ti, Michael Gelman, Niklas Kroner, John Leahy, Cyril Monnet, David Munro, Dirk Niepelt, Ricardo Reis, Fabian Seyrich, Hannes Twieling and seminar and conference participants at the NBER Summer Institute 2024 Impulse and Propagation Mechanisms Program, SED 2024, the Midwest Macro 2024, Texas A\&M University, the University of Bern, the International Workshop on Macroeconomic Regime Changes at LSE, the Workshop on Macroeconomics and Survey Data at the National Bank of Belgium, and the VIMM seminar for helpful comments and suggestions. First Version: August 2023.}}
\author{Oliver Pfäuti}

\date{August 2024 \vspace{0.1cm}\\  \href{https://opfaeuti.github.io/website/IAT.pdf}{\textcolor{blue}{ \hspace{0.2cm}Link to most recent version}}}
\maketitle

\begin{abstract}
At the outbreak of the recent inflation surge, the public’s attention to inflation was low but increased quickly once inflation started to rise. In this paper, I quantify when and by how much the public's attention to inflation changes, and derive the macroeconomic implications of these attention changes. I estimate an attention threshold at an inflation rate of about 4\%, and that attention doubles when inflation exceeds this threshold. Adverse supply shocks become more inflationary in times of high attention, and the increase in people's attention to inflation in 2021 accounts for half of the subsequent supply-driven inflation. I develop a model accounting for the attention threshold and show that shocks that are usually short lived lead to a persistent surge in inflation if they induce an increase in people's attention. The attention threshold further lengthens the last mile of disinflation after an inflation surge, and leads to an asymmetry in the dynamics of inflation.
\vspace*{0cm}\newline\noindent
{\footnotesize {\textbf{JEL Codes:} E3, E4, E5, E7\newline
}}
\end{abstract}
\clearpage\newpage
\section{Introduction}
\begin{figure}[!ht]
\caption{Inflation, attention to inflation, and inflation expectations}
\label{fig:moti}\vspace{0.15cm} \centering%
\begin{tabular}{cc}
(a) Inflation and Google Trends & (b) Inflation and inflation expectations  \\ \includegraphics[width=.47\textwidth]{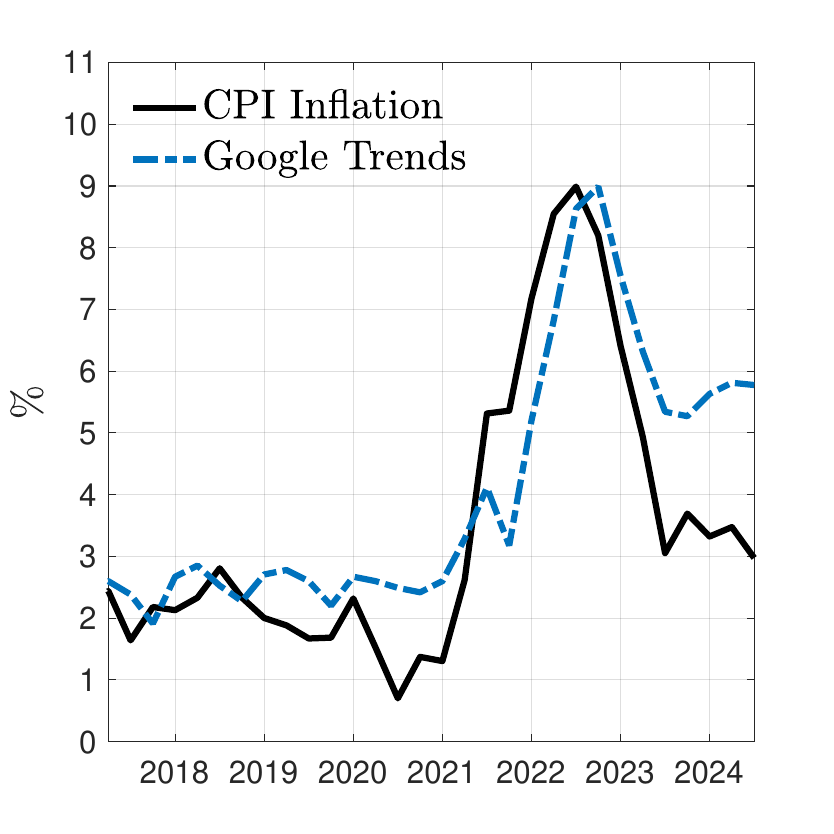} & 
\includegraphics[width=.47\textwidth]{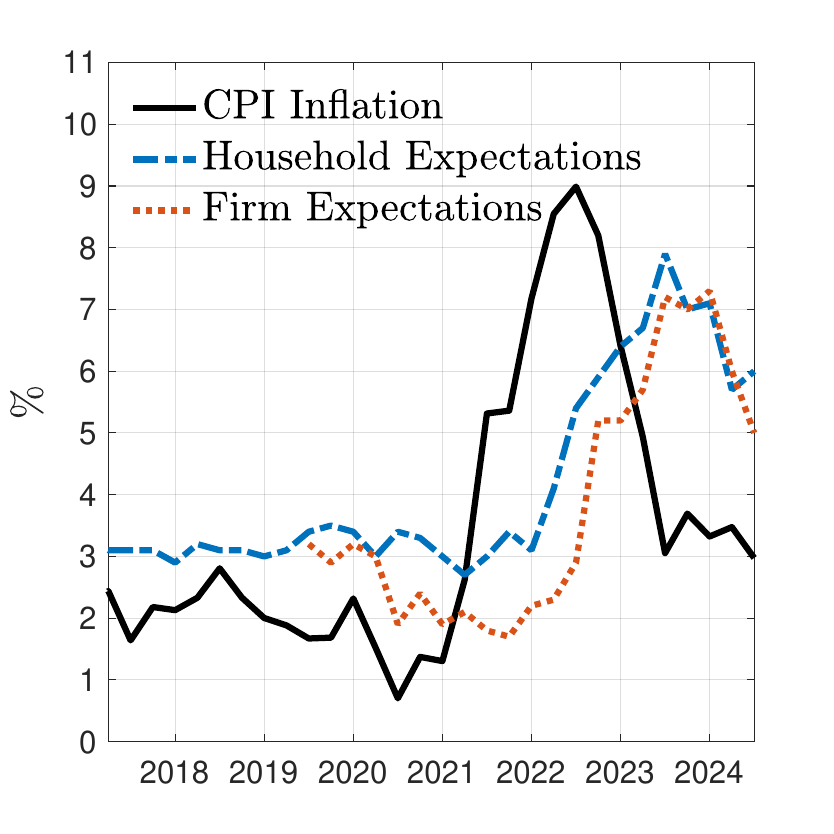} 
\end{tabular}%

\begin{minipage}{1\textwidth}
\footnotesize{\emph{Notes}: Panel (a) shows U.S.\ CPI inflation (black-solid line) together with Google Trends searches of the word `inflation' in the U.S. Google Trends are normalized to have the same maximum as CPI inflation. Panel (b) shows U.S.\ CPI inflation (black-solid line) together with one-year-ahead inflation expectations for U.S.\ households (blue-dashed line) and firms (red-dotted line). Expectations are shifted such that the vertical distance to the black line captures the forecast errors. Household expectations are from the Survey of Consumers from the University of Michigan and firm expectations from the Survey of Firms' Inflation Expectations, provided by the Cleveland Fed.
}%
 \end{minipage}
\end{figure}

\vspace{-0.2cm}
Inflation is back. After decades of low and stable inflation, inflation surged in many advanced economies during the recovery phase of the pandemic. Inflation turned out to be higher and more persistent than many expected.\footnote{For example, Federal Reserve Chair Jerome H.\ Powell said in his 2021 Jackson Hole speech that inflation concerns are ``likely to prove temporary'' (see \href{https://www.federalreserve.gov/newsevents/speech/powell20210827a.htm}{\textcolor{blue}{https://www.federalreserve.gov/newsevents/speech/powell20210827a.htm}}).} With inflation rising, the public's attention to inflation increased as well (panel (a) in Figure \ref{fig:moti}), and as inflation peaked, households' and firms' inflation expectations also started to pick up (panel (b) in Figure \ref{fig:moti}). During the build up of the inflation surge, however, inflation expectations were substantially below actual inflation. Only as inflation started to come down, inflation expectations started to overshoot inflation and these expectations remained high while inflation stayed persistently above its target of 2\%.

In this paper, I show that accounting for the observed changes in people's attention to inflation provides an explanation for the recent inflation surge as well as the interplay between inflation and inflation expectations observed in panel (b) of Figure \ref{fig:moti}. When attention to inflation increases when inflation exceeds a certain threshold, inflation expectations become more sensitive to inflation in times of high inflation and thus, reinforce the inflation surge and render it more persistent. As inflation falls back below the threshold, however, attention decreases and inflation expectations become more sluggish, leading to a persistent overshooting of inflation expectations and a longer `last mile' of disinflation back to target.

To arrive at these results, I propose a way to estimate attention to inflation and an inflation attention threshold at which attention changes, as well as the degrees of attention across the different attention regimes. Attention captures how strongly agents update their short-run inflation expectations in response to forecast errors.\footnote{For example, \cite{werning2022expectations} discusses why short-run expectations are more important than long-run expectations for inflation. In \cite{pfauti2021inflation}, I show that measuring attention to inflation by estimating how agents update their short-run expectations correlates strongly with other measures of attention.} In this framework, agents may become more attentive because their prior uncertainty increases, because the cost of being attentive decreases, or because the perceived persistence of inflation increases. I provide evidence for all three channels.

I estimate the attention threshold and the attention levels in the two regimes jointly using survey data from the University of Michigan's Survey of Consumers for the period 1978 to 2024. I find an inflation attention threshold at an annualized inflation rate of about 4\%, and that attention is about 0.18 when inflation is below the threshold and practically doubles to 0.35 when inflation is above the threshold. An attention level of 0.35 means that following a 1pp forecast error, agents increase their inflation expectations by 0.35pp. A Markov-switching model, in which I do not need to assume what determines the threshold, confirms these findings. Using regional variation in inflation and inflation expectations, or when relying on the panel data from the Survey of Consumer Expectations from the New York Fed for the period after 2013, I obtain very similar results. I find that the data suggests that there is only one attention threshold, that the threshold specification is preferred to smooth changes in attention, and that once I control for this threshold, I do not find any evidence that attention increases with inflation \textit{within} regime.

The attention regimes matter for the dynamics of inflation and inflation expectations. 
Using the high-frequency identification of \cite{kanzig2021macroeconomic}, I use oil supply news shocks as my proxy for supply shocks and I find that inflation increases more than twice as much in response to these shocks when they hit in the high-attention regime compared to the low-attention regime. A one standard deviation shock that pushes inflation up by about 20 basis points on average across regimes, leads to a peak increase in inflation of about 40 basis points when the shock hits in the high-attention regime. These effects are persistent as they only diminish after about one to two years. Using regional data and Google Trends data as a proxy for people's attention, I further disentangle the effects of attention from other changes in the economy that may arise from higher inflation, and find that attention indeed leads to a stronger inflation response to supply shocks even when controlling for inflation directly.

The attention threshold I estimate implies that the U.S.\ economy entered the high-attention regime most recently in early 2021. To quantify the importance of this increased attention for the recent inflation surge, I first show that the oil supply news shocks explain about 55\% of the inflation surge from the time the U.S.\ economy entered the high-attention regime in early 2021 until the peak in mid-2022. Without the attention increase, however, the implied inflation increase would have only been half as strong.\footnote{I show that also other shocks, i.e., the \textit{inflation shock} from \cite{angeletos2020business} as well as monetary policy shocks, become more inflationary in the high-attention regime, indicating that shocks generally become more inflationary when attention to inflation is high. However, for the recent inflation surge I focus on oil supply news shocks as these shocks are available until the end of 2022. In the model, however, I show that demand shocks can lead to similar inflation dynamics as supply shocks, consistent with my empirical findings.} These results suggest that people's higher attention to inflation is not merely a byproduct but an important driver of the recent inflation surge and the increase in attention substantially amplified the already inflationary supply shocks and rendered inflation more persistent.

To understand these findings as well as their implications for monetary policy, I develop a New Keynesian model accounting for the attention threshold and the changing levels of attention. 
Attention changes occur because the shock volatility increases once inflation exceeds the threshold. These regime-dependent shock volatilities are calibrated to match the empirically-estimated shock volatility and the increase in attention across regimes jointly. I find that in response to an inflationary supply shock that pushes inflation above the attention threshold, inflation keeps on increasing through the attention increase.  As attention increases, agents update their inflation expectations more strongly in response to the increase in inflation. These higher inflation expectations then fuel further inflation increases, leading to even higher inflation expectations. Thus, what normally would be a transitory inflation shock can become a very persistent one due to the heightened attention. As the shock dies out, inflation starts to decline after some time. However, inflation may remain elevated for a substantial period of time, because once it falls back below the threshold, people pay less attention to inflation, and hence, only slowly revise their inflation expectations downward. As their prior expectations are now relatively high due to the experienced high inflation period, expectations remain persistently high, leading to a slow decline of actual inflation. The inflation attention threshold and changing degrees of attention hence offer an explanation for why inflation followed a hump-shaped pattern and inflation expectations initially fell short of the actual inflation surge, but then surpassed it once inflation started to decline, as documented in panel (b) of Figure \ref{fig:moti}.
Therefore, the changes in attention to inflation can explain why inflation came down relatively quickly during the second half of 2022 but remained stubbornly above the Fed's inflation target even in early 2024.

That inflation expectations stay persistently high, even when inflation has already fallen back below the threshold, further gives rise to a heightened risk of another subsequent inflation surge. The higher inflation expectations keep actual inflation higher for longer, and therefore, closer to the attention threshold. Thus, a subsequent inflationary shock is more likely to push inflation back above the threshold and therefore, leading to another episode of persistently high inflation.

The attention threshold induces an asymmetry in the dynamics of inflation: attention increases when inflation is particularly high but remains constant when inflation is particularly low. Relative to the model abstracting from the attention threshold, this asymmetry increases the risk of periods of persistently high inflation rates but leaves the risk of deflation largely unchanged. This inflation asymmetry induced by the attention threshold also offers a potential explanation for why we did not observe long lasting deflationary periods, for example, during the Great financial crisis \citep{coibion2015phillips}. In fact, annualized quarter-on-quarter CPI inflation was negative only 7\% of the time between 1978 and 2023, whereas it exceeded the attention threshold about 30\% of the time during the same period.

This asymmetry also matters for the normative implications of the attention threshold. When monetary policy sets the nominal interest rate following a relatively dovish Taylor rule, the associated central bank losses are significantly larger compared to optimal policy rules or a strict inflation targeting rule. The reason is that under a standard Taylor rule, the economy spends a substantial amount of time in the high-attention regime in which inflation is high and volatile. Being frequently in the high-attention regime also increases the \textit{average level of inflation} due to the asymmetry of the threshold which is costly for the central bank. By following a more hawkish monetary policy rule, the central bank can mitigate these losses.

\paragraph{Related literature.} This paper contributes to the ongoing debate about the drivers of the recent inflation surge and offers an alternative and novel perspective by highlighting the role of people's heightened attention during that period. \cite{shapiro2022decomposing} decomposes demand- and supply-driven inflation and estimates a similar role for supply factors as I do. \cite{gagliardone2023oil} estimate a New Keynesian model to account for the recent inflation dynamics and attribute large parts of it to oil supply shocks and easy monetary policy. \cite{bernanke2023caused} find that shocks to prices given wages, including shocks to energy prices and sectoral shortages, played a key role for the recent inflation spike. \cite{cerrato2022inflation} and \cite{benigno2023s} find evidence for a non-linear Phillips Curve that is steeper when labor markets are tight. Relatedly, \cite{ball2022understanding} find that the tightening of labor markets was a main driver of the rise in core inflation, and \cite{lorenzoni2023wage} focus on wage-price spirals. \cite{crump2024unemployment} document a prominent role of expected unemployment gaps for the observed inflation dynamics, \cite{erceg2024monetary} focus on non-linear wage and price Phillips curves coupled with indexation and learning about the persistence of the shocks.
\cite{amiti2023inflation} focus on the role of supply chain disruptions, labor supply constraints and the shift of consumption from services to goods, whereas \cite{bianchi2022inflation} and \cite{bianchi2023fiscal} attribute the inflation surge to unfunded fiscal shocks or changes in people's beliefs about the fiscal framework. \cite{cavallo2023large} and \cite{blanco2024inflation} argue that the frequency of price adjustments has increased.
In contrast to all of these papers, I focus on the role of changes in people's attention to inflation. Attention serves as a propagation mechanism rather than as an exogenous impulse and my results show that the increase in attention in early 2021 roughly doubled the inflationary effects of supply shocks between early 2021 and the end of 2022.\footnote{See \cite{binder2022expected} for a narrative history of how inflation and inflation expectations evolved jointly during several recessions and periods of high inflation since the Great Depression.} Additionally, my findings offer an explanation for the interplay between inflation and inflation expectations (panel (b) in Figure \ref{fig:moti}) as well as for the long last mile of disinflation.

\cite{pfauti2021inflation} documents that before the recent inflation surge, attention to inflation was at a historical low.\footnote{For the years before the recent inflation surge, \citet{candia2021inflation,candia2023perceived} and \citet{coibion2020average} show that U.S.\ firms as well as households are usually poorly informed about and inattentive to inflation and monetary policy (see also \cite{weber2022subjective} for a recent review). \cite{goldstein2023tracking} also finds that inattention varies over time. Different forms of changing attention are considered, e.g., in \cite{coibion2015information} who show state-dependence in the degree of information rigidity (focusing on GDP growth), or in \cite{kim2023learning} who examine how repeat participation in surveys that ask about inflation expectations may lead to higher attention to inflation, or in \cite{flynn2022attention} who show---by using a textual proxy for firms' attention toward macroeconomic conditions---that attention is counter-cyclical, or \cite{gallegos2023inflation} who shows that firms' less sluggish inflation expectations after the Great inflation period offer a potential explanation for the decrease in the persistence of inflation as found, e.g., in \cite{benati2008investigating}. \cite{york2023household} exploits differences in survey dates to examine how inflation expectations respond to macroeconomic news releases.} \cite{bracha2023inflation} find that when inflation increases, attention to inflation increases as well. \cite{korenok2022inflation} find, for a large number of countries, that people's attention to inflation increases with inflation only after inflation exceeds a certain threshold. Using Google search data for the period from 2004 to 2022, they estimate this attention threshold to be at an annualized inflation rate of 3.55\% for the U.S. \cite{buelens2023googling} uses internet search behavior of households in the Euro Area to show that this form of attention to inflation positively depends on the level of inflation. \cite{cavallo2017inflation} show that survey respondents in high-inflation environments (Argentina) respond less to information about inflation than in low-inflation environments (United States), which is consistent with higher attention to inflation in high-inflation environments. \cite{weber2023tell} confirm, using a range of randomized control trials spanning over several years and different countries, that attention of households and firms is indeed higher in times of high inflation. \cite{kroner2023inflation} focuses on financial markets and shows that attention---measured as asset price responses to inflation news---is higher in times of high inflation. My key innovation relative to these papers is that I provide estimates of the attention threshold and the attention levels in the two regimes in a way that directly maps into otherwise standard macroeconomic models, and I quantify the macroeconomic implications of these attention changes. 

I further contribute to the literature on the state dependency of shocks \citep{auerbach2012fiscal,auerbach2012measuring,caggiano2014uncertainty,ramey2018government,tenreyro2016pushing,aastveit2023inflation,jo2022state}. The two papers most closely related are \cite{ascari2022non} and \cite{joussier2023energy}. \cite{ascari2022non} show that the inflationary effects of monetary policy shocks are larger in times of high trend inflation. \cite{joussier2023energy} find that French firms pass through energy price shocks more to their prices in times of high inflation.\footnote{Based on the present paper and \cite{korenok2022inflation}, \cite{joussier2023energy} use Google Trends data to proxy for attention to energy prices and energy cost and show that the pass through of energy shocks to prices in France is stronger in times of higher attention to energy prices.} I contribute to that literature by showing that attention to inflation matters for the transmission of supply and monetary shocks to aggregate inflation and inflation expectations.

The theoretical insights in this paper contribute to a growing literature on the role of changes in attention and the degree of anchoring of inflation expectations. 
\cite{evans199515} propose a model in which agents choose how forward-looking they want to be and show that inflation remains stable when agents are not very forward looking. \cite{turen2023state} shows how a model with state-dependent attention can reconcile several facts about the microeconomic evidence on price setting.\footnote{Focusing on endogenous but constant attention, \cite{mackowiak2009optimal} show how firms set prices optimally information is costly, and \cite{paciello2014exogenous} how the optimal monetary policy changes when firm managers decide optimally how much attention they want to pay to aggregate conditions compared to a setting where their attention is exogenous (see also \cite{sims2003implications,sims2010rational}). \cite{reis2006inattentive,reis2006inattentivep} study how decisions of consumers and producers change when updating their information set infrequently, and \cite{lorenzoni2009theory} focuses on dispersed information.} \cite{gati2020monetary} studies how changes in the degree of anchoring of long-run inflation expectations affect the optimal monetary policy. In her paper, anchoring changes continuously, thus, can also jump in response to large shocks. In contrast, attention in my model only changes across but not within regime, which is consistent with what I find in the data, namely, that attention is constant within regimes. \cite{carvalho2020anchored} focus on discrete changes in anchoring of long-run inflation expectations, but do so in a partial equilibrium setting. In \cite{pfauti2021inflation}, I study the implications of low attention for optimal monetary policy when the zero lower bound poses a constraint to monetary policy. In contrast to that paper, I allow here for an inflation attention threshold, whereas \cite{pfauti2021inflation} compares economies with different---but time-invariant---degrees of attention.
\cite{hazell2022slope} show that the greater inflation stability after 1990 is mostly due to more firmly-anchored long-run inflation expectations. Related, \cite{afrouzi2020dynamic} show that in a model of dynamic rational inattention the stance of monetary policy affects the slope of the Phillips curve (see also \cite{jorgensen2019anchored}). The main contribution of the present paper to this literature is that I estimate and incorporate an inflation attention threshold into a general equilibrium model and show how accounting for such a threshold can help us better understand how inflation surges may happen and how they can sustain themselves, as well as the monetary policy implications that follow from such attention-fueled inflation surges.

\paragraph{Outline.} In Section \ref{sec:data}, I estimate the inflation attention threshold and attention across regimes. In Section \ref{sec:empirical_irfs}, I show that the attention regime matters for the inflationary effects of shocks, and quantify the role of attention in the recent inflation surge. I introduce the New Keynesian model with limited attention in Section \ref{sec:model}, and the model's positive results in Section \ref{sec:results}. In Section \ref{sec:welfare}, I discuss the threshold's normative implications and Section \ref{sec:conclusion} concludes.

\section{Attention and the Inflation Attention Threshold}\label{sec:data}\vspace{-0.2cm}
In this section, I derive a model of optimal attention choice which I then bring to the data to estimate the inflation attention threshold and the different degrees of attention in the two attention regimes.  

\subsection{Attention to inflation}\vspace{-0.2cm}
In order to estimate attention to inflation as well as a threshold of when attention changes, I extend the method I develop in \cite{pfauti2021inflation} by allowing for state dependent parameters.\footnote{The method proposed in \cite{pfauti2021inflation} builds on \cite{mackowiak2023rational} and \cite{vellekoop2019inflation}, see also \cite{weber2023tell} for a similar model but with the objective to interpret their RCT findings (see also \cite{mackowiak2024rct}).}  
The agent believes that inflation in the next period, $\pi'$, depends on inflation today, $\pi$, as follows
\begin{equation}
\pi' =   \rho_{\pi,r} \pi + \nu, \notag
\end{equation}
where $\nu \sim i.i.N.(0,\sigma^2_{\nu})$, and $\rho_{\pi,r}\in[0,1]$ denotes the perceived persistence of inflation which may depend on the state or regime $r$ that the economy is in.\footnote{Agents' subjective model does not necessarily need to be consistent with the actual law of motion (see \cite{andre2022subjective} and \cite{macaulay2022shock} for empirical evidence that the subjective models agents hold may not necessarily be consistent with the actual behavior of the economy or with experts' subjective models).} 
Inflation in the current period is unobservable, so before forming an expectation about future inflation, the agent needs to form expectations about today's inflation. I denote this nowcast $\widetilde{\pi}$, and the resulting forecast about next period's inflation $\pi^e = \rho_{\pi,r}\widetilde{\pi}$. The assumption that inflation expectations solely depend on perceived inflation is supported by a large empirical literature documenting that perceived inflation is indeed the key predictor for inflation expectations (see, e.g., \cite{jonung1981perceived} and \cite{weber2022subjective} for households and \cite{candia2023perceived} for firms). That said, however, I also allow for a multivariate expectations formation process in the estimation later.
Given these beliefs, the full-information forecast $\pi^{e*}$ is 
$\pi^{e*} \equiv \rho_{\pi,r} \pi.$
But because $\pi$ is not perfectly observable, the actual forecast will deviate from this full-information forecast. These deviations are costly and the agent acquires information in order to prevent these deviations.
Thus, the agent's choice is not only about how to form her expectations given certain information, but about how to choose this information optimally, while taking into account how this will later affect her forecast. That is, she chooses the form of the signal $s$ she receives about current inflation.\footnote{Since acquiring information is costly, it cannot be optimal to acquire different signals that lead to an identical forecast. Due to this one-to-one relation of signal and forecast, we can directly work with the joint distribution of $\pi^e$ and $\pi$, $f(\pi^e, \pi)$, instead of working with the signal.}

Let $U(\pi^e, \pi)$ denote the negative of the loss that is incurred when the agent's forecast deviates from the forecast under full information, and $C(f)$ the cost of information. Then, the agent's problem is given by
\begin{align}
&\max_{f} \int U(\pi^e, \pi)f(\pi^e, \pi)d\pi d\pi^e - C(f) \label{pr1}\\
\text{subject to } & \int f(\pi^e, \pi) d\pi^e = g(\pi), \text{ for all } \pi, \notag
\end{align}
where $g(\pi)$ is the agent's prior, which is assumed to be Gaussian; $\pi \sim N\left(\widehat{\pi},\sigma^2_{\pi,r}\right)$, where I also allow prior uncertainty $\sigma^2_{\pi,r}$ to be state or regime dependent.
$C(.)$ is the cost function that captures how costly information acquisition is. It is linear in mutual information $I(\pi;\pi^e)$, i.e., the expected reduction in entropy of $\pi$ due to knowledge of $\pi^e$:
\begin{equation}
C(f) = \frac{1}{\lambda_r} I(\pi;\pi^e) = \frac{1}{\lambda_r}\left(H(\pi)-E\left[H(\pi|\pi^e)\right]\right), \notag
\end{equation}
where $H(x) = -\int f(x)log(f(x))dx$ is the entropy of $x$ and $\frac{1}{\lambda_r}$ measures the cost of information, which may also depend on the current state.

The objective function $U(.)$ is assumed to be quadratic:
\begin{equation}
U(\pi^e,\pi) = -\chi\left(\rho_{\pi,r}\pi-\pi^e\right)^2, \notag
\end{equation}
where $\chi$ measures the stakes of making a mistake.

In this setup, Gaussian signals are optimal (and in fact the unique solution, see \citet{matvejka2015rational}) and take the form
\begin{equation}
s = \pi+\varepsilon, \notag
\end{equation}
with $\varepsilon\sim i.i.N.(0,\sigma^2_{\varepsilon,r})$. 
The problem \eqref{pr1} now reads
\begin{equation}
\max_{\sigma^2_{\pi|s}\leq \sigma^2_{\pi,r}}E_{\pi}\left[E_s\left[-\chi\rho_{\pi,r}^2\left(\pi-E[\pi|s]\right)^2\right]\right] - \frac{1}{\lambda_r} I(\pi;\pi^e)= \max_{\sigma^2_{\pi|s}\leq \sigma^2_{\pi,r}} \left(-\chi\rho_{\pi,r}^2\sigma^2_{\pi|s}-\frac{\frac{1}{\lambda_r}}{2} log\frac{\sigma^2_{\pi,r}}{\sigma^2_{\pi|s}}\right). \label{eq:pr2}
\end{equation}
The optimal forecast is given by $\pi^e = \rho_{\pi,r} E\left[\pi|s\right]$, and Bayesian updating implies
\begin{equation}
\pi^e = \rho_{\pi,r}\left(1-\gamma_{\pi,r}\right)\widehat{\pi}_r+\rho_{\pi,r}\gamma_{\pi,r} s,\label{updating_theo}
\end{equation}
where $\gamma_{\pi,r} = 1-\frac{\sigma^2_{\pi|s}}{\sigma^2_{\pi,r}}\in[0,1]$ measures the agent's attention to inflation in state or regime $r$, and $\widehat{\pi}_r$ denotes the prior mean of $\pi$. When the agent is inattentive, she obtains relatively noisy signals and thus, puts little weight on these signals, reflected in a small $\gamma_{\pi,r}$. Therefore, lower attention, i.e., a lower $\gamma_{\pi,r}$, implies that the agent updates her expectations to a given signal $s$ less strongly.

Now, since the agent {chooses} the level of attention, we can re-formulate \eqref{eq:pr2} as
\begin{equation}
\max_{\gamma_{\pi,r}\in[0,1]}\left(-\chi\rho_{\pi,r}^2(1-\gamma_{\pi,r})\sigma^2_{\pi,r}-\frac{\frac{1}{\lambda_r}}{2}log\frac{1}{1-\gamma_{\pi,r}}\right), \label{eq:maxi}
\end{equation}
which yields the \textit{optimal} level of attention
\begin{equation}
\gamma_{\pi,r} = max\left\{0,1-\frac{\tfrac{1}{\lambda_r}}{2\chi\rho_{\pi,r}^2\sigma^2_{\pi,r}}\right\}. \label{opt_gamma1}
\end{equation}
Expression \eqref{opt_gamma1} illustrates that attention to inflation depends on the economic environment. I focus on the case in which attention fluctuates between two regimes $r\in\{L,H\}$ depending on the current level of inflation. Later, I will test and confirm that the data indeed prefers this two-regime specification vis-a-vis specifications with multiple thresholds or other forms of changing degrees of attention. Such a regime shift occurs if either (i) the cost of information acquisition and processing decreases as inflation exceeds a certain threshold $\Bar{\pi}$, (ii) the perceived persistence increases, or (iii) when $\sigma_{\pi,r}^2$ increases with the threshold:
\begin{equation*}
     \text{(i) } \frac{1}{\lambda_r} = \begin{cases}
        \frac{1}{\lambda_L}, \text{ if } \pi < \Bar{\pi}\\
        \frac{1}{\lambda_H}, \text{ if } \pi \geq \Bar{\pi} 
    \end{cases}, \text{ or (ii) }     \rho_{\pi,r} = \begin{cases}
        \rho_{\pi,L}, \text{ if } \pi < \Bar{\pi}\\
        \rho_{\pi,H}, \text{ if } \pi \geq \Bar{\pi} 
    \end{cases} 
    , \text{ or (iii) }     \sigma^2_{\pi,r} = \begin{cases}
        \sigma^2_{\pi,L}, \text{ if } \pi < \Bar{\pi}\\
        \sigma^2_{\pi,H}, \text{ if } \pi \geq \Bar{\pi},
    \end{cases}
\end{equation*}
with at least one of the following holding with inequality: $\frac{1}{\lambda_L} \geq \frac{1}{\lambda_H}$, $\rho_{\pi,L} \leq \rho_{\pi,H}$ and $\sigma^2_{\pi,L} \leq \sigma^2_{\pi,H}$. 
Later, I also discuss that all of the three possible drivers of changing degrees of attention are supported in the data, but that given the evidence in \cite{weber2023tell}, channel (iii) is the most likely one and will be the one I focus on in the New Keynesian model later. The expression for optimal attention \eqref{opt_gamma1} has the main purpose of providing testable predictions of what drives attention, and helps guiding the model calibration, but note, that my (empirical and model-implied) results do not depend on the particular analytical expression \eqref{opt_gamma1} or the exact driver of the observed attention changes.

The question arises why to use the level of inflation as the threshold-defining variable. The main reason why I use this specification is that the level of inflation is very salient and easier to detect in real time, in contrast to changes in the persistence or volatility of inflation. Also in the empirical exercises that follow, using proxies for changes in the persistence or volatility of inflation perform worse than the level-threshold specification, mainly because measures of persistence and volatility are noisier and therefore, the model predicts very frequent jumps between regimes and ends up performing quite poorly in terms of the overall fit of the data. Nonetheless, it is important to note that it is not the higher level of inflation \textit{per se} that triggers the increase in attention, but what comes with this higher inflation, such as, higher prior uncertainty. To see this, imagine an economy with an always high but perfectly stable inflation rate. In that case, agents could simply set their prior mean beliefs to the high level of inflation and be completely inattentive, they would still always perfectly forecast inflation in that case and therefore, the agents would have no incentive to be more attentive.

\paragraph{Dynamic model.} To bring the attention model to the data, I now introduce dynamics. 
The agent believes that inflation $\pi$ follows 
\begin{equation}
\pi_{t} = (1-\rho_{\pi,r})\underline{\pi}_r+\rho_{\pi,r}\pi_{t-1}+\nu_t, \notag
\end{equation}
where $\underline{\pi}_r$ is the agent's long-run belief about inflation which may also be regime dependent. I assume that the error term $\nu_t$ is normally distributed with mean zero and variance $\sigma^2_{\nu}$. 
The agent receives a signal about inflation of the form
\begin{equation}
s_{t} = \pi_t + \varepsilon_{t}, \notag
\end{equation}
where the noise $\varepsilon_{t}$ is normally distributed with variance $\sigma^2_{\varepsilon,r}$. As in the static model, higher attention is reflected in less noise, i.e., a lower $\sigma^2_{\varepsilon,r}$.

I assume that agents always use the steady state Kalman filter in each regime $r$, so that optimal updating in a given regime $r$ is given by
\begin{equation}
\Tilde{E}_t\pi_{t+1} = (1-\rho_{\pi,r})\underline{\pi}_r+\rho_{\pi,r}\Tilde{E}_{t-1}\pi_{t} + \rho_{\pi,r}\gamma_{\pi,r}\left(\pi_t - \Tilde{E}_{t-1}\pi_{t}\right)+u_{t}, \label{updating}
\end{equation}
where $\gamma_{\pi,r}$ captures how attentive the agent is.\footnote{Assuming that agents always use the steady state Kalman filter is a standard assumption in the rational inattention literature and basically means that the agent receives all her signals before forming her expectations (see, e.g., \cite{mackowiak2009optimal, mackowiak2018dynamic, mackowiak2023rational}). This leaves conditional second moments time-invariant and thus, the optimal level of attention constant. An important recent paper that relaxes this assumption is \citet{afrouzi2020dynamic}. However, they focus on the linear-quadratic-Gaussian case which in my case is not applicable due to the non-linearity introduced by the attention threshold.} 
Explicitly accounting for the attention threshold $\Bar{\pi}$, agents form their one-period ahead inflation expectations according to:
\begin{equation}
    \tilde{E}_t\pi_{t+1} = \begin{cases}
        (1-\rho_{\pi,L})\underline{\pi}_L+\rho_{\pi,L}\Tilde{E}_{t-1}\pi_{t} + \rho_{\pi,L}\gamma_{\pi,L}\left(\pi_t - \Tilde{E}_{t-1}\pi_{t}\right)+u_{t}, \text{ when } \pi_{t-1} < \Bar{\pi} \\
        (1-\rho_{\pi,H})\underline{\pi}_H+\rho_{\pi,H}\Tilde{E}_{t-1}\pi_{t} + \rho_{\pi,H}\gamma_{\pi,H}\left(\pi_t - \Tilde{E}_{t-1}\pi_{t}\right)+u_{t}, \text{ when } \pi_{t-1} \geq \Bar{\pi}.
    \end{cases} \label{eq:lom_pi}
\end{equation}
Note, that I allow the long-run mean belief, the perceived persistence and attention to differ across regimes.
Later on, I test for multiple thresholds, changes of attention within regime as well as that attention may depend linearly on the level of inflation or inflation volatility. I find however that the data prefers specification \eqref{eq:lom_pi}. Additionally, I will also allow people's inflation expectations to depend on unemployment and their unemployment expectations. My results are robust to this alternative specification.

\subsection{Estimating attention and the attention threshold}\vspace{-0.2cm}

\paragraph{Data.} As my baseline measure of inflation expectations, I rely on the Survey of Consumers from the University of Michigan. I use average and median household inflation expectations for the period 1978M1-2024M5. For the period 2013-2024, I also use individual inflation expectations from the Survey of Consumer Expectations from the New York Fed. Even though I focus on inflation and inflation expectations over one quarter, I use monthly data to increase the number of observations.
{As a robustness check, however, I also consider expectations at quarterly frequency. One advantage of using quarterly observations is that the Survey of Consumers provides mean expectations going back to 1960Q2.} 
Both surveys ask consumers for their price growth expectations one-year ahead: $\Tilde{E}_t\pi_{t+12}$.
I transform them into one-quarter-ahead forecasts as follows:
$\Tilde{E}_t\pi_{t+3} \equiv\frac{\Tilde{E}_t\pi_{t+12}}{4}$. Using $\Tilde{E}_t\pi_{t+3} \equiv \left[\left(1+{\Tilde{E}_t\pi_{t+12}}\right)^{1/4}-1\right]$ instead does not change the results. 
Computing one-quarter-ahead expectations allows me to compare the results directly to the model which is calibrated at quarterly frequency.  For actual inflation, I use the monthly CPI inflation rate from the FRED database and to be consistent with the model, I focus on quarter-on-quarter inflation: $\pi_{t} \equiv \frac{P_t-P_{t-3}}{P_{t-3}}$. As I show in Appendix \ref{ap:data}, the results are qualitatively similar when using one-year-ahead expectations and year-on-year inflation as the actual measure of inflation. I also discuss heterogeneity in attention across demographic groups in Appendix \ref{ap:data}.

\paragraph{Estimating attention and the attention threshold.}
In order to estimate the two attention levels $\gamma_{\pi,r}$ for $r \in \{L,H\}$, as well as the attention threshold $\Bar{\pi}$, I estimate the following threshold regression:
\begin{align}
\begin{split}
\tilde{E}_t\pi_{t+3} &= \mathds{1}_{\pi_{t-1} \leq \Bar{\pi}}\left(\beta_{0,L} + \beta_{1,L} \tilde{E}_{t-3}\pi_{t} +\beta_{2,L} \left(\pi_t - \tilde{E}_{t-3}\pi_{t}\right)\right)\\& +(1-\mathds{1}_{\pi_{t-1} \leq \Bar{\pi}})\left(\beta_{0,H} + \beta_{1,H} \tilde{E}_{t-3}\pi_{t} +\beta_{2,H} \left(\pi_t - \tilde{E}_{t-3}\pi_{t}\right)\right)+\epsilon_{t}, \label{reg1_thresh}
\end{split}
\end{align}
where $\beta_{0,r} = (1-\rho_{\pi,r})\underline{\pi}_r$ denotes the intercept in regime $r\in\{L,H\}$, $\beta_{1,r} = \rho_{\pi,r}$ the perceived persistence, and $\tfrac{\beta_{2,r}}{\beta_{1,r}} = \gamma_{\pi,r}$ the attention level in regime $r$. I therefore not only allow the attention parameters to differ across regimes, but also the perceived persistence of inflation as well as agents' long-run beliefs about inflation. $\mathds{1}_{\pi_{t-1} \leq \Bar{\pi}}$ is the indicator function that equals one when inflation in the previous month $\pi_{t-1}$ was below the threshold $\Bar{\pi}$ and zero otherwise.
The threshold value is then estimated jointly with the regression coefficients by minimizing the sum of squared residuals obtained for all possible thresholds (see, e.g., \cite{gonzalo2002estimation} and  \cite{hansen2011threshold}). Note, that I do not impose that the attention level in the regime in which inflation is above the threshold needs to be higher than in the regime with inflation below the threshold.

\paragraph{Estimation results.} Table \ref{tab:empirics} shows the estimation results. For the baseline specification, I estimate an attention threshold of $\hat{\Bar{\pi}} = 3.91\%$. Attention in the regime in which inflation is below this threshold is $\widehat{\gamma}_{\pi,L} = 0.18$, and attention in the high-inflation regime is equal to $\widehat{\gamma}_{\pi,H} = 0.35$. Thus, attention in the high-inflation regime is higher than in the low-inflation regime, and I therefore refer to the regime with inflation above the threshold as the \textit{high-attention regime}. The threshold value of 3.91\% implies that the U.S.\ economy spent about 30\% of the time between 1978 and 2024 in the high-attention regime. Most recently, inflation exceeded this threshold in early 2021. Google searches also started to increase around that time (see panel (a) in Figure \ref{fig:moti} in the Introduction). 

To test for multiple threshold, I use the Bayesian Information Criterion to select the numbers of thresholds, and I find that the data prefers the specification with only one threshold.\footnote{For example, the BIC increases from -534.12 to -524.70 when going from one to two thresholds and to -512.03 when going to three thresholds. The BIC is given by $T \ln{(SSR/T)} + k\ln{(T)}$, where $k$ is the number of parameters, $T$ the number of periods, and $SSR$ the sum of squared residuals. As a robustness check, I also consider the Hannan-Quinn IC and find that also the HQIC selects the model with one threshold only.}

\begin{table}[h]
 \caption{Estimated attention levels and the attention threshold}\label{tab:empirics}
 \centering
\begin{tabular}{lcccc}
\hline \hline
& Threshold $\hat{\Bar{\pi}}$ & Low Att.\ $\widehat{\gamma}_{\pi,L}$ & High Att.\ $\widehat{\gamma}_{\pi,H}$ & $p$-val.\ $H_0: \gamma_{\pi,L} = \gamma_{\pi,H}$\\\hline\vspace{-0.4cm}\\
Mean expectations & 3.91\% & 0.18 &  0.35  & 0.000 \\\vspace{-0.5cm}\\
s.e.\ &  & (0.018) & (0.042) \\\vspace{-0.5cm}\\
Median expectations & 4.44\% & 0.13 & 0.21 & 0.014\\\vspace{-0.5cm}\\
s.e.\ &  & (0.018) & (0.027) \\ 
Quarterly frequency & 3.21\% & 0.14 & 0.38 & 0.000 \\
s.e.\ & & (0.033) & (0.076) 
\\\vspace{-0.4cm}\\
\hline \hline
\end{tabular}\\\vspace{.1cm}
\begin{minipage}{1\textwidth}
\footnotesize{Notes: This table shows the results from regression \eqref{reg1_thresh}, where $\hat{\Bar{\pi}}$ denotes the estimated threshold, $\widehat{\gamma}_{\pi,L}$ and $\widehat{\gamma}_{\pi,H}$ the estimated attention levels when inflation is below or above the threshold, respectively. The last column shows the $p$-value for the null hypothesis that the two attention levels are equal. Standard errors are robust with respect to heteroskedasticity.
}%
 \end{minipage}
\end{table}

The results for median expectations are similar. Attention in both regimes tends to be somewhat lower when using median expectations and the attention threshold higher. Median expectations tend to be less volatile than average expectations, therefore, it is not surprising that the estimated attention levels---capturing the updating gains---are lower for median expectations. Nevertheless, in both cases, the null hypothesis that the two attention levels across regimes are equal is rejected at most conventional significance levels (with $p$-values of 0.000 and 0.014, respectively, see last column). Again, I find that the Bayesian information criterion selects having one threshold only.

The last two rows in Table \ref{tab:empirics} show the results when using observations at quarterly frequency for the period 1960Q2-2023Q2. We see that the estimated threshold is somewhat lower at 3.21\%. The estimated attention levels within regime are very similar to the baseline monthly specification and again, the difference in attention across regimes is highly statistically significant.

When I use average inflation of the last three months as the threshold-defining variable instead of the last month's inflation rate, I estimate a threshold at 3.45\% and attention levels of $\widehat{\gamma}_{\pi,L} = 0.19$ and $\widehat{\gamma}_{\pi,H} = 0.33$.

A drawback of the Survey of Consumers is that it surveys households at most twice. Therefore, it is not possible to examine how individual households update their expectations and how they change their updating behavior with the level of inflation. Focusing on mean (and median) expectations helps to overcome this issue but it also ignores potentially important cross-sectional information. 
Therefore, I also estimate the attention levels using individual consumer inflation expectations from the Survey of Consumer Expectations. When last month's inflation was below the threshold of 3.91\%, I estimate an attention level of 0.08. When inflation was above the threshold, attention increases to 0.32, and I reject the nullhypothesis that the two estimates are equal ($p$-value of 0.000). Thus, I obtain similar, though slightly lower, attention parameters than in my baseline estimates even though this survey is only available since 2013.

When using the firm expectations from the Survey of Firms' Inflation Expectations (SoFIE) from the Cleveland Fed---with the caveat that that data is only available at quarterly frequency and only since 2018Q2, resulting in only 23 observations---I estimate that attention of firms has increased from practically 0 up to 0.48 as inflation exceeded the threshold of 3.91\%. 

I also consider mean expectations from the Survey of Professional Forecasters from the Philadelphia Fed. CPI inflation expectations from that survey are available at quarterly frequency since 1981Q3. I estimate an attention threshold at 3.92\% and thus, very close to the one for households. The estimated attention levels are $\widehat{\gamma}_{\pi,L} = 0.07$ and $\widehat{\gamma}_{\pi,H} = 0.17$ (with an associated $p$-value that the two are equal of 0.008).

\paragraph{A multivariate system of inflation expectations.}
As the perceived law of motion of inflation is an AR(1) process, inflation expectations under limited attention only depend on prior beliefs and inflation itself. In principle, however, it could be possible that inflation expectations also depend on other variables such as unemployment.\footnote{\cite{kamdar2024attention}, for example, show that consumers tend to expect disinflation when expecting an economic expansion.} To take this into account, I therefore now estimate the regression
\begin{align}
\tilde{E}_t\pi_{t+1} &= \mathds{1}_{\pi_{t-1} \leq {\Bar{\pi}}}\Bigg[\beta_{0,{L}} + \beta_{1,{L}} \tilde{E}_{t-1}\pi_{t} +\beta_{2,{L}} \left(\pi_t - \tilde{E}_{t-1}\pi_{t}\right) \\& +(1-\mathds{1}_{\pi_{t-1} \leq {\Bar{\pi}}})\Bigg[\beta_{0,{H}} + \beta_{1,{H}} \tilde{E}_{t-1}\pi_{t} +\beta_{2,{H}} \left(\pi_t - \tilde{E}_{t-1}\pi_{t}\right)\Bigg]\\
&+\beta_{3}\tilde{E}_{t-1} U_t + \beta_{4}\left( U_t - \tilde{E}_{t-1} U_{t}\right)+\tilde{\epsilon}_{t},
\end{align}
where $\tilde{E}_{t-1} U_t$ denote expectations about unemployment changes and $U_t$ are unemployment changes (see Appendix \ref{app:unemp} for further details).
I find that both, $\beta_3$ and $\beta_4$, are insignificant ($\hat{\beta}_3=-0.13$ and $\hat{\beta}_4=0.04$ with $p$-values of 0.332 and 0.337, respectively). The estimated attention-to-inflation parameters are also barely affected. In fact, I estimate attention in the low regime to be equal to 0.18 and 0.34 in the high regime.

\paragraph{Regional variation.} The Survey of Consumers also provides the regions consumers are in, which allows me to leverage regional variation to test for the robustness of my results. In particular, I use the four census regions West, North East, North Central (or Midwest) and South. I control for region fixed effects and estimate that attention is 0.22 with a standard error of 0.023 (standard errors are robust with respect to heteroskedasticity as well as serial and cross-sectional correlation) when last-period's inflation is below the threshold of 3.91\%. Attention increases to 0.42 (s.e.\ of 0.051) when inflation exceeds the threshold. The two estimates are highly statistically significantly different from each other ($p$-value of 0.000). These results indicate that my baseline results remain robust when considering regional variation.

\paragraph{Markov-switching model.} Instead of estimating regression \eqref{reg1_thresh}, I also estimate a Markov-switching dynamic regression model in which the long-run inflation beliefs $\underline{\pi}$, the perceived persistence $\rho_{\pi}$ and attention to inflation $\gamma_{\pi}$ are allowed to differ across unobserved states. That implies that I do not need to take a stand on what defines the regimes, in contrast to equation \eqref{reg1_thresh} where I assumed that it is last period's inflation rate that determines the regimes (see Chapter 22 of \cite{hamilton1994time} for a textbook treatment).

The implied attention estimates are 0.19 and 0.33 for the two states and thus, close to the estimates from regression \eqref{reg1_thresh} of 0.18 and 0.35. When I allow the variance to change with the state, I estimate attention parameters of 0.2 and 0.32. The implied probability of staying in the low-attention regime from the Markov-switching model is 70\% which aligns well with the fact that annualized quarter-on-quarter CPI inflation was below the threshold 70\% of the time.

\paragraph{Smooth changes in attention.} My baseline regression imposes that attention to inflation takes on only two values and jumps from one to the other when inflation exceeds the threshold or falls back below it. Another possibility is that attention changes smoothly with inflation or inflation volatility. To test for this, I estimate the two alternative regressions:
\begin{align}
    &\widetilde{E}_t\pi_{t+3} = \beta_0 + \beta_1 \widetilde{E}_{t-3}\pi_t + \beta_2\left(\pi_t - \widetilde{E}_{t-3}\pi_t\right) + \beta_3\pi_{t-1}\left(\pi_t - \widetilde{E}_{t-3}\pi_t\right) + \varepsilon_t \label{lin1}\\
    &\widetilde{E}_t\pi_{t+3} = \delta_0 + \delta_1 \widetilde{E}_{t-3}\pi_t + \delta_2\left(\pi_t - \widetilde{E}_{t-3}\pi_t\right) + \delta_3\left(\pi_{t-1}-\pi_{t-2}\right)^2\left(\pi_t - \widetilde{E}_{t-3}\pi_t\right) + \varepsilon_t,\label{lin2}
\end{align}
where $\beta_3$ captures that attention may change with the past level of inflation and $\delta_3$ that it may change with inflation volatility, here measured as the squared change in past inflation. 

Again, I use the Bayesian Information Criterion to select which model fits the data best and I find that the data prefers the threshold specification \eqref{reg1_thresh} over the two specifications \eqref{lin1} and \eqref{lin2}. Additionally, even though the estimated $\beta_3$ and $\delta_3$ are statistically significantly different from 0, they are economically close to 0 (0.009 and -0.002, respectively). In the following, I also test whether a combination of such a 'smooth-attention model' with my threshold model would perform better.

\paragraph{Attention changes within regime.} In Table \ref{tab:empirics}, we saw that attention increases when inflation exceeds the threshold of 3.91\%. But what about changes within regime? To look at this, I estimate a time series of attention. In particular, I estimate the regression
\begin{equation}
\tilde{E}_t\pi_{t+3} = \beta_0 + \beta_1 \tilde{E}_{t-3}\pi_{t} +\beta_2 \left(\pi_t - \tilde{E}_{t-3}\pi_{t}\right)+\epsilon_{t}, \notag
\end{equation}
using a rolling-windows approach, where each window has a length of two or five years, respectively.\footnote{To be consistent with the theory, I impose that attention is between 0 and 1, see expression \eqref{opt_gamma1}. In the case of five-year windows, all estimated attention parameters are between 0 and 1 without having to impose it.} I denote the estimated time series of attention parameters by $\widehat{\gamma}_{\pi,t}$, and I compute the window-specific average of the monthly quarter-on-quarter inflation rate, ${\pi}^{avg}_t$. To then test whether attention within regime is higher when inflation is higher, I estimate the following regression
\begin{equation}
\widehat{\gamma}_{\pi,t} = \delta_0 + \delta_1 \mathds{1}_{{\pi}^{avg}_t \geq 3.91} + \delta_2 \pi_{t-1} + \delta_3 \mathds{1}_{{\pi}^{avg}_t \geq 3.91}\pi_{t-1}+\varepsilon_t, \label{eq:att_within_r}
\end{equation}
where $\mathds{1}_{{\pi}^{avg}_t \geq 3.91}$ is an indicator that equals one when in period $t$ average inflation over the last twelve months is above 3.91\% and zero otherwise. Thus, $\delta_1$ tells us the difference of attention across regimes, $\delta_2$ the effect of last-periods inflation on attention and $\delta_3$ the additional effect of inflation on attention in the high-inflation regime.

Table \ref{tab:attention_within_r} shows the results. We see that attention is significantly higher in the high-inflation regime, as indicated by the estimate of $\delta_1$. Yet, inflation does not have any additional significant effect on attention when accounting for the threshold, as depicted by the last two columns. This holds regardless of whether we use two-year or five-year windows.

\begin{table}[h]
 \caption{Attention changes within regime}\label{tab:attention_within_r}
 \centering
\begin{tabular}{lccc}
\hline \hline\vspace{-0.4cm}\\
& $\widehat{\delta}_1$ & $\widehat{\delta}_2$ & $\widehat{\delta}_3$ \\\hline\vspace{-0.4cm}\\
2-year windows & $0.193^{***}$   & -0.002 & 0.022 \\\vspace{-0.5cm}\\
s.e.\ & (0.067) & (0.005) & (0.032)\\\vspace{-0.5cm}\\
5-year windows  & $0.119^{***}$ & -0.004 & 0.001    \\\vspace{-0.5cm}\\
s.e.\ & (0.030) & (0.003) & (0.005) 
\\\vspace{-0.4cm}\\
\hline \hline
\end{tabular}\\\vspace{.1cm}
\begin{minipage}{1\textwidth}
\footnotesize{Notes: This table shows the results from regression \eqref{eq:att_within_r} when using two- or five-year windows. Standard errors are robust with respect to heteroskedasticity and serial correlaton (\cite{newey1987simple} with 12 lags). Significance levels: $^{*}$: $p$-value $<$ 0.1, $^{**}$: $p$-value $<$ 0.05, $^{***}$: $p$-value $<$ 0.01.
}%
 \end{minipage}
\end{table}

An alternative way to test for changes of attention in addition to the jumps across regimes is to estimate the regression
\begin{align*}
\tilde{E}_t\pi_{t+1} = &\mathds{1}_{\pi_{t-1} \leq {\Bar{\pi}}}\left(\beta_{0,{L}} + \beta_{1,{L}} \tilde{E}_{t-1}\pi_{t} +\beta_{2,{L}} \left(\pi_t - \tilde{E}_{t-1}\pi_{t}\right)\right) \\
&+(1-\mathds{1}_{\pi_{t-1} \leq {\Bar{\pi}}})\left(\beta_{0,{H}} + \beta_{1,{H}} \tilde{E}_{t-1}\pi_{t} +\beta_{2,{H}} \left(\pi_t - \tilde{E}_{t-1}\pi_{t}\right)\right)  \\& {+\beta_3 \left(\pi_t - \tilde{E}_{t-1}\pi_{t}\right)\cdot \pi_{t-1} + \beta_4\tilde{E}_{t-1}\left[\pi_{t}\right]\cdot \pi_{t-1} }+\tilde{\epsilon}_{t},
\end{align*}
where $\beta_3$ and $\beta_4$ capture that the perceived persistence and the attention parameter may depend on last-period's inflation in addition to the changes across regimes. I estimate $\hat{\beta}_3 = 0.0037$ (with a standard error of 0.0033 and a p-value of 0.26), and $\hat{\beta}_4 = -0.0031$ (with a standard error of 0.0057 and a p-value of 0.60). When ignoring the effects coming from $\beta_3$ and $\beta_4$, the attention parameters are $\hat{\gamma}_{\pi,L} =\tfrac{\widehat{\beta}_{2,L}}{\widehat{\beta}_{1,L}} = 0.20$ and $\hat{\gamma}_{\pi,H} =\tfrac{\widehat{\beta}_{2,H}}{\widehat{\beta}_{1,H}} = 0.31$. When accounting for the changes in attention through $\beta_3$ and $\beta_4$ the implied attention parameters are $\hat{\gamma}_{\pi,L} = \tfrac{\widehat{\beta}_{2,L} + \hat{\beta}_3\cdot 2}{\widehat{\beta}_{1,L}+ \hat{\beta}_4\cdot 2} = 0.21$ at $\pi_{t-1} = 2\%$ and $\hat{\gamma}_{\pi,H} = \tfrac{\widehat{\beta}_{2,H} + \hat{\beta}_3\cdot 5}{\widehat{\beta}_{1,H}+ \hat{\beta}_4\cdot 5}= 0.34$ at $\pi_{t-1} = 5\%$. Overall, these estimated attention parameters are quantitatively close to the ones from my baseline estimation of 0.18 and 0.35. As furthermore the interaction terms are insignificant, these findings suggest that the threshold model of attention captures the data well.

\paragraph{What drives attention?}

Expression \eqref{opt_gamma1} predicts that attention to inflation is higher when the cost of information is lower or when the perceived persistence or prior uncertainty are higher. Consistent with the first one---that the cost of information is lower---I show in Appendix \ref{app:news} that news coverage of inflation is substantially higher in times of higher inflation. Thus, information about inflation is more prevalent in times of higher inflation, which lowers the cost of acquiring information about inflation, leading to an increase in people's attention.\footnote{Consistent with this, \cite{bracha2023inflation} show that higher inflation rates indeed lead to more media reporting about inflation, and \cite{lamla2014role} find that more intensive news reporting about inflation improves the accuracy of consumers' inflation expectations, consistent with agents being more attentive. \cite{schmidtinflation} show that during episodes of intensive newspaper coverage of inflation, news reporting has strong effects on inflation expectations but not during other episodes. \cite{larsen2021news} find that news media coverage predicts households’ inflation expectations, and \cite{nimark2019news} show that major events (such as strong inflation increases) lead to a shift in the news focus towards these events. } \cite{korenok2022inflation} also consider news coverage of inflation and estimate a threshold between 3.77-3.94\%, which is very close to the threshold I estimate. They also estimate attention directly, using Google search data, and estimate for the U.S.\ that once inflation exceeds the threshold of 3.55\% that attention increases with inflation. One reason why I may find a slightly higher threshold could be that people might start to google for information about inflation at lower levels of inflation already, but only start incorporating that information in their expectations once inflation further increases. Additionally, I focus on the time starting in 1978, whereas their sample is restricted to 2004-2022 because the Google search data is not available for the years before 2004.\footnote{\cite{fasani2024nonlinearities} estimate a threshold VAR to identify shocks to long-horizon inflation expectations (shocks to the Fed's inflation target). They also identify two inflation regimes and estimate the median threshold to be at an inflation rate of 3.9\%, very similar to my findings.} One caveat of this channel, however, is that a decrease in the cost of information should reduce people's posterior uncertainty in times of high inflation, which, however, \cite{weber2023tell} show is not the case in the data.

Not only is news coverage of inflation higher in times of higher inflation, the perceived persistence, $\rho_{\pi,r}$, also tends to be higher when inflation is above the threshold. In my baseline estimation, for example, the estimated $\beta_1$ increases from 0.78 to 0.85 when inflation exceeds the threshold. \cite{weber2023tell} also find that the perceived inflation persistence has increased with inflation. They also show, however, that in their RCTs treatment effects when providing information about future inflation should not change when the increase in attention is solely driven by changes in $\rho_{\pi}$. The empirical findings, however, show that they do change, thus, suggesting that the perceived persistence is not the key driver of attention. 

The last possibility, that $\sigma^2_{\pi}$ is higher when inflation is above the threshold, is supported by the data. For example, I find that quarter-on-quarter inflation volatility in my sample increases from 2.04\% (annualized) to 3.21\%, and the volatility of inflation expectations from 0.78\% to 2.42\%. Consistent with that, \cite{weber2023tell} also arrive at the conclusion that changes in $\sigma_{\pi}^2$ are a likely driver of changes in people's attention to inflation across inflation regimes.

\section{Inflation Dynamics Across Attention Regimes}\label{sec:empirical_irfs}\vspace{-0.2cm}
In this section, I show that negative supply shocks become more inflationary when attention is high, and that the increase in attention in early 2021 roughly doubled the inflationary effects of supply shocks in the subsequent inflation surge.

\paragraph{Supply shocks.}
As my empirical measure of supply shocks, I use the oil supply news shocks from \cite{kanzig2021macroeconomic} for the period 1975M1-2023M12. In a first step, oil surprises are identified by using institutional features of the Organization of the Petroleum Exporting Countries (OPEC) and high-frequency data on variation in oil futures prices around OPEC announcements. In a second step, the resulting surprises are then used as an external instrument in an oil VAR, to identify a structural oil supply news shock. In the following, I show the responses to a negative shock that pushes up oil prices and lowers oil production. I refer to these shocks as oil news shocks, cost-push shocks or supply shocks interchangeably. 

Later on, I show that the following results also hold when focusing on other shocks. Given that the oil supply news shocks cover most of the recent inflation surge---while the other shocks do not---I use the oil supply news shocks as my baseline.

\paragraph{Attention regimes.}
Given the results in Section \ref{sec:data}, I define the high-attention regime to be periods in which inflation in the past month exceeded 3.91\%. As an alternative indicator for the attention regimes, I also use Google Trends data, with the drawback that this data is only available since 2004. However, it will allow me to disentangle attention from other changes in the economy that tend to arise when inflation is higher (e.g., changes in the frequency of price adjustments, see, e.g., \cite{blanco2024inflation}). The Google data is normalized such that the month with the most Google searches of the word ``inflation'' is assigned a value of 100 and all the other months are expressed relative to that month. I assign months to the high-attention regime when Google searches of inflation in that month exceed the $75^{th}$ percentile.

\begin{table}[h]
 \caption{Shock properties across regimes}\label{tab:shock_stats}
 \centering
\begin{tabular}{lccc}
\hline \hline
Regime & Mean & Standard deviation  \\ 
\hline 
\underline{Inflation as regime-defining variable} \\
High  & -0.029 & 0.576   \\
Low &  0.015 & 0.567  \\
\underline{Google Trends as regime-defining variable} \\
High  & -0.011 & 0.786   \\
Low &  0.006 & 0.610 
\\\vspace{-0.4cm}\\
\hline \hline
\end{tabular}\\\vspace{.1cm}
\begin{minipage}{1\textwidth}
\footnotesize{Notes: This table shows the mean and the standard deviation for the oil supply news shocks across the two attention regimes. The upper part of the table shows the case where the previous month's inflation is the threshold-defining variable and the lower part the case where Google Trends are used to identify the two regimes.
}%
 \end{minipage}
\end{table}

Table \ref{tab:shock_stats} lists the mean and the standard deviation of the shocks for the high-attention and low-attention regime, separately. The upper part of the table shows these statistics for the case in which the regimes are defined based on whether the previous month's inflation rate was below or above the 3.91\% threshold, and the lower part of the table for the case where the regimes are defined based on Google Trends data. We see that in both cases, the shock series have similar first and second moments across regimes. When using Google Trends as the regime-defining variable, the volatility of the supply shocks is slightly higher. In the model calibration, I will exploit this as one potential explanation for the increase in people's attention. That said, however, the following empirical results remain robust even when excluding shocks that are larger than one standard deviation. Thus, the following results are unlikely to be driven by differences in the shock series across regimes.\footnote{Excluding large shocks also helps to address the issues that may arise with state-dependent local projections when the state is endogenous \citep{gonccalves2024state}.}

\paragraph{The role of the attention regime for the propagation of supply shocks to inflation.}
To examine whether the effects of negative supply shocks on inflation differ across regimes, I estimate the local projection \citep{jorda2005estimation}:
\begin{equation}
    y_{t+j} = \mathds{1}_{H}\left(\alpha_{j}^H + \beta^H_j\varepsilon_t \right) + \left(1 - \mathds{1}_{H} \right) \left(\alpha_{j}^L + \beta^L_j\varepsilon_t \right) + \Gamma' X_{t} + u_{t+j}, \label{eq:baseline_lp}
\end{equation}
where $y_{t+j}$ denotes inflation (or inflation expectations) at time $t+j$, $\mathds{1}_{H}$ is an indicator function that equals one when the economy is in the high-attention regime at the time of the shock and 0 else, $\varepsilon_t$ denotes the shock at time $t$ and $X_t$ are controls. In my baseline estimation, I use four lags of the shock, four lags of the unemployment rate, of inflation and of inflation expectations as controls. In Appendix \ref{app:robustness_irfs}, I show that the results remain robust when using other controls.

\begin{figure}[ht]
\caption{Inflation response to an oil supply news shock}
\label{fig:emp_irfs_inflation1}\vspace{0.15cm} \centering%
\begin{tabular}{cc}
(a) High-Attention Regime & (b) Low-Attention Regime \\ 
\includegraphics[width=.44\textwidth]{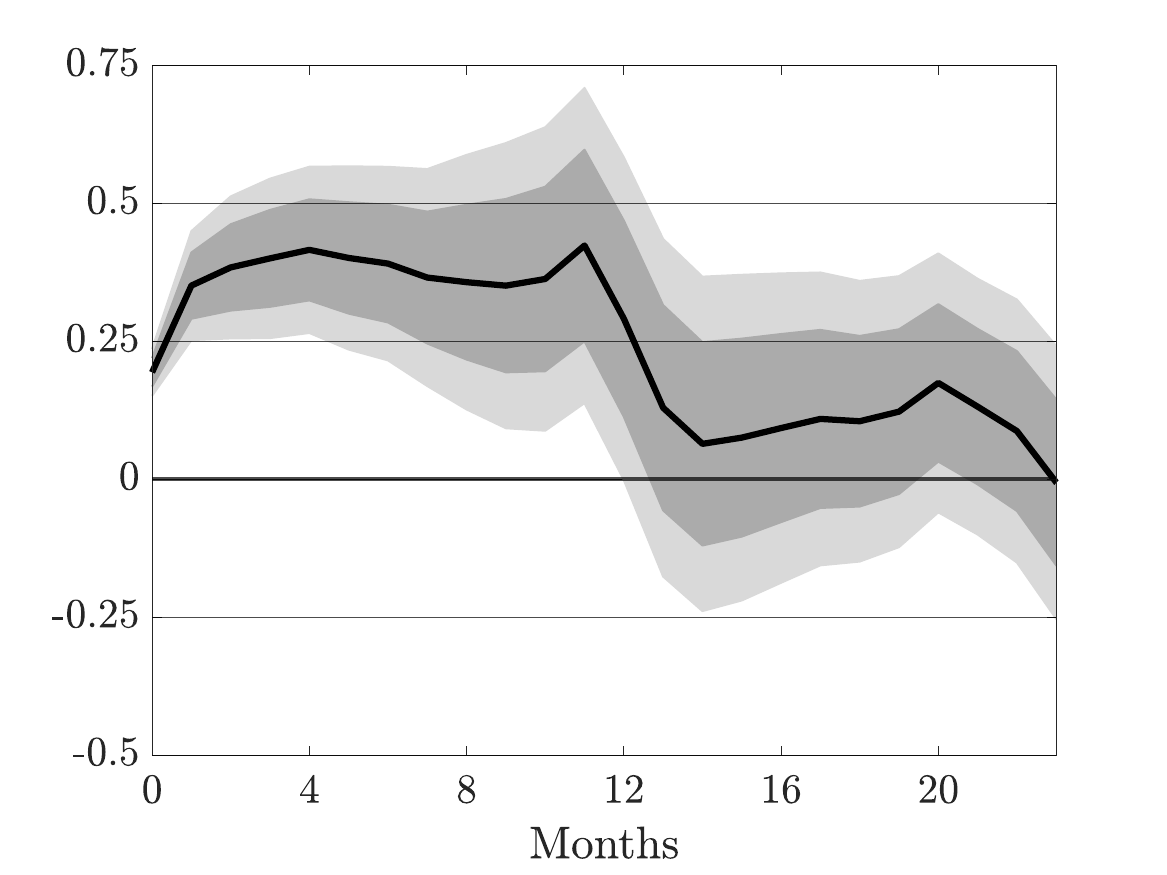} &    \includegraphics[width=.44\textwidth]{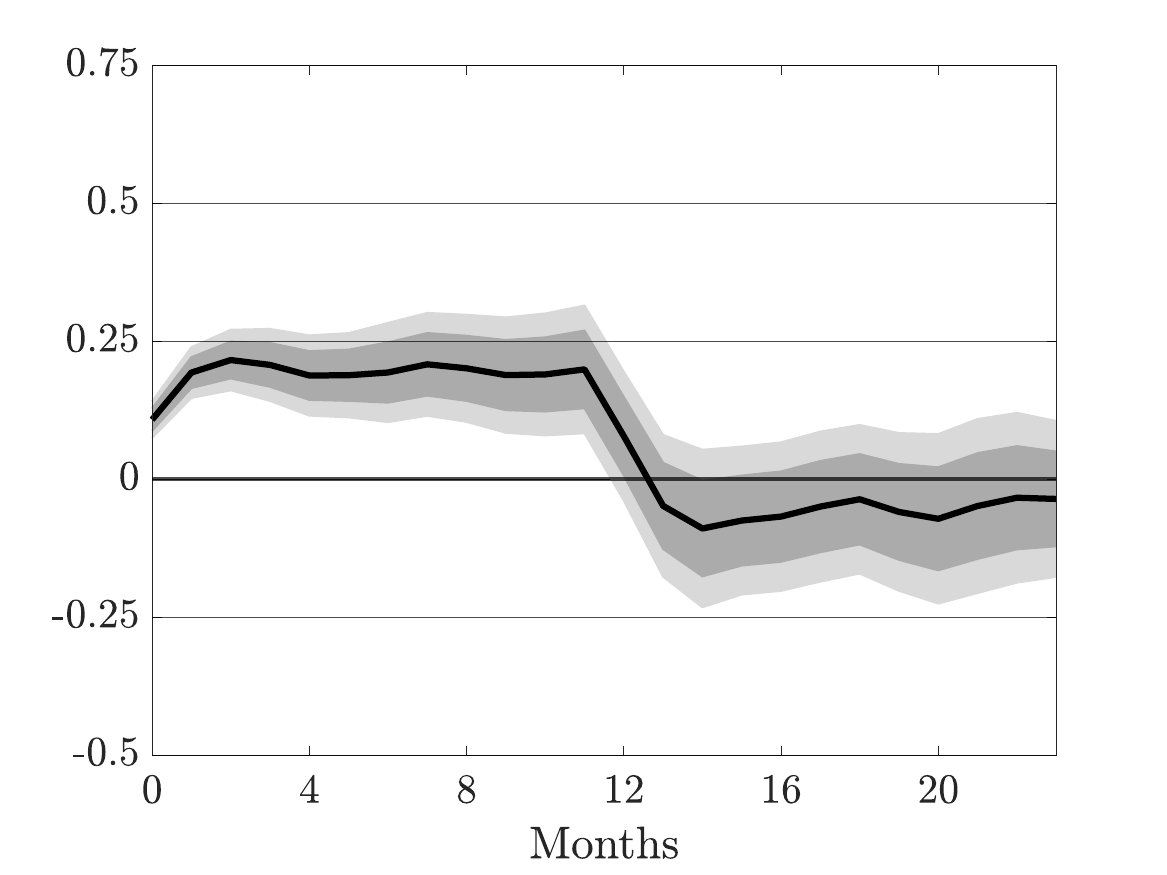} \\
(c) Average effect   & (d) Difference \\\includegraphics[width=.44\textwidth]{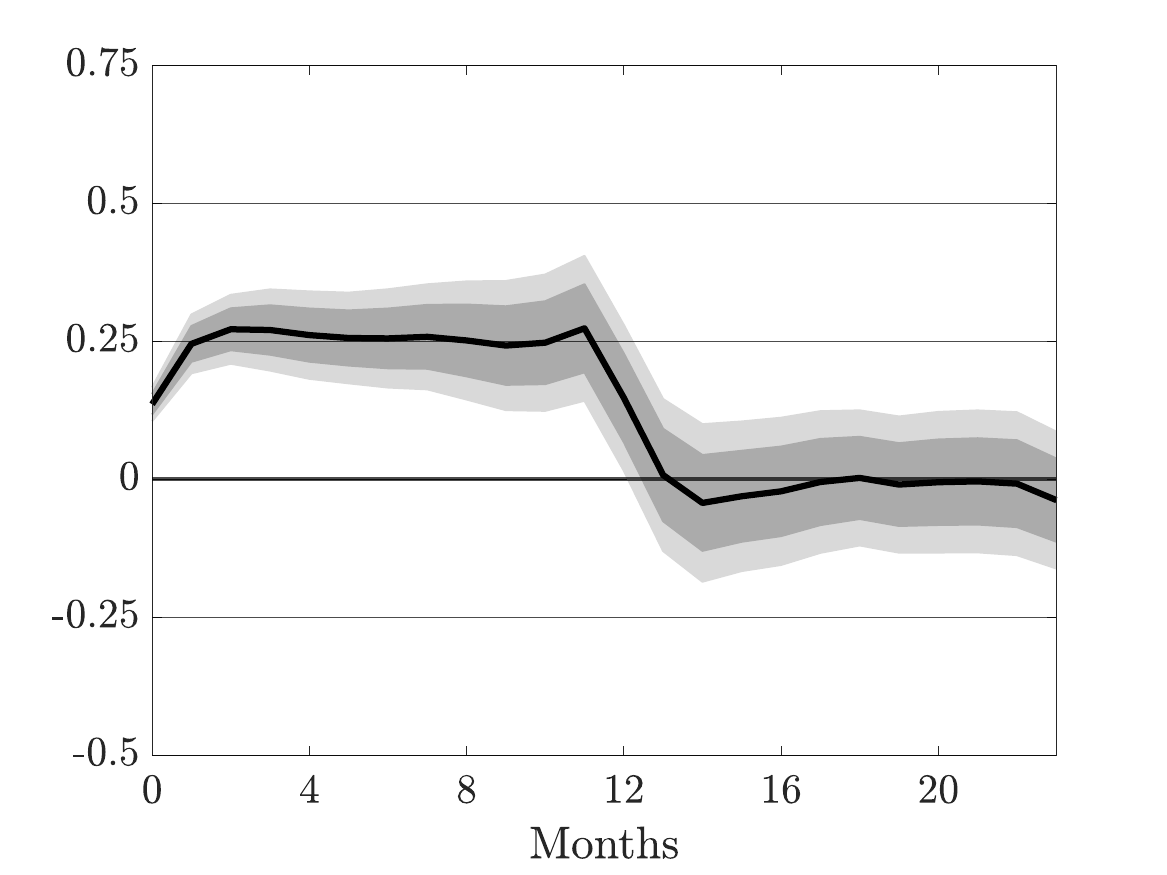} & \includegraphics[width=.44\textwidth]{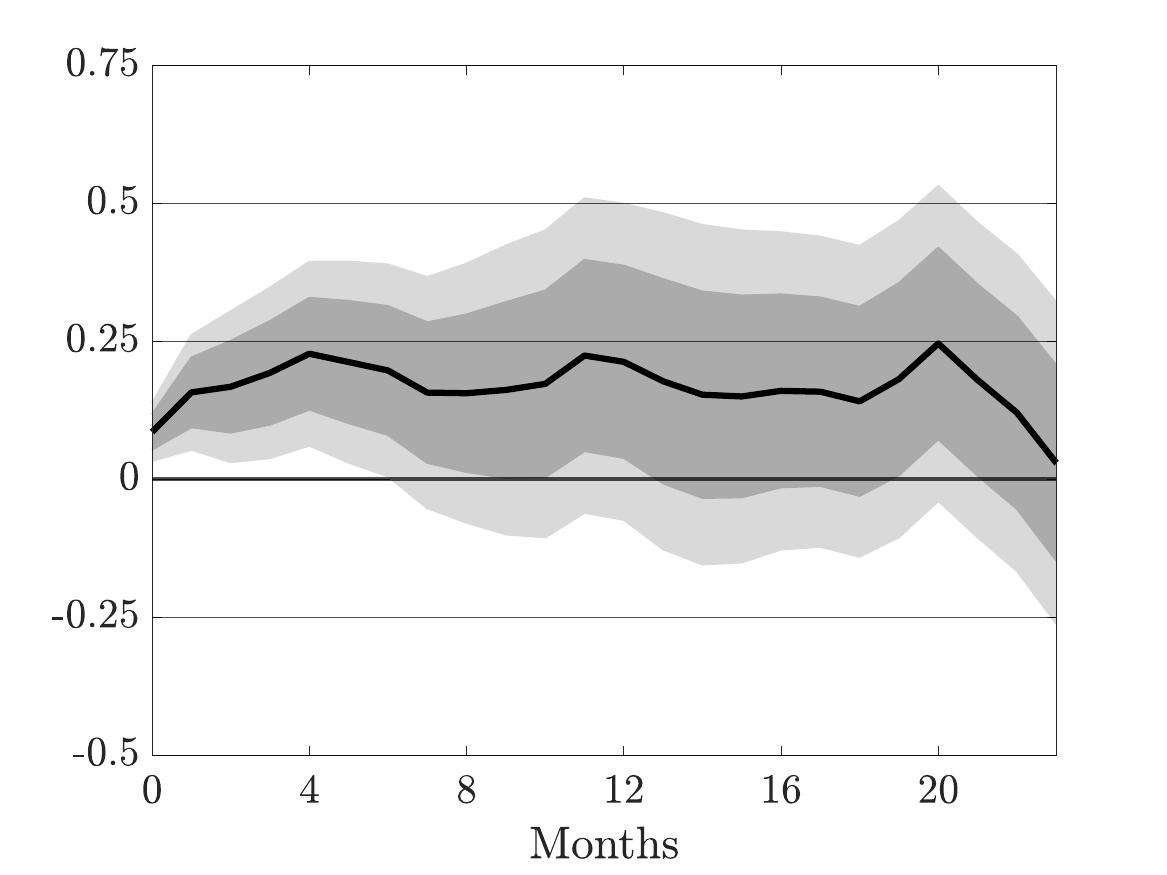}
\end{tabular}%

\begin{minipage}{1\textwidth}
\footnotesize{\emph{Notes}: This figure shows the inflation response to an oil supply news shock (equation \eqref{eq:baseline_lp}) in the high-attention regime (panel (a)), the low-attention regime (panel (b)), on average across regimes (panel (c)), and the difference between the two regimes (panel (d)). The dark shaded areas depict the 68\% confidence bands and the light-shaded area the 90\% confidence bands. Standard errors are robust with respect to serial correlation and heteroskedasticity (\cite{newey1987simple} with 12 lags). 
}%
 \end{minipage}
\end{figure}

Figure \ref{fig:emp_irfs_inflation1} shows the estimation results of regression \eqref{eq:baseline_lp}. Panel (a) depicts the inflation response to a negative oil news shock in the high-attention regime, panel (b) in the low-attention regime, panel (c) shows the average effect, and panel (d) the difference between the effects in the high-attention regime and the low-attention regime. In all cases, I consider a one-standard deviation shock. The dark shaded areas depict the 68\% confidence bands and the light-shaded areas the 90\% confidence bands. Standard errors are robust with respect to serial correlation and heteroskedasticity (\cite{newey1987simple} with 12 lags). 

Inflation increases about twice as much in the high-attention regime compared to the low-attention regime. These differences are quite persistent and statistically significant at the 10\% significance level in the first six months. Thus, when the economy is hit by a one-standard deviation supply shock when people's attention to inflation is high, inflation increases on average by about 40 basis points. In contrast, when attention to inflation is low at the time of the shock, inflation only increases by about 20 basis points. Figure \ref{fig:emp_irfs_inflation_google} in Appendix \ref{app:robustness_irfs} shows that the results are very similar when using Google Trends as the regime-defining variable.

\paragraph{Regional variation.}
As in Section \ref{sec:data}, I now also use regional variation of the four consensus regions. In particular, I use region-specific CPI inflation as the dependent variable. Additionally, I control for region fixed effects as well as four lags of region-specific inflation. 

\begin{figure}[!ht]
\caption{Inflation response to an oil supply shock using regional variation}
\label{fig:emp_irfs_inflation_region}\vspace{0.15cm} \centering%
\begin{tabular}{cc}
(a) High-Attention Regime  & (b) Low-Attention Regime  \\ 
\includegraphics[width=.44\textwidth]{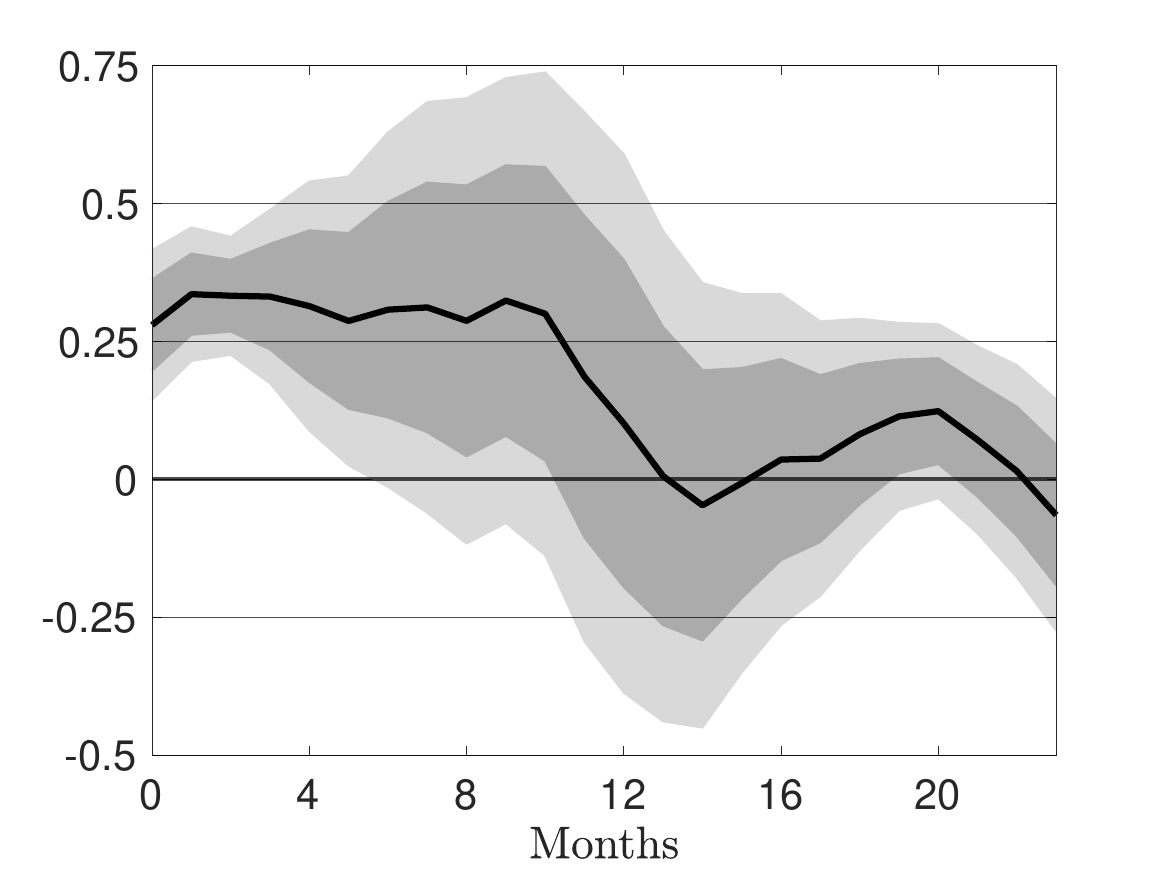} &    \includegraphics[width=.44\textwidth]{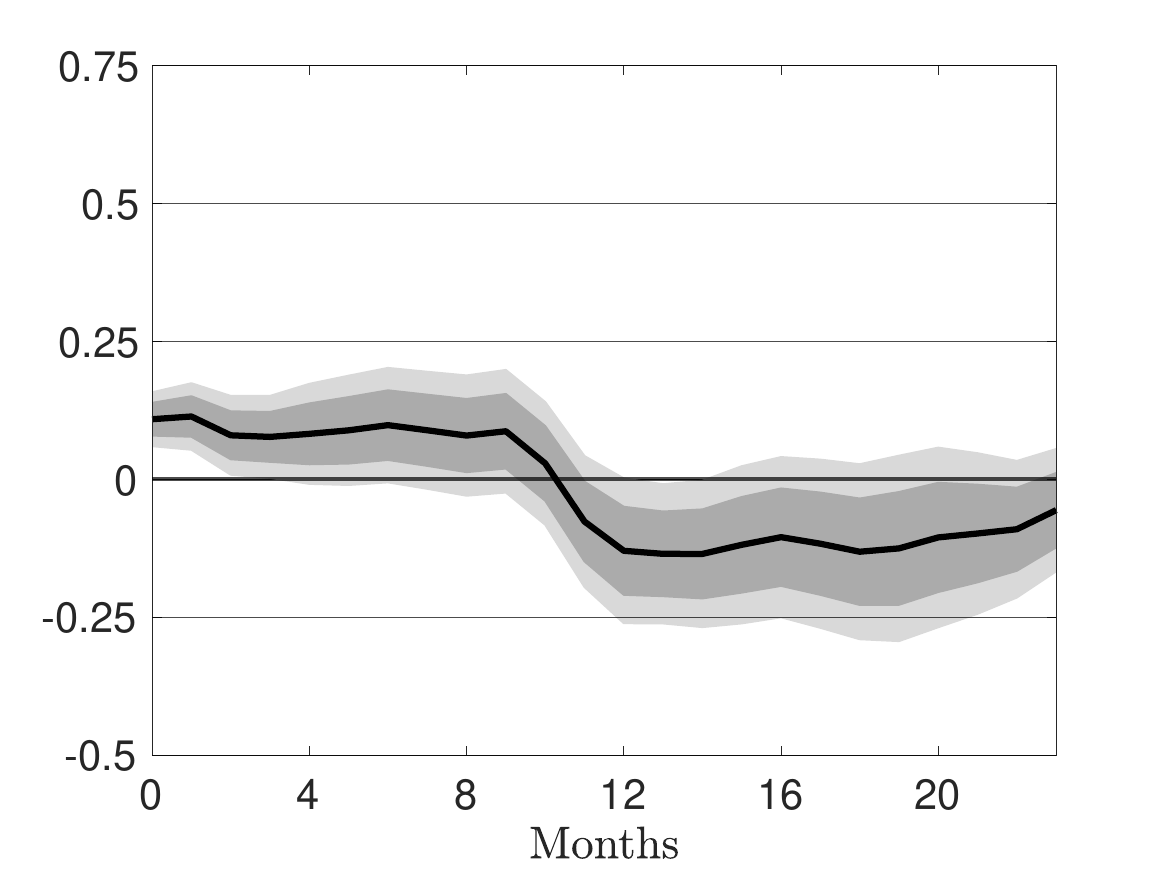} \\
(c) Average effect  & (d) Difference \\ \includegraphics[width=.44\textwidth]{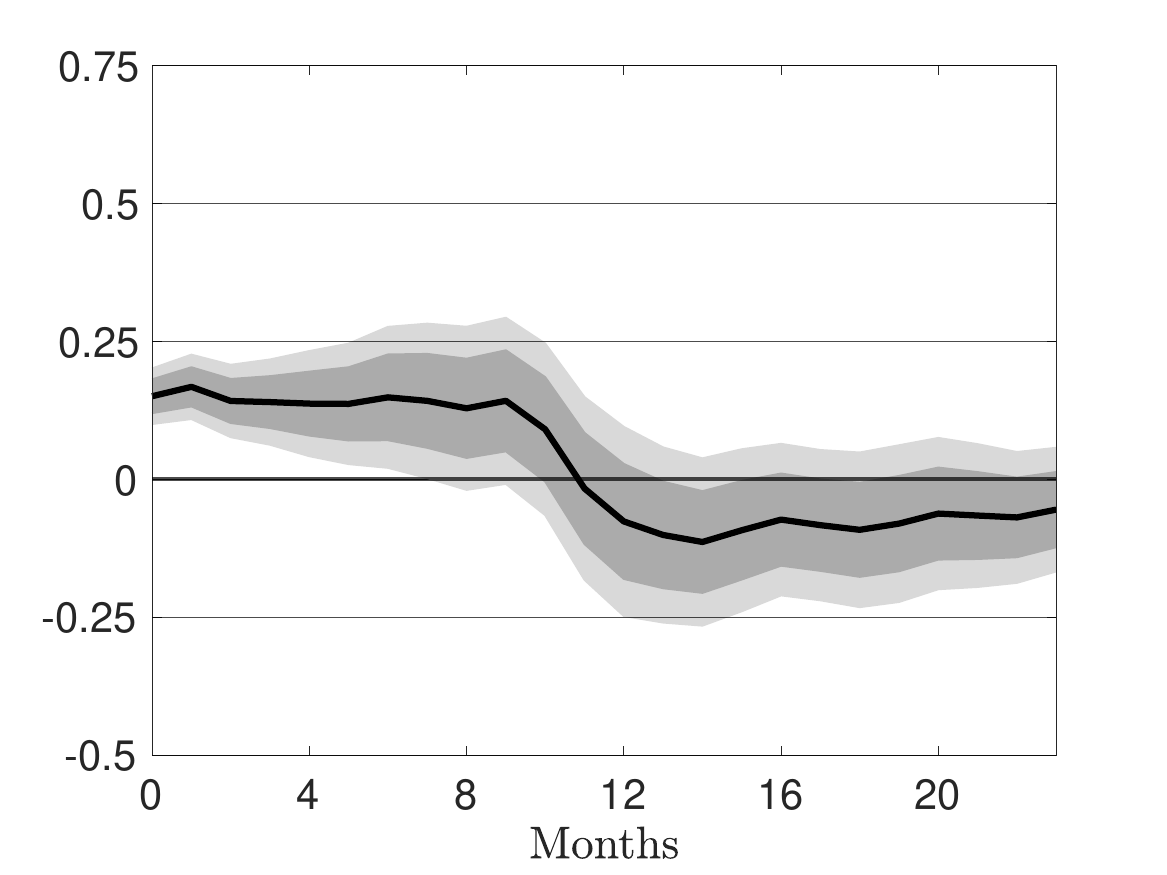} & \includegraphics[width=.44\textwidth]{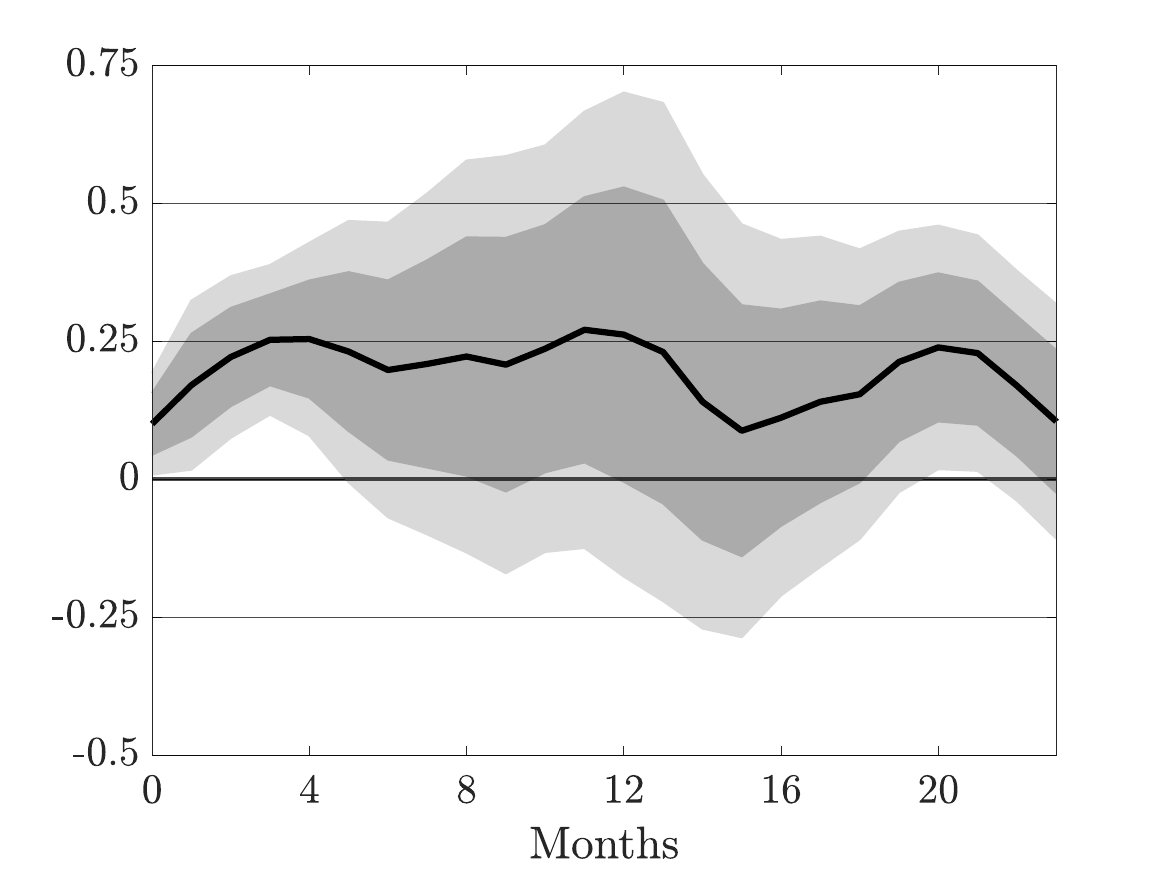}
\end{tabular}%

\begin{minipage}{1\textwidth}
\footnotesize{\emph{Notes}: This figure shows the inflation response to an oil supply news shock in the high-attention regime (panel (a)), the low-attention regime (panel (b)), on average across regimes (panel (c)), and the difference between the two regimes (panel (d)) when using regional variation and region-specific CPI inflation. The dark shaded areas depict the 68\% confidence bands and the light-shaded area the 90\% confidence bands. Standard errors are robust with respect to heteroskedasticity as well as serial and cross-sectional correlation. 
}%
 \end{minipage}
\end{figure}

Figure \ref{fig:emp_irfs_inflation_region} shows the results. We see that the results are very similar to my baseline specification even though the overall effects are slightly smaller and the confidence bands for the high-attention regime wider. 

\begin{figure}[!ht]
\caption{Disentangling attention and inflation}
\label{fig:emp_irfs_inflation_disentangle}\vspace{0.15cm} \centering%
\begin{tabular}{cc}
(a) High-Attention Regime  & (b) Low-Attention Regime  \\ 
\includegraphics[width=.44\textwidth]{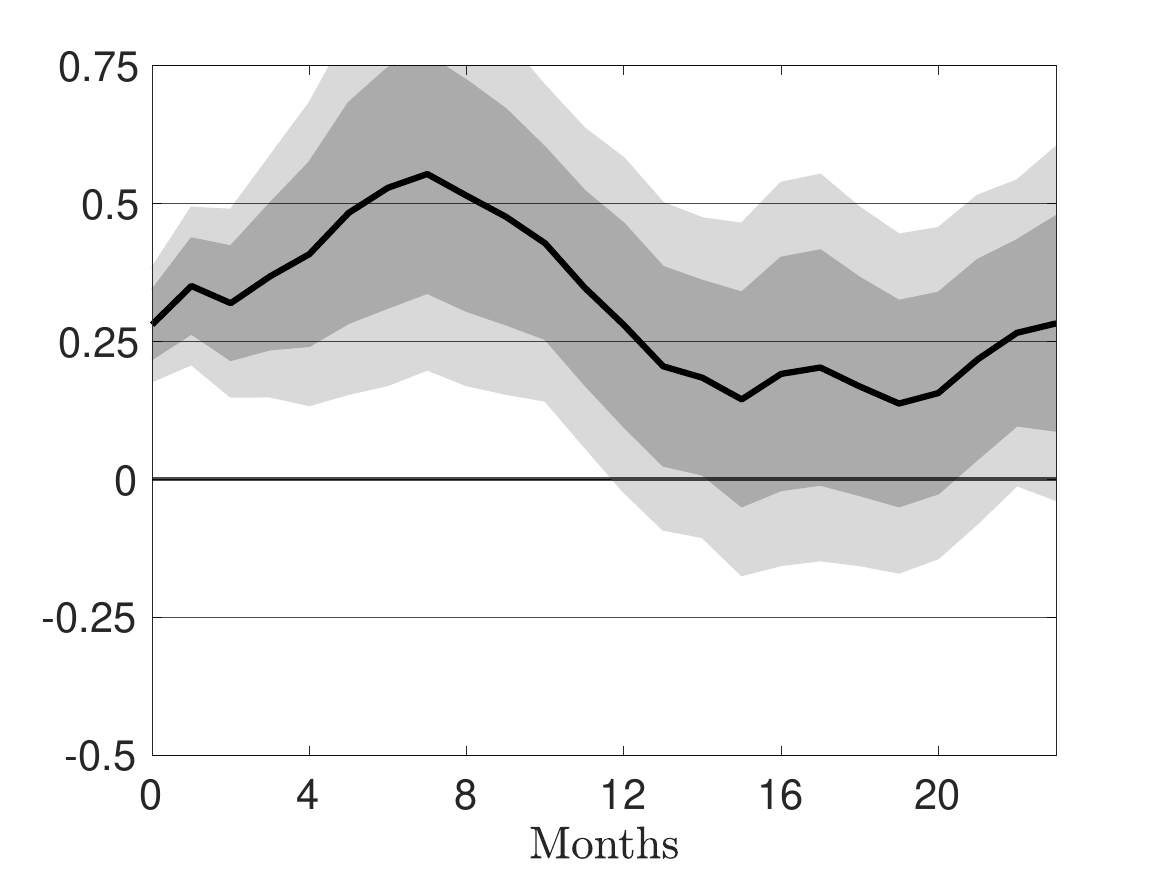} &    \includegraphics[width=.44\textwidth]{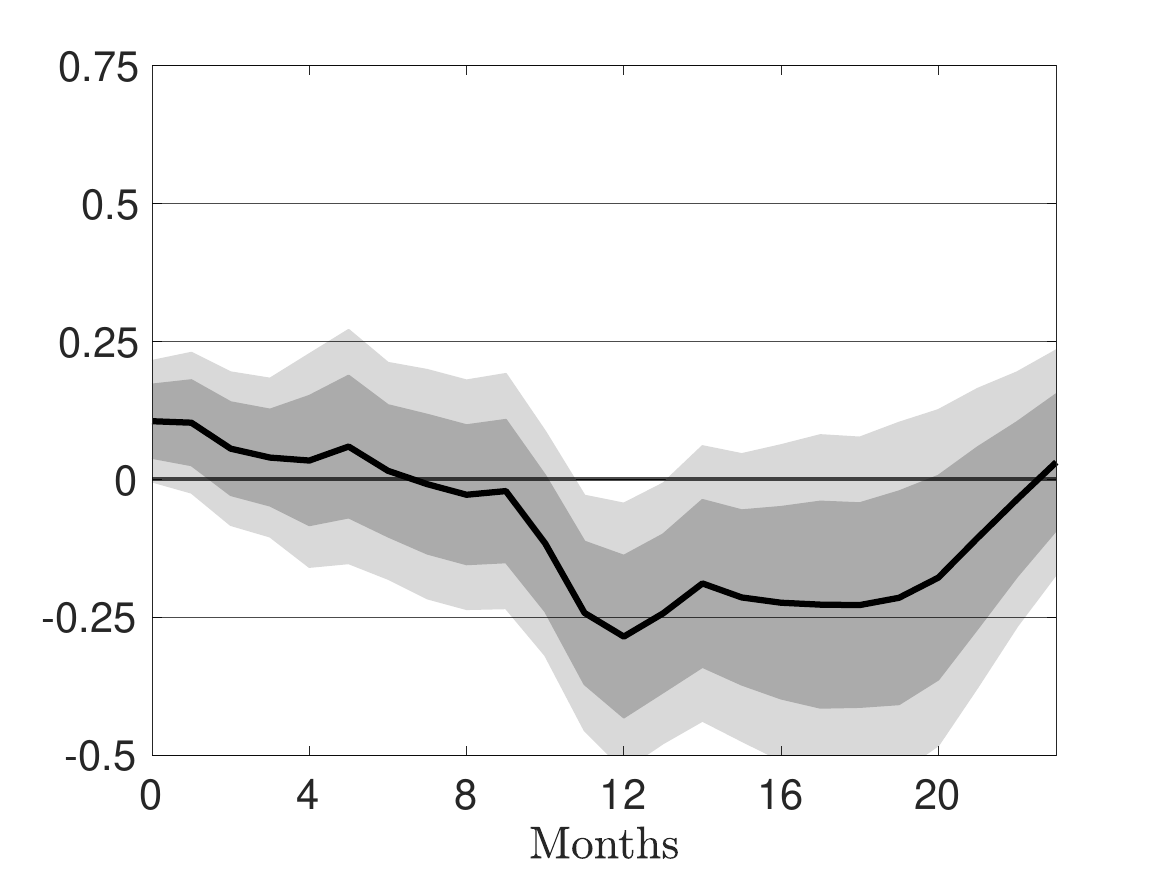} \\
(c) Interaction with inflation  & (d) Google with time FE \\ \includegraphics[width=.44\textwidth]{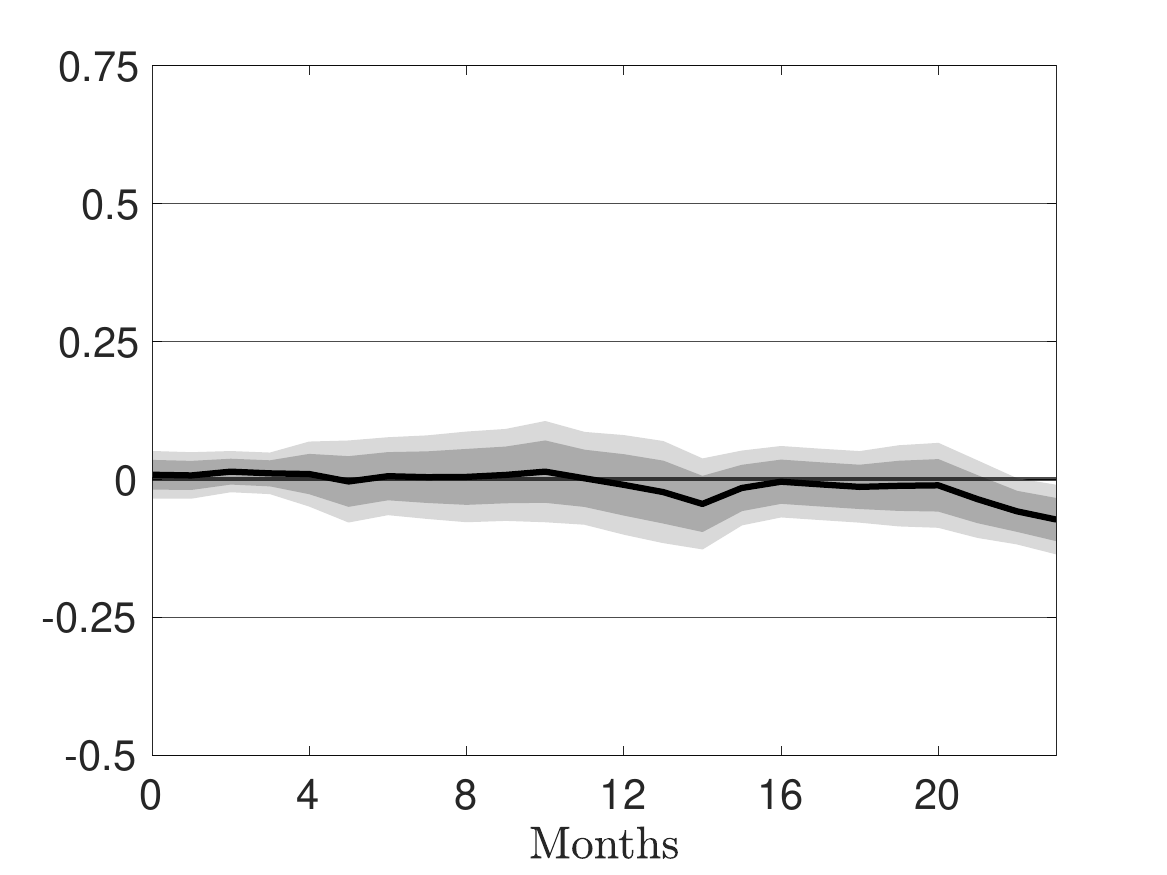} & \includegraphics[width=.44\textwidth]{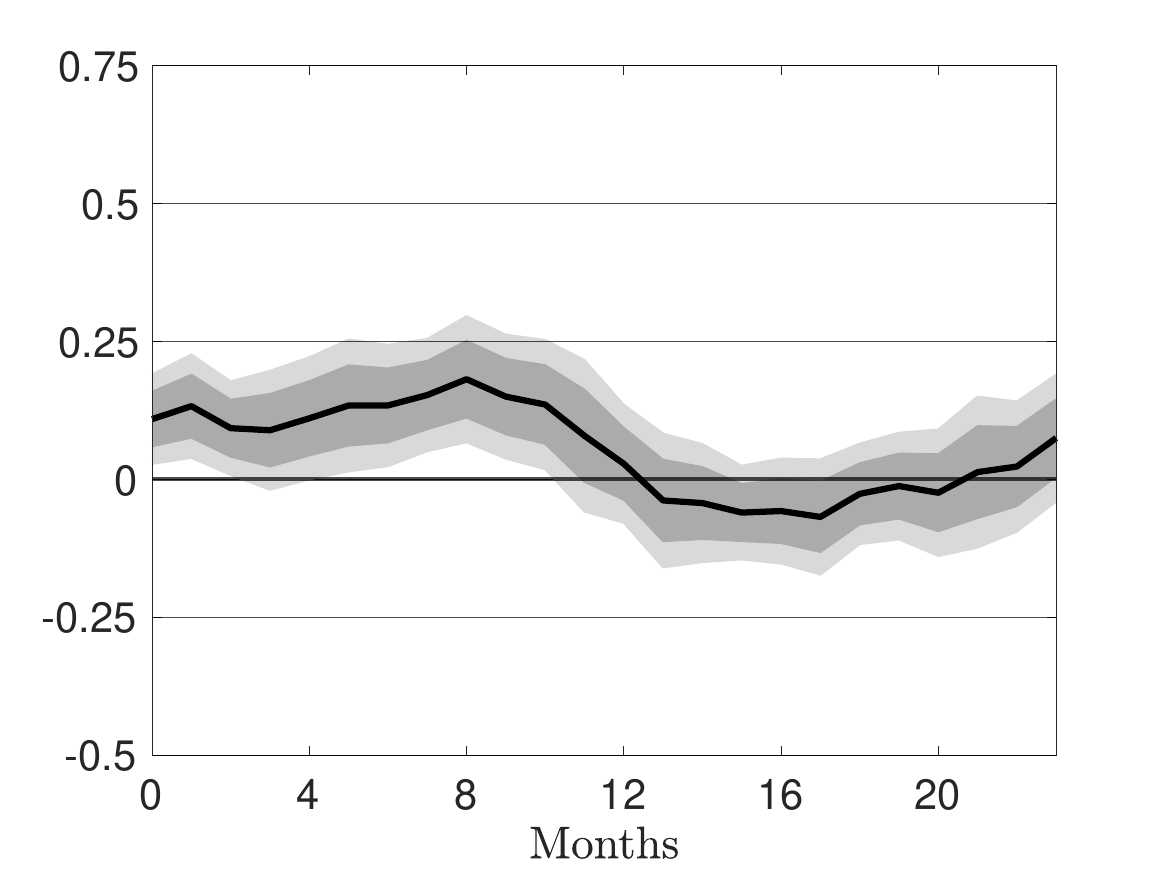}
\end{tabular}%

\begin{minipage}{1\textwidth}
\footnotesize{\emph{Notes}: This figure shows the inflation response to an oil supply news shock in the high-attention regime (panel (a)) and the low-attention regime (panel (b)) when using regional CPI data and when controlling for the interaction of the shock with last-month's inflation (which is shown in panel (c)). The attention regimes are based on Google Trends. Panel (d) shows the interaction term of Google Trends with the shock (for a one standard deviation shock and Google Trends being one standard deviation above its mean) when controlling for regional CPI inflation and time and region fixed effects. Standard errors are robust with respect to heteroskedasticity as well as serial and cross-sectional correlation.
}%
 \end{minipage}
\end{figure}

The regional data also helps me to disentangle attention from other phenomena that arise with higher inflation. To do so, I take two approaches. In the first one, I use Google Trends as the attention-regime defining variable for the four different regions.\footnote{I take the average within each month across all states within a particular consensus region in order to aggregate Google Trends at the consensus region level.} To control for other phenomena apart from attention that may arise with inflation, I further include an interaction term of the shock with lagged inflation, and I include region fixed effects. Panels (a)-(c) in Figure \ref{fig:emp_irfs_inflation_disentangle} show the results. In particular, we see that adverse oil supply news shocks have a strong and persistent effect on inflation when attention is high, whereas the effects in times of low attention are substantially weaker. In addition, the interaction term with inflation is practically zero at all horizons. These results therefore suggest that attention to inflation renders supply shocks more inflationary even when controlling for inflation itself. The results in panels (a) and (b) also suggest that the inflation response is \textit{qualitatively} different across the two regimes. While inflation peaks on impact in the low-attention regime, it shows more of a hump-shaped response in the high-attention regime.

In the second approach, I control for time and region fixed effects and for regional-specific inflation. However, since the regimes based on Google Trends overlap strongly across regions---and are therefore largely absorbed by time fixed effects---I instead include an interaction term of last month's Google Trends searches with the shock rather than focusing on regimes. Panel (d) in Figure \ref{fig:emp_irfs_inflation_disentangle} shows that this interaction term is highly significant in the first 10 months (the coefficient is normalized to show the additional effect of a one standard deviation shock when Google Trends is one standard deviation above its mean). These results indicate that the different inflation responses across attention regimes are unlikely to be solely driven by aggregate phenomena that may confound the results, such as different policy responses across regimes, a general increase in the frequency of price adjustments, or an increase in the slope of the aggregate Phillips curve. Instead, the significant effects arising from more Google searches of inflation indicate that higher attention indeed drives at least partly the different inflation responses observed across regimes.

\paragraph{Other shocks.} So far, I focused on the implications of supply shocks on inflation. In Appendix \ref{app:robustness_irfs}, I show that other shocks become more inflationary in the high-attention regime as well. In figure \ref{fig:emp_irfs_infshock}, I show that inflation responds twice as much to the \textit{main inflation shock} from \cite{angeletos2020business} and that these differences are highly persistent. In figure \ref{fig:emp_irfs_inflation_mp}, I show that inflation also responds more strongly in the high-attention regime to monetary policy shocks identified using a high-frequency identification (the shocks are taken from \cite{jarocinski2020deconstructing} and are purged from the information effects of monetary policy statements). These findings suggest that higher attention to inflation not only renders supply shocks more inflationary, but shocks more generally become more inflationary. The predictions of the model derived later will be consistent with these findings.

\paragraph{Robustness.}
In Appendix \ref{app:robustness_irfs}, I discuss a number of robustness checks. In particular, I show that my results are not driven by different oil price responses (figures \ref{fig:emp_irfs_realoilp} and \ref{fig:emp_irfs_nominal_oilp}) or a higher sensitivity of macroeconomic variables in general (figures \ref{fig:emp_irfs_unemp_nocovid} and \ref{fig:emp_irfs_unemp}). 
The results are very similar when excluding the Covid period (figure \ref{fig:emp_irfs_inflation_noCovid}), or when excluding the Great Inflation period (figure \ref{fig:emp_irfs_inflation_after1990}). Similarly, the results remain robust when using different control variables (figure \ref{fig:emp_irfs_inflation_noctrls}), or when focusing on the cumulative changes in the price level rather than the cumulative changes in inflation compared to the initial inflation rate (figure \ref{fig:emp_irfs_oilprice}). Overall, the results in this section show that inflation responds 2-3 times as strongly to inflationary supply shocks when the public's attention to inflation is high compared to periods in which it is low.

\subsection{The Role of Attention in the Post-Covid Inflation Surge}
\begin{figure}[h]
\caption{Supply shocks, attention and the post-Covid inflation surge}
\label{fig:pi_postcovid}\vspace{0.05cm} \centering%
\begin{tabular}{c}
\includegraphics[width=.9\textwidth]{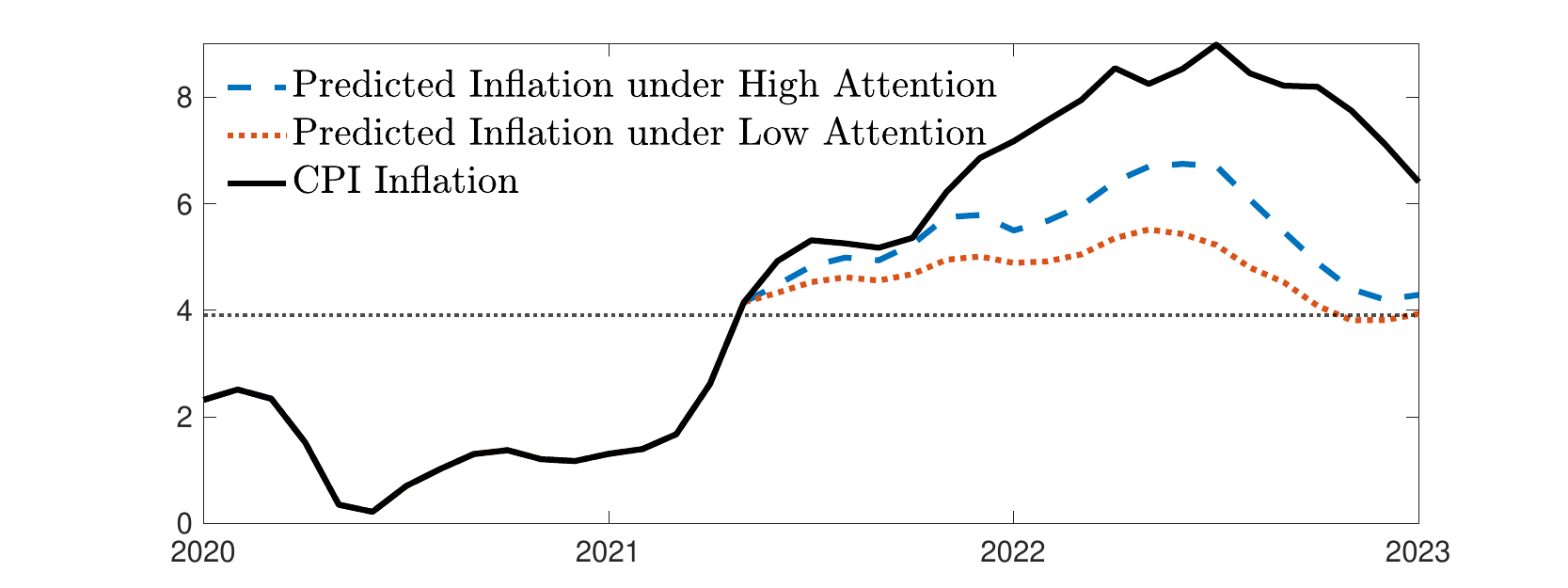}  
\end{tabular}%

\begin{minipage}{1\textwidth}
\footnotesize{\emph{Notes}: This figure shows the actual U.S.\ CPI inflation from 2020 until 2023, as well as the counterfactual inflation dynamics arising only from oil supply news shocks in the economy accounting for the increase in attention in April 2021 (blue-dashed line) and in the economy ignoring the increase in attention (orange-dotted line).
}%
 \end{minipage}
\end{figure}

\vspace{-0.2cm}
I now show that the increase in people's attention to inflation played a quantitatively important role in the post-Covid inflation surge in the United States.\footnote{A similar approach is, e.g., used in \cite{coibion2012effects} who analyzes the historical contribution of monetary policy shocks to industrial production, unemployment and inflation, or by \cite{mitra2023imperfect} who examines the role of noise shocks for the dynamics of labor markets.} The black-solid line in Figure \ref{fig:pi_postcovid} shows the evolution of CPI inflation for the period 2020-2023. 

To quantify the importance of the change in attention and of oil supply shocks, I estimate the counterfactual evolution of CPI inflation when only considering oil supply news shocks starting in April 2021, which is when inflation exceeded 3.91\% and thus, the economy entered the high-attention regime.\footnote{I ignore all previous shocks, i.e., I compute the counterfactuals as if the economy was in steady state in April 2021 and that there were no shocks before that period. Thus, the reported results in this section are likely to be at the lower end of the actual importance of attention for the inflation dynamics, given that inflation was already increasing before and that the supply shocks in the first three months of 2021 were all negative (i.e., inflationary). Consistent with early 2021 being the time attention increased, \cite{hilscher2022likely} document a rise in the probability of persistent high inflation for the U.S.\ in mid-2021. \cite{reis2022burst} also documents that household inflation expectations in mid-2021 were not very well anchored any more.} 
I do this twice: once, taking the increase in attention into account, and once, without the increase in attention. For the case with the increase in attention, I therefore feed in the identified oil supply shocks from April 2021 until the end of 2022 using the estimated impulse response functions from panel (a) of Figure \ref{fig:emp_irfs_inflation1}. The blue-dashed line shows the implied inflation rate. We see that the shape of the implied inflation rate is similar to the actual inflation rate, and they both peak in mid-2022. Furthermore, even though I only consider oil supply shocks and abstract from all other potential exogenous drivers of inflation, the implied inflation rate increases from the initial value of 4.1\% up to almost 7\%. Thus, the oil supply shocks account for about 55\% of the additional increase from early 2021 until mid-2022. This is consistent with the findings in \cite{shapiro2022decomposing} who estimates the supply-driven part of (PCE) inflation in 2022 to be of similar magnitude.\footnote{Even though I focus on supply shocks, the increase in people's attention to inflation in 2021 likely also increased the inflationary effects of demand shocks, such as monetary stimulus, given that I also find demand shocks to be more inflationary in times of higher attention (Figure \ref{fig:emp_irfs_inflation_mp}).}

How much of this is due to the increase in attention? The red-dotted line shows that if there would not have been an increase in people's attention, that is when relying on the estimated impulse response functions from the low-attention regime (panel (b) in Figure \ref{fig:emp_irfs_inflation1}), the implied inflation increase from April 2021 until mid-2022 would have only been about 1.5 percentage points (compared to the 2.7pp.\ when accounting for people's change in attention). These results suggest that the rise in the public's attention to inflation was quantitatively important in driving inflation further up in response to supply shocks during the recent inflation surge, and that without the increase in people's attention, supply-driven inflation would have only been about half as high as it was. 

To better understand the underlying mechanisms of these results as well as why the increase in people's attention may explain the interplay between inflation and inflation expectations (as shown in panel (b) in Figure \ref{fig:moti}), I now introduce the attention threshold and changing degrees of people's attention to inflation into an otherwise standard New Keynesian model.

\section{A Monetary Model with the Attention Threshold}\label{sec:model}\vspace{-0.2cm}
In this section, I develop a monetary model which builds on the textbook New Keynesian model \citep{woodford2003interest,gali2015monetary} but households and firms pay only limited attention to inflation.

\subsection{Households} \vspace{-0.2cm}
There is a representative household obtaining utility from consumption and disutility from working, with lifetime utility
\begin{equation}
\Tilde{E}_{0}\sum_{t=0}^{\infty }\beta^{t}Z_t\left[ \log\left(C_t\right)-\Xi H_t\right] ,  \label{utilorig}
\end{equation}
where $C_{t}$ is consumption of the final good, $H_{t}$ is hours worked, $\beta$ is the household's time discount factor, and $\tilde{E}_t$ denotes the household's subjective expectations operator based on information available in period $t$. $Z_t$ are exogenous preference shocks. The parameter $\Xi$ is the utility weight on hours worked.

Households maximize their lifetime utility subject to the flow budget constraints 
\begin{align}
& C_{t}+B_{t}= w_tH_t+\frac{1+i_{t-1}}{1+\pi _{t}}B_{t-1}+\frac{T_{t}}{P_{t}}, \quad\text{ for all } t, \label{BC_final}
\end{align}%
where $B_t$ is the real value of government bonds, $w_t$ the real wage, $\pi_t$ is the net inflation rate, and $i_t$ the nominal interest rate. $T_t$ denotes lump-sum taxes and transfers from the government.

Maximizing \eqref{utilorig} subject to \eqref{BC_final} yields the Euler equation
\begin{equation}
    Z_tC_t^{-1} = \beta(1+i_t)\Tilde{E}_t\left[Z_{t+1}C_{t+1}^{-1}\frac{1}{1+\pi_{t+1}}
\right], \label{eq:euler_model}
\end{equation}
and the labor-leisure condition
\begin{equation}
w_{t}= \Xi C_t. \label{eq:ll}
\end{equation}%
The labor-leisure equation \eqref{eq:ll} shows that consumption fluctuations are solely driven by aggregate wage fluctuations.

\subsection{Firms}\vspace{-0.2cm}
The firm sector is held by risk-neutral managers that discount future profits by $\beta$ and they have a mass of zero, such that their consumption is 0 and all their profits go to the households, as in \cite{bayer2020shocks}.  

\paragraph{Final good producer.} There is a representative final good producer that aggregates the intermediate goods $Y_t(j)$ to a final good $Y_t$, according to
\begin{equation}
    Y_t = \left( \int_0^1 Y_t(j)^{\frac{\epsilon-1}{\epsilon}} dj\right)^{\frac{\epsilon}{\epsilon -1}},
\end{equation}
with $\epsilon > 1$. Nominal profits are given by
$
    P_t \left( \int_0^1 Y_t(j)^{\frac{\epsilon-1}{\epsilon}} dj\right)^{\frac{\epsilon}{\epsilon -1}} - \int_0^1P_t(j)Y_t(j)dj,
$
and profit maximization gives rise to the demand for each variety $j$:
\begin{equation}
    Y_t(j) = \left( \frac{P_t(j)}{P_t} \right)^{-\epsilon} Y_t.
\end{equation}
Thus, demand for variety $j$ is a function of its relative price, the price elasticity of demand $\epsilon$ and aggregate output $Y_t$. The aggregate price level is given by
\begin{equation}
    P_t = \left( \int_0 ^1 P_t(j)^{1-\epsilon}dj \right)^{\frac{1}{1-\epsilon}}.
\end{equation}

\paragraph{Intermediate producers.} Intermediate producer of variety $j$ produces output $Y_t(j)$ using labor $H_t(j)$ as its only input
\begin{equation}
    Y_t(j) = H_t(j).
\end{equation}
All intermediate producers pay the same wage $w_t$ and a sales tax (or subsidy) $\tau_t$, which in steady state is set such that profits in steady state are 0.\footnote{Therefore, we have $1-\tau = \frac{\epsilon}{\epsilon - 1}$ in steady state.} These taxes are given back to firms in a lump-sum fashion, denoted $t^F_t(j)$. Taxes are assumed to be constant in the efficient economy, i.e., absent price rigidities, but fluctuate around their steady state in the economy with price rigidities in order to give rise to exogenous cost-push shocks. 

Each intermediate firm has two managers: one is responsible for the firm's forecasts and the other one sets the price of firm $j$ given these forecasts, similar to the setup in, e.g., \cite{adam2011inflation}. I first discuss the problem of the price setter and discuss the forecaster's problem later.

When adjusting the price, the firm is subject to a \cite{rotemberg1982sticky} price-adjustment friction.\footnote{Thus, I abstract from changes in the frequency of price adjustments. \cite{cavallo2023large} and \cite{blanco2024inflation} argue that the frequency of price adjustments has increased during the recent inflation surge. While it is difficult to empirically disentangle attention to inflation from changes in price-setting behavior (which itself might change due to a change in attention of price setters), my results using Google Trends data as well as the survey evidence from \cite{schwartzman2024inflation} suggest that changes in price-setting behavior cannot explain the whole inflation dynamics and that attention indeed played an important role.} 
The per-period profits (in real terms) are given by
\begin{equation}
    (1-\tau_t)P_t(j)\left( \frac{P_t(j)}{P_t}\right)^{-\epsilon} \frac{Y_t}{P_t} - w_t H_t(j) - \frac{\psi}{2}\left(\frac{P_t(j)}{P_{t-1}(j)} - 1\right)^2Y_t + t^F_t(j),
\end{equation}
where $\psi \geq 0$ captures the price-adjustment cost parameter. The price setter sets the price to maximize profits
\begin{align}
    \Omega_0(j) = \Tilde{E}^j_0\sum\limits_{t=0}^{\infty}\beta^t\left[(1-\tau_t)P_t(j)\left( \frac{P_t(j)}{P_t}\right)^{-\epsilon} \frac{Y_t}{P_t} - mc_t H_t(j) - \frac{\psi}{2}\left(\frac{P_t(j)}{P_{t-1}(j)} - 1\right)^2Y_t + t^F_t(j)\right], \notag
\end{align}
where $mc_t = w_t$ denotes the real marginal cost which is the same for every firm. Using the production function to substitute for $H_t(j)$ and the demand for firm $j$'s product from the final goods producer, the corresponding first order condition is then given by
\begin{align*}
        (1-\tau_t) (\epsilon-1)P_t(j)^{-\epsilon}\frac{Y_t}{P_t^{1-\epsilon}}  = &\epsilon mc_t \left(\frac{P_t(j)}{P_t} \right)^{-\epsilon -1} \frac{Y_t}{P_t} - \psi\left(\frac{P_t(j)}{P_{t-1}(j)} -1\right)\frac{Y_t}{P_{t-1}(j)}\\
&+ \beta \psi \tilde{E}^j_t\left[\left(\frac{P_{t+1}(j)}{P_t(j)}-1\right)\frac{P_{t+1}(j)}{P_t(j)}\frac{Y_{t+1}}{P_t(j)} \right].
\end{align*}
Defining $T_t \equiv 1-\tau_t$, it follows that after a linearization around the zero-inflation steady state, firm $j$ sets its price according to
\begin{equation}
    \widehat{p}_t(j) = \frac{1}{\psi+\epsilon}\left[\psi \widehat{p}_{t-1}(j) + \epsilon\left( \widehat{mc}_t - \widehat{T}_t + \widehat{p}_t\right) + \beta\psi \tilde{E}^j_t\pi^j_{t+1} \right],\label{eq:ptj}
\end{equation}
where hatted variables denote log deviations of the respective variables from their steady state values (see Appendix \ref{app:model} for all derivations).
Therefore, prices may only differ across firms $j$ due to differences in forecasts $\tilde{E}^j_t\pi^j_{t+1}$ or differences in past prices $\widehat{p}_{t-1}(j)$.

\paragraph{Government.} The government imposes a sales tax $\tau_t$ on sales of intermediate goods, issues government bonds, and pays lump-sum
taxes and transfers $T_{t}$ to households and $t^F_t(j)$ to firms. The real government budget constraint is given
by 
\begin{equation*}
B_{t}=B_{t-1}\frac{1+i_{t-1}}{\Pi_t}+\frac{T_{t}}{P_{t}}-\tau Y_{t} + t^f_t.
\end{equation*}%
Lump-sum taxes and transfers are
set such that they keep real government debt constant at the initial level $%
B_{-1}/P_{-1}$, which I set to zero. 

The monetary authority sets the nominal interest rate according to a (linearized) Taylor rule:
\begin{equation}
    \tilde{i}_t = \rho_i \tilde{i}_{t-1} + (1-\rho_i)\left(\phi_{\pi}\pi_t + \phi_x \widehat{x}_t \right),
\end{equation}
where $\tilde{i}_t$ denotes the nominal interest rate in deviations from its steady state, $\rho_i$ captures interest-rate smoothing, $\phi_{\pi}$ and $\phi_x$ pin down the response coefficients with respect to inflation and the output gap, respectively. I discuss other rules for monetary policy in Section \ref{sec:welfare}.

\subsection{Subjective expectations under limited attention}\label{sec:subj}\vspace{-0.2cm}

The subjective expectations of households and firm managers are modelled in such a way that there is a direct mapping from the empirical results of Section \ref{sec:data} to the model.

\paragraph{Households.} Households have to form expectations about inflation and consumption. According to the labor-leisure equation \eqref{eq:ll}, consumption fluctuations are solely driven by aggregate wage fluctuations. Households therefore use information about aggregate wages to form their consumption expectations.
Households believe that wages and inflation both follow a random walk:
\begin{align*}
    \pi_t &= \pi_{t-1} + \nu_{\pi,t} \\
    \widehat{w}_t &= \widehat{w}_{t-1} + \nu_{w,t},
\end{align*}
where $\nu_{\pi,t}$ and $\nu_{w,t}$ are normally distributed with mean zero, and independent from each other. I discuss the case in which households hold rational expectations about consumption later. The random walk assumption ensures that long-run expectations align with the actual long-run variables.\footnote{Assuming an AR(1) with $\rho_{\pi} < 1$ does not qualitatively change the results.} They observe the aggregate shocks $\tau_t$ and $Z_t$ perfectly. 

At the time the household forms her expectations about future wages and inflation, she does not perfectly observe their current realizations, capturing that the household is not at all times perfectly monitoring all consumption expenditures and wage incomes of the members of the household  and that inflation is not perfectly observable in real time. Instead, the household only sees noisy signals of the form
\begin{align*}
    s_{\pi,t} &= \pi_t + \varepsilon_{\pi,r,t} \\
    s_{w,t} &= \widehat{w}_t + \varepsilon_{w,t}, 
\end{align*}
with normally distributed noise terms $\varepsilon_{\pi,r,t}$ and $\varepsilon_{w,t}$. I assume that signals are public signals and every agent in the economy receives the same signals. This implies that all agents have the same information and hence, hold the same expectations.

As detailed in Appendix \ref{app:model}, potential output, i.e., output under flexible prices, is constant and  consumption in log-deviations from steady state is equal to the output gap $\widehat{x}_t$ in equilibrium (and equal to wages in deviations from steady state), $\widehat{c}_t =  \widehat{w}_t = \widehat{x}_t$.\footnote{Note, that the price adjustment costs do neither affect the steady state nor the linearized resource constraint when linearized around the zero-inflation steady state, such that $\widehat{y}_t = \widehat{c}_t$.} It follows that if we assume initial values $\tilde{E}_{-1}\widehat{c}_{0} = \tilde{E}_{-1}\widehat{w}_{0} = \tilde{E}_{-1}\widehat{x}_{0}$, we have $\tilde{E}_t\widehat{c}_{t+1} = \tilde{E}_t\widehat{w}_{t+1} = \tilde{E}_t\widehat{x}_{t+1}$ for all $t$. This holds true, even if the household does not know that consumption and wages are equal to the output gap in equilibrium.

Expectations about the output gap, $\widehat{x}_t$, then evolve as follows:
\begin{equation}
    \tilde{E}_t\widehat{x}_{t+1} =  \tilde{E}_{t-1}\widehat{x}_{t} + \gamma_x\left(\widehat{x}_t - \tilde{E}_{t-1}\widehat{x}_{t} \right), \label{eq:lom_x}
\end{equation}
where $\tilde{E}_t\widehat{x}_{t+1}$ denotes the agent's expectations of the one-period-ahead output gap. The parameter $\gamma_x$ denotes the optimal level of attention to the output gap, based on the agent's subjective model of how the output gap evolves. A higher $\gamma_x$ denotes a higher attention level. If $\gamma_x = 0$, the agent is completely inattentive and just sticks to her prior belief $\tilde{E}_{t-1}\widehat{x}_{t}$, whereas $\gamma_x = 1$ captures the case of full attention in which case the agent believes $\tilde{E}_t\widehat{x}_{t+1} = \widehat{x}_t $, which is the full-information belief of someone who believes that the output gap follows a random walk. As I discuss in more detail in the calibration section later on, I do not find any differences in attention to unemployment changes (which I use as a proxy for the output gap) in the data. I therefore impose that $\gamma_x$ does not change across regimes in the baseline case. I discuss the cases with regime-dependent $\gamma_x$ and with rational consumption expectations later.

Inflation expectations follow the law of motion derived in equation \eqref{eq:lom_pi} for $\rho_{\pi} = 1$, so that they are given by
\begin{equation}
    \tilde{E}_t\pi_{t+1} = \begin{cases}
        \Tilde{E}_{t-1}\pi_{t} + \gamma_{\pi,L}\left(\pi_t - \Tilde{E}_{t-1}\pi_{t}\right), \text{ when } \pi_{t-1} < \Bar{\pi} \\
        \Tilde{E}_{t-1}\pi_{t} + \gamma_{\pi,H}\left(\pi_t - \Tilde{E}_{t-1}\pi_{t}\right), \text{ when } \pi_{t-1} \geq \Bar{\pi},
    \end{cases} \label{eq:lom_pi_mode}
\end{equation}
where $\gamma_{\pi,r}$ captures the optimal level of attention to inflation in regime $r\in\{L,H\}$.\footnote{Without loss of generality, I do not explicitly model noise shocks. Accounting for noise shocks would introduce an additional exogenous shock, but does not affect my analysis of how supply and demand shocks are transmitted. } 
I leverage the expression \eqref{opt_gamma1} for optimal attention to inflation
\begin{equation}
\gamma_{\pi,r} = max\left\{0,1-\frac{\tfrac{1}{\lambda_r}}{2\chi\rho_{\pi,r}^2\sigma^2_{\pi,r}}\right\} \notag
\end{equation}
to calibrate the model. Defining the cost of information relative to the stakes $\chi$, $\tilde{\lambda} \equiv \frac{1}{2\lambda  \chi}$ and imposing $\rho_{\pi,r} = 1$, I obtain 
\begin{equation}
\gamma_{\pi,r} = max\left\{0,1-\frac{\tilde{\lambda}_r}{\sigma^2_{\pi,r}}\right\}. \label{gamma_rho1}
\end{equation}

As my proxy for $\sigma^2_{\pi,L}$, I use the model-implied variance of inflation in regime $L$ given the rest of the calibration (discussed in Section \ref{sec:calib}). I set $\tilde{\lambda}$ such that $\gamma_{\pi,L} = 0.18$ as found in the data. In order to pin down $\gamma_{\pi,H}$, I then increase the exogenous volatility of the cost-push shocks whenever inflation exceeded the attention threshold in the previous period. I then solve for the fixed point such that $\gamma_{\pi,H} = 0.36$ and that the model-implied variance of inflation is equal to the $\sigma^2_{\pi,H}$ implied by equation $\eqref{gamma_rho1}$ for $\gamma_{\pi,H} = 0.36$ and the cost of information $\tilde{\lambda}$ obtained before.

As an alternative, the attention change may arise due to a change in the exogenous cost of information, $\tilde{\lambda}_H < \tilde{\lambda}_L$, which could reflect the fact that news coverage of inflation is substantially higher in times of high inflation, thus, lowering the cost of information in these periods. All the following results are very similar under that alternative assumption.

\paragraph{Firm managers.}
Since there are no idiosyncratic shocks, I assume that the forecasting manager of firm $j$ uses expectations about aggregate inflation to form her expectations about firm $j$'s future price change, i.e., $\tilde{E}^j_t \pi_{t+1}(j) = \tilde{E}^j_t\pi_{t+1} $, and that they form expectations about inflation in the same way as households. Consistent with these assumptions, \cite{yotzov2024speed} show that firms’ expected own-price growth is strongly positively correlated with changes in CPI inflation, and \cite{mcclure2022macroeconomic} who show that managers and non-managers hold similar average inflation and unemployment expectations and respond similarly to information treatments.\footnote{\cite{mcclure2022macroeconomic} further show that managers' expectations indeed affect their economic decisions.} 
As for households, all forecasters receive the same public signal about current inflation, from which it follows that $\tilde{E}^j_t \pi_{t+1}(j) = \tilde{E}^j_t\pi_{t+1} = \tilde{E}_t\pi_{t+1} $.

Assuming that firms all start out with the same initial price, i.e., $\widehat{p}_{-1}(j) = \widehat{p}_{-1}$ for all $j\in [0,1]$, it follows that all firms set the same price (as they do under rational expectations). Thus, firm-specific inflation is indeed equal to aggregate inflation, which confirms the forecaster's belief that the two are equal, and hence, the forecaster has no reason to deviate from these beliefs.

Given these assumptions, equation \eqref{eq:ptj} can be written as
\begin{equation}
    \pi_t = \frac{\epsilon}{\psi}\left(\widehat{mc}_t - \widehat{T}_t \right) + \beta \tilde{E}_t \pi_{t+1}.
\end{equation}
From the labor-leisure equation and the production function, we have $\widehat{mc}_t = \widehat{y}_t$, and since potential output is constant we have $\widehat{y}_t = \widehat{x}_t$. Defining cost-push shocks as $u_t \equiv -\frac{\epsilon}{\psi}\widehat{T}_t$ and $\kappa \equiv \frac{\epsilon}{\psi}$, we arrive at the linearized New Keynesian Phillips Curve:
\begin{equation}
    \pi_t =  \beta \Tilde{E}_t\pi_{t+1} +\kappa \widehat{x}_t  + u_t. \label{eq_nkpc_1}
\end{equation}

\subsection{Equilibrium}\vspace{-0.2cm}
The model can be summarized by three equilibrium equations when expressed in log-deviations from the zero-inflation steady state (see Appendix \ref{app:model}):
\begin{align}
\pi_t &= \beta  \tilde{E}_t\pi_{t+1} + \kappa \widehat{x}_t + u_t, \label{eq:nkpc} \\
\widehat{x}_t &= \tilde{E}_t\widehat{x}_{t+1} - \left(\tilde{i}_t-\tilde{E}_t\pi_{t+1}-r^*_t\right), \label{eq:euler}
\\
\tilde{i}_t &= \rho_i \tilde{i}_{t-1} + (1-\rho_i)\left(\phi_{\pi}\pi_t + \phi_x \widehat{x}_t \right), \label{eq:taylor}
\end{align}
as well as the two law of motions for output gap expectations and inflation expectations, equations \eqref{eq:lom_x} and \eqref{eq:lom_pi_mode}. 
Equations \eqref{eq:nkpc}-\eqref{eq:taylor} are identical to the 3-equation representation of the standard New Keynesian model (see \cite{gali2015monetary}) except for the expectations formation of agents.

Equation \eqref{eq:nkpc} is the New Keynesian Phillips curve, representing the supply side of the economy, and equation \eqref{eq:euler} denotes the aggregate Euler (or IS) equation, which together with monetary policy (equation \eqref{eq:taylor}) pins down aggregate demand. The natural interest rate $r^*_t$ is the real rate that prevails in the economy with fully flexible prices, and solely depends on the exogenous shocks $Z_t$. It follows an AR(1) process with persistence $\rho_r\in[0,1]$ and innovations $\varepsilon^r\sim i.i.N.(0,\sigma^2_r),$ independent of $\varepsilon^u$. I will refer to shocks to $r^*_t$ as demand shocks. The nominal interest rate $\tilde{i}_t$ and the natural rate are both expressed in absolute deviations of their respective steady state values, $\underline{i}$ and $\underline{r}^*$, with $\underline{i} = \underline{r}^*$, as the model is linearized around the zero-inflation steady state.

\subsection{Calibration}\label{sec:calib}\vspace{-0.2cm}
Most of the calibration is standard. I set the discount factor $\beta$ to target a steady state natural rate of $1\%$ (annualized), and the slope of the Phillips curve $\kappa$ to 0.057. The Taylor rule coefficients are set to $\rho_i = 0.7$, $\phi_{\pi} = 2$ and $\phi_x = 0.125$. 

I assume that both shocks follow an AR(1) process with persistence 0.8. To calibrate the standard deviation of the cost-push shocks at quarterly frequency, I sum up all monthly oil supply news shocks within a quarter and use the resulting standard deviation as the exogenous shock volatility when inflation is below the threshold, $\sigma_{u,L}$. This results in $\sigma_{u,L} = 0.3\%$. I assume that demand shocks have the same volatility, $\sigma_r = \sigma_{u,L}$ and I assume that the demand shock volatility is the same in both regimes. Given my empirical estimates, I set $\gamma_{\pi,L} = 0.18$ and the threshold to $\Bar{\pi} = 3.91\%$ (annualized). In order to calibrate the cost-push shock volatility in the high-attention regime $\sigma_{u,H}$, I solve for the fixed-point problem discussed in Section \ref{sec:subj} by targeting $\gamma_{\pi,H} = 0.35$. To do so, I simulate the model for 10,000 periods and vary $\sigma_{u,H}$ until the fixed point is reached. This results in $\sigma_{u,H} = 1.23\cdot \sigma_{u,L}$. That said, however, the subsequent results are very similar when imposing that the change in attention arises from a change in the cost of information and therefore, keeping the shock volatility constant across regimes.

To calibrate the attention parameter with respect to the output gap $\gamma_x$, I follow the same procedure as for inflation but focus on expectations about unemployment changes (see Appendix \ref{app:unemp}). This results in $\gamma_x = 0.25$ for both regimes, and hence, I impose that $\gamma_x$ does not change across regimes and set it to $\gamma_x = 0.25$.

As a sanity check of the proposed empirical approach to estimate the attention threshold and the degrees of attention in a setup in which inflation and inflation expectations are determined jointly and different type of shocks are hitting the economy, I estimate the threshold regression \eqref{reg1_thresh} on model-simulated data and find that the regression results are exactly equal to the calibrated parameters.

\section{Inflation Surges}\label{sec:results}\vspace{-0.2cm}
In this section, I show how the model with two inflation-attention regimes can jointly generate persistent heightened inflation periods, a long last mile of disinflation, forecast error dynamics mirroring the ones recently observed in the United States, and lead to an asymmetry in the dynamics of inflation. 
In Appendix \ref{sec:example}, I show in a slightly stylized version of the model how the model can be represented in a standard AS-AD framework in which crossing the attention threshold leads to a steepening of both the AS and the AD curve---thus, the attention threshold gives rise to a \textit{dynamic} non-linearity.
Additionally, the heightened attention amplifies the inflationary effects of shocks.

\subsection{Crossing the attention threshold}\label{sec:irfs}
\begin{figure}[h]
\caption{Inflation dynamics}
\label{fig:pi_irfs}\vspace{0.05cm} \centering%
\begin{tabular}{c}
\includegraphics[width=.9\textwidth]{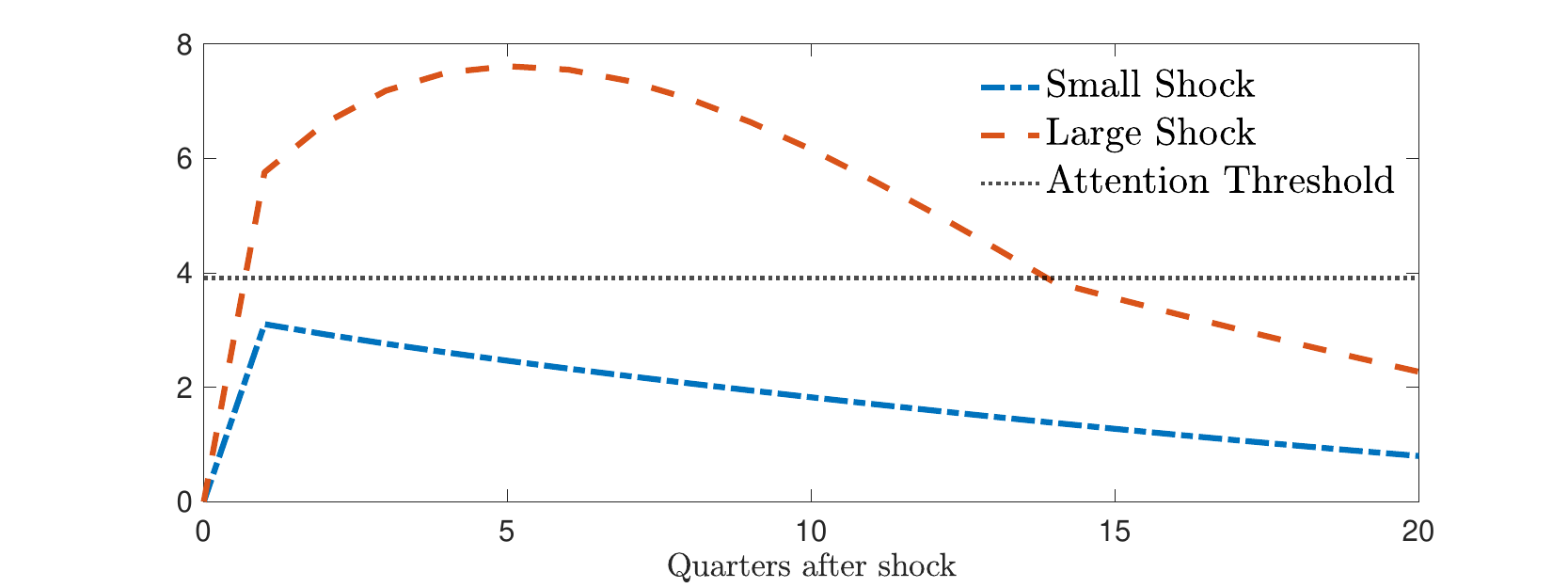}  
\end{tabular}%

\begin{minipage}{1\textwidth}
\footnotesize{\emph{Notes}: The blue-dashed-dotted line shows the (annualized) inflation response to a cost-push shock that does not push inflation above the attention threshold, and the red-dashed line shows the inflation dynamics for a cost-push shock that is large enough to push inflation above the threshold.
}%
 \end{minipage}
\end{figure}

\vspace{-0.2cm} 
Figure \ref{fig:pi_irfs} shows the inflation dynamics following a cost-push shock that pushes inflation in the first period above the attention threshold $\Bar{\pi}$ (the red-dashed line shows the inflation response, and the black-dotted line the attention threshold), as well as following a shock that does not push inflation above the threshold (blue-dashed-dotted line).

The two cases exhibit qualitatively different inflation dynamics. When attention remains low, which happens in response to the relatively small shock, inflation peaks on impact and then gradually returns back to its initial value of zero. After the larger shock, however, inflation keeps on increasing for about five periods before it peaks and then starts to decrease thereafter. While this decrease happens relatively fast at first, it slows down once inflation falls back below the attention threshold, and thus, inflation remains elevated quite persistently. These self-reinforcing dynamics in the first few periods would not be present in the case of deflationary shocks because attention to inflation stays constant in that scenario. Therefore, the model with the attention threshold offers a simultaneous explanation for the recent inflation surge as well as the 'missing deflation puzzle' in the aftermath of the Great Financial crisis (see, e.g., \cite{coibion2015phillips}). Additionally, the different inflation dynamics also mirror the ones discussed in Figure \ref{fig:emp_irfs_inflation_disentangle}, namely that inflation peaks on impact when attention is low whereas it is hump shaped in times of high attention.\footnote{Figure \ref{fig:pi_irfs_fire} shows the response of inflation when the shock hits in times of high-attention and illustrates that the inflation response is indeed hump shaped in that case.}

To understand the inflation surge in the model with the attention threshold, both regime switches that take place are key. The first regime switch occurs because the shock impulse is large enough to push inflation above the attention threshold. Thus, in the second period, agents are more attentive to inflation and hence, increase their inflation expectations more strongly in response to the forecast errors they make. For a given nominal rate, households now perceive the real rate to be lower and thus, increase their consumption in response. The attention regime change also matters for the supply side. Firms increase their inflation expectations more strongly which leads them to increase their prices more strongly. On top of that, the equilibrium inflation response to the cost-push shock in a given state of the economy is higher in the high-attention regime.
Hence, inflation keeps on increasing, further fueling higher inflation expectations, leading to additional inflation increases.

As the shock slowly dies out, inflation eventually starts to decline. This initial period of disinflation happens relatively quickly. The reason is that once inflation actually starts to come down, households and firm managers are still highly attentive and hence, revise their expectations downwards strongly. Thus, actual inflation falls relatively quickly.

However, once inflation falls back below the threshold $\Bar{\pi}$---and the second regime switch occurs---the speed of disinflation decreases. When the economy enters the low-attention regime, agents decrease their attention to inflation. Thus, they mainly stick to their prior beliefs and only slowly update their expectations. Their priors are now relatively high because of the high inflation period that agents just lived through, and therefore, their inflation expectations remain persistently high which hinders actual inflation from decreasing quickly.\footnote{\cite{lebow2024inflation} also show that the perceptions of inflation among households came down more slowly than actual inflation.}

\begin{figure}[h]
\caption{Inflation and inflation expectation dynamics}
\label{fig:fe}\vspace{.5cm} \centering%
\begin{tabular}{c}
Model \vspace{-0.1cm} \\
\includegraphics[width=.9\textwidth]{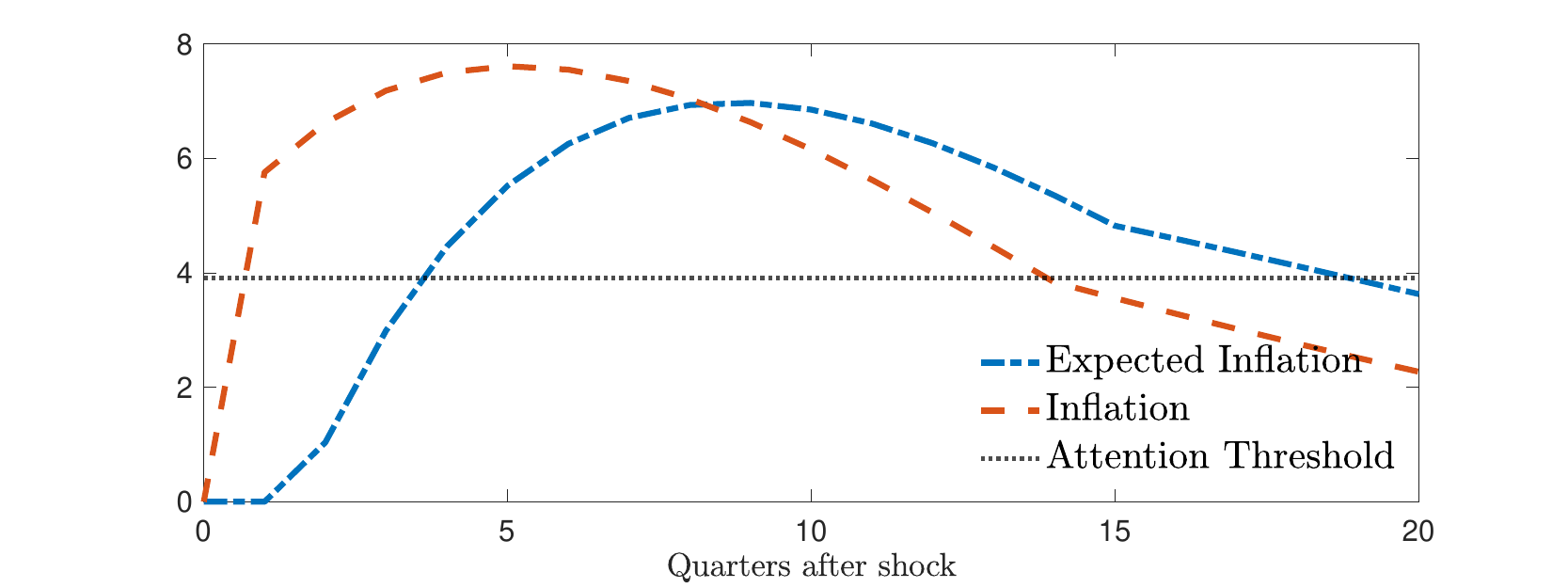} \vspace{0.2cm}\\ Data \vspace{-0.1cm} \\ \includegraphics[width=.9\textwidth]{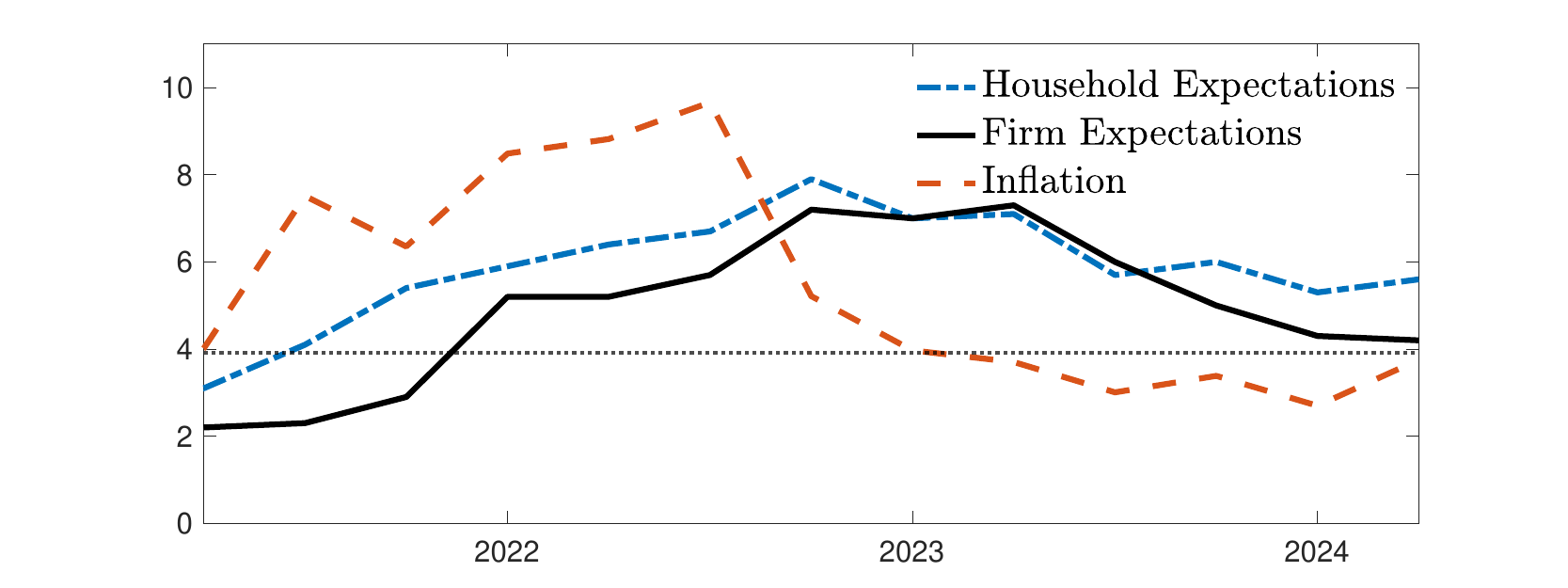} 
\end{tabular}%

\begin{minipage}{1\textwidth}
\footnotesize{\emph{Notes}: The upper panel shows the dynamics of inflation and inflation expectations after a cost-push shock that pushes inflation above the attention threshold. The lower panel plots quarter-on-quarter (annualized) CPI inflation and average inflation expectations for the period 2021Q1-2024Q1, as well as the estimated attention threshold at 3.91\%. In both panels, the inflation expectations are shifted such that the vertical distance between the two shows the respective forecast errors.
}%
 \end{minipage}
\end{figure}

To clearly see these patterns of inflation and inflation expectations, the upper panel in figure \ref{fig:fe} plots expected and actual inflation, jointly. Inflation expectations are shifted such that the vertical distance between the two lines captures the forecast errors. Initially, expected inflation does not quite catch up with actual inflation, leading to positive forecast errors. After some time---around 8 quarters after the shock---when inflation is already decreasing, expected inflation surpasses actual inflation. Hence, forecast errors become negative.

These patterns are consistent with the behavior of inflation and inflation expectations during the recent inflationary period, as documented in panel (b) in Figure \ref{fig:moti} in the Introduction. The lower panel in figure \ref{fig:fe} shows the same time series but for quarter-on-quarter inflation to make them consistent with the model (Figure \ref{fig:emp_irfs_expinf} in Appendix \ref{app:robustness_irfs} shows these patterns in the data conditional on an oil supply news shock).\footnote{\cite{angeletos2020imperfect} show that this initial undershooting followed by a delayed overshooting of expectations is a general pattern in response to shocks in the data.} The figure shows annualized quarter-on-quarter CPI inflation (red-dashed line) and average inflation expectations from the Michigan Survey (blue-dashed-dotted line) and from the Survey of Firms' Inflation Expectations (black-solid line) for the period from 2021Q1 until 2024Q1. Consistent with the model, inflation peaks about a year and a half after exceeding the threshold. During this inflation surge, inflation expectations lagged actual inflation. As inflation peaked and began to decline, however, inflation expectations started to surpass actual inflation after about two years after inflation exceeded the attention threshold. Consistent with the model, inflation peaks at a higher value than inflation expectations. The model's dynamics are also consistent with the empirical findings in \cite{blanco2022dynamics} who document, for a number of countries and different periods of inflation surges, that (i) inflation stays persistently high after the initial surge and (ii) that short-run inflation expectations initially fall short of actual inflation.

That inflation expectations stay persistently high may lead to a prolonged period of inflation stubbornly above the central bank's target even though the initial dis-inflationary phase happens relatively quickly. Put differently, once attention to inflation decreases again, the higher prior expectations after the inflation surge may render the `last mile' of inflation back to target more arduous. This also implies that the persistently higher inflation expectations give rise to a heightened risk of another subsequent inflation surge. The higher inflation expectations keep actual inflation higher for longer, and therefore, closer to the attention threshold. Thus, a subsequent inflationary shock is more likely to push inflation back above the threshold and therefore, leading to another episode of persistently high inflation.

In Appendix \ref{app:additonal_model}, I show that these inflation dynamics are unique to the model with the attention threshold. In particular, the models under full-information rational expectations (FIRE) or in which attention remains low predict that inflation peaks on impact and then comes down relatively quickly. Additionally, models with FIRE could not account for the observed dynamics of forecast errors. A model variant in which attention to inflation is always high, on the other hand, can produce the initial hump-shaped response but then predicts a fast decline back to target, i.e., this model would not explain why we see such a long last mile of disinflation. Furthermore, a model in which attention is always high would have predicted a very pronounced deflationary period after the Great Financial crisis. 

In Appendix \ref{app:additonal_model}, I further discuss (i) the inflation dynamics following a demand shock (Figure \ref{fig:pi_irfs_r}), (ii) what the model predicts when attention to the output gap decreases in times of high attention to inflation (Figure \ref{fig:pi_irfs_ychange}), and (iii) how the inflation dynamics look like when consumption expectations are formed under FIRE. In cases (i) and (ii), the inflation surge is somewhat more persistent but otherwise very similar to the baseline case of the discussed cost-push shock with constant attention to the output gap, whereas in (iii) the model produces slightly less endogenous persistence.

\subsection{Asymmetry}\label{sec:asy}\vspace{-0.2cm}
The attention threshold induces an asymmetry in the dynamics of inflation. When inflation exceeds the threshold, attention increases, leading to a period of persistently high inflation. If, however, inflation is particularly low, attention remains unchanged. For example, when a shock pushes inflation from 0\% to -4\%, attention does not change. Therefore, we would not see long-lived periods of pronounced deflation after such a shock, consistent with the fact the U.S.\ economy did not experience any long lasting deflationary periods since the second World War. These asymmetries matter for the overall properties of inflation. In particular, ignoring the attention threshold results in underpredicting the risk of persistently high inflation rates. 

\begin{figure}[h]
\caption{The attention threshold and inflation asymmetry}
\label{fig:dist_pi}\vspace{0.05cm} \centering%
\begin{tabular}{c}
\includegraphics[width=.6\textwidth]{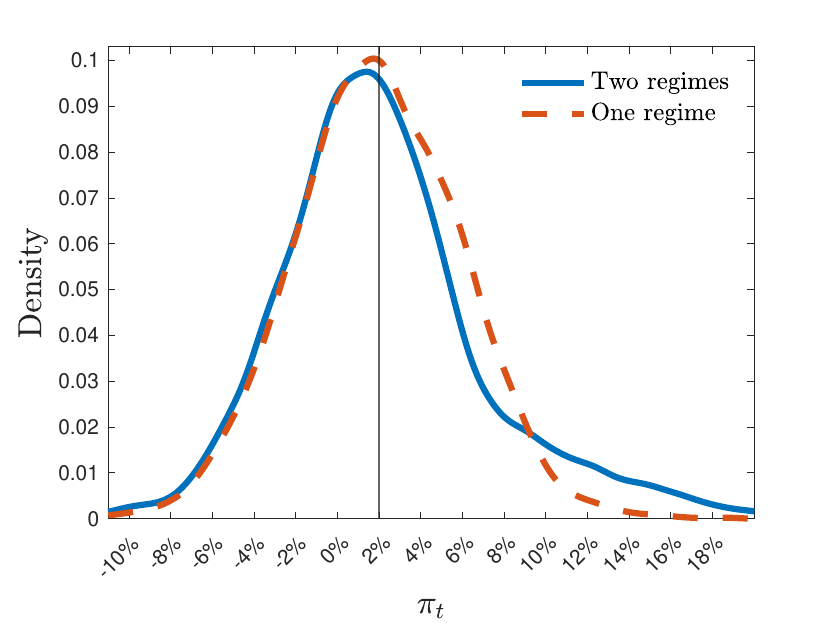}  
\end{tabular}%

\begin{minipage}{1\textwidth}
\footnotesize{\emph{Notes}: This figure shows the distribution of inflation (annualized) for the model with the attention threshold (solid-blue line) and the one without the attention threshold (dashed-red line).
}%
 \end{minipage}
\end{figure}

Figure \ref{fig:dist_pi} plots the distribution of inflation for the model with the attention threshold (blue-solid line) and for the limited-attention model without the threshold (red-dashed line). These figures are obtained by simulating the model for 10,000 periods by hitting it with random normally-distributed cost-push and demand shocks.

The model with the attention threshold has a substantially thicker right tail than the one without the attention threshold, as is the case empirically.\footnote{Note, that even though the volatility of cost-push shocks is slightly higher in the higher-attention regime, the asymmetry also arises when shocks are drawn from the same distribution in both regimes due to the attention threshold.} The probability that inflation exceeds 8\% (annualized) is about 8.2\% in the model with the threshold, and thus substantially closer to its empirical counterpart of 12\% than in the model without the threshold in which inflation exceeds 8\% only 2\% of the time.  When it comes to periods of deflation, the two models exhibit very similar properties. For example, both models predict inflation to fall below -8\% in only 2-3\% of the periods, compared to 0.5\% empirically, and the median in both models is very close to 0.\footnote{The mean is above 0 in the model with the threshold, whereas it is practically 0 in the model without the threshold. I discuss this in more detail in Section \ref{sec:welfare} where I derive the normative implications of these results.} Thus, inflation in the model without the attention threshold is symmetric around its steady state value of 0, whereas it features a substantially higher risk of periods of high inflation in the model with the threshold. 

\section{Monetary Policy Implications}\label{sec:welfare}\vspace{-0.2cm}
I now characterize the normative implications of the inflation attention threshold and the corresponding changes in attention. To compare different policy rules, I define the central bank's loss function as
\begin{equation}
-\frac{1}{2}E_0\sum\limits_{t=0}^{\infty}\beta^t \left[\pi^2_t+\Lambda \widehat{x}_t^2\right], \label{lossfct}
\end{equation}
where $\Lambda$ is the relative weight of the output gap, which I set to $\Lambda = 0.007$ as in \citet{adam2006optimal}.
The policymaker wants to maximize \eqref{lossfct} subject to the private sector's optimality conditions, characterized by equations \eqref{eq:nkpc} and \eqref{eq:euler}. But I assume that the planner wrongly believes that the private sector holds rational expectations. The problem from the policymaker's perspective is then identical to the one in \cite{clarida1999science} who show that the optimal policy is given by 
\begin{equation}
    \pi_t + \tfrac{\Lambda}{\kappa}\left(\widehat{x}_t - \widehat{x}_{t-1}\right) = 0 \notag
\end{equation}
if the policymaker can fully commit to its policy. If the policymaker cannot commit, the optimal policy rule under discretion is given by
\begin{equation}
    \pi_t + \tfrac{\Lambda}{\kappa}\widehat{x}_t = 0. \notag
\end{equation}

In the following, I compare the implications of these two policy rules to the Taylor rule \eqref{eq:taylor} as well as to a Taylor rule without interest rate smoothing and no response to the output gap, i.e., to \eqref{eq:taylor} where $\rho_i = \phi_x = 0$, as well as to a strict-inflation targeting regime in which inflation is kept at zero at all times. Table \ref{tab:rules} summarizes these different policy rules.

\begin{table}[h]
 \caption{Monetary policy rules}\label{tab:rules}
 \centering
\begin{tabular}{lll}
\hline \hline
 Nr.\ & Name & Equation\\\hline\vspace{-0.4cm}\\
(1) & Taylor rule with smoothing & $\tilde{i}_t = \rho_i \tilde{i}_{t-1} + (1-\rho_i)\left(\phi_{\pi}\pi_t + \phi_x \widehat{x}_t 
\right)$\\\vspace{-0.4cm}\\
(2) & Taylor rule without smoothing & $\tilde{i}_t = \phi_{\pi}\pi_t  $\\\vspace{-0.4cm}\\
(3) & Optimal RE commitment policy & $\pi_t + \tfrac{\Lambda}{\kappa}\left(\widehat{x}_t - \widehat{x}_{t-1}\right) = 0$ \\\vspace{-0.4cm}\\
(4) & Optimal RE discretionary policy & $\pi_t + \tfrac{\Lambda}{\kappa} \widehat{x}_t = 0$ \\\vspace{-0.4cm}\\
(5) & Strict inflation targeting & $\pi_t = 0$ \\\vspace{-0.4cm}\\
\hline \hline
\end{tabular}%
\end{table}

\begin{figure}[h]
\caption{Implications of different monetary policy rules}
\label{fig:wf}\vspace{0.05cm} \centering%
\begin{tabular}{ccc}
& (a) Central bank loss \\
\multicolumn{3}{c}{\includegraphics[width=.9\textwidth]{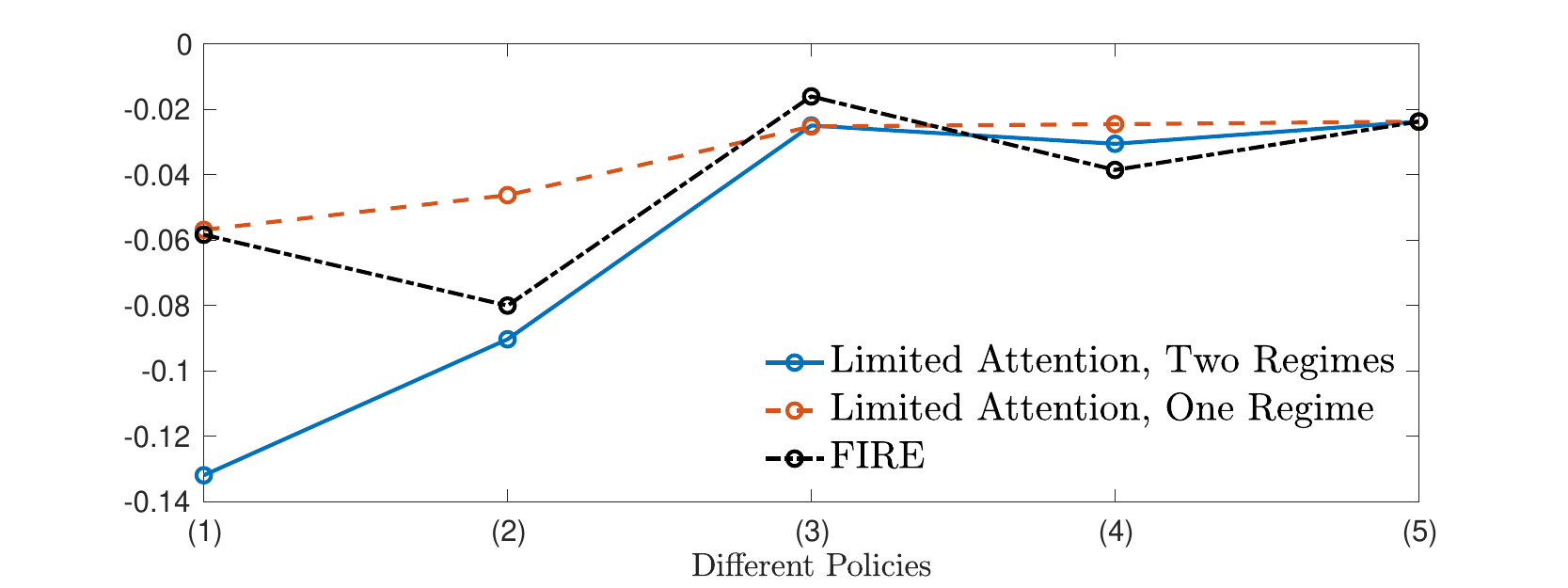}  }
\\
(b) Inflation volatility & (c) Average inflation & (d) Frequency high-attention \\
\includegraphics[width=.3\textwidth]{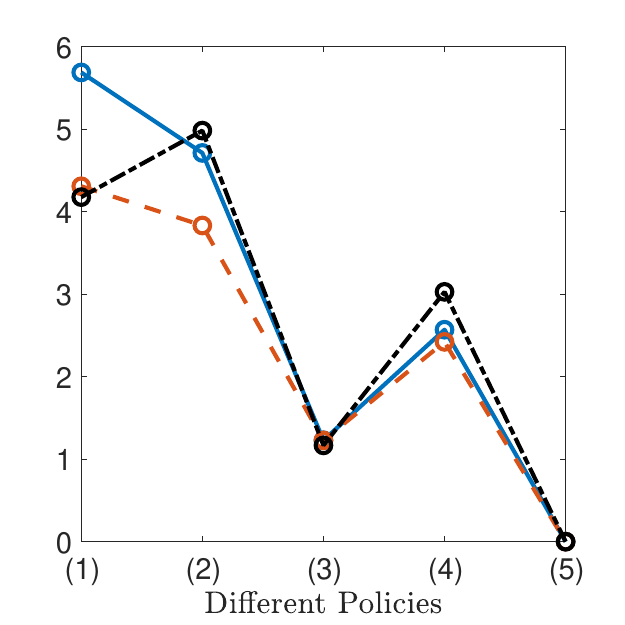} &    \includegraphics[width=.3\textwidth]{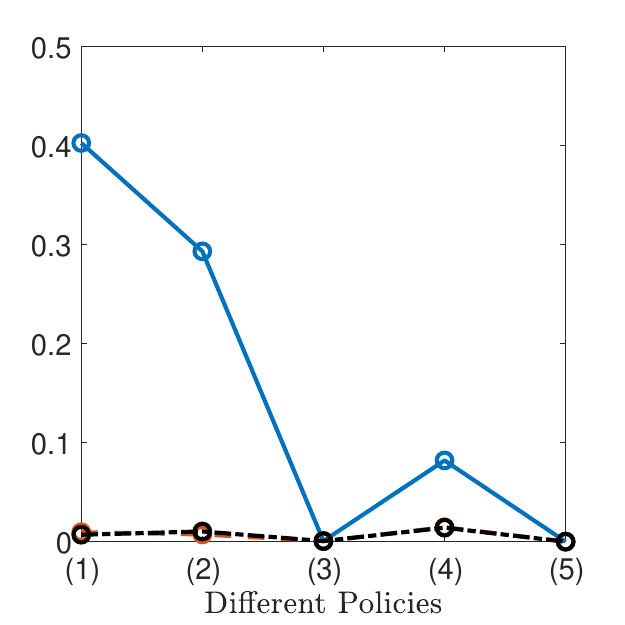} & \includegraphics[width=.3\textwidth]{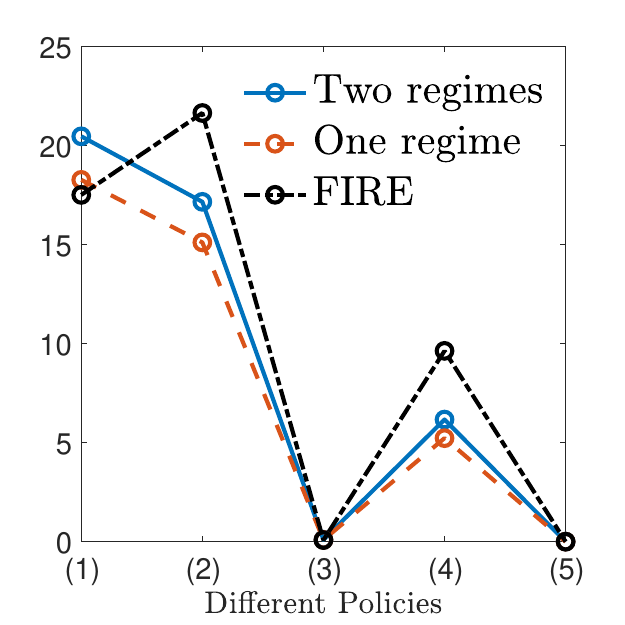}

\end{tabular}%

\begin{minipage}{1\textwidth}
\footnotesize{\emph{Notes}: This figure shows the central bank loss (using equation \eqref{lossfct} and shown in panel (a)), inflation volatility (panel (b)), average inflation (panel (c)), and the frequency of being in the high-attention regime (panel (d)) for the five different monetary policy rules from Table \ref{tab:rules}.
}%
 \end{minipage}
\end{figure}

I then simulate the economy for 10,000 periods for each of the different policy rules. 
Panel (a) in Figure \ref{fig:wf} plots the central bank's losses \eqref{lossfct} for the model with the attention threshold (blue-solid line), with limited attention but absent the attention threshold (red-dashed line) and the model under FIRE (black-dashed-dotted line) for the 5 different policy rules. Panel (b) shows the respective inflation volatilities, panel (c) the average level of inflation, and panel (d) the frequency of how often inflation is above the threshold of 4\%.

There are two main takeaways from Figure \ref{fig:wf}. First, following simple Taylor rules (policy rules (1) and (2)) leads to substantially larger central bank losses compared to optimal policy rules or a strict inflation targeting rule in the model with the attention threshold, especially for the case with interest-rate smoothing.\footnote{\cite{gati2020monetary} shows that in her model with varying degrees of long-run-expectations anchoring, losses of Taylor rules with a low inflation-response coefficient may go to infinity, as in this case the model becomes unstable and inflation (and the output gap and interest rates) becomes explosive. Similarly, \cite{benchimol2023optimal} find that under inattention to prices, central bank losses under Taylor rules are higher than with other policy rules.} The reason is that the economy spends a substantial amount of time in the high-attention regime in which inflation is high and volatile (see panel (d) in figure \ref{fig:wf}). Due to the asymmetry of the attention threshold the {average level of inflation} is higher when the economy is in the high-attention regime frequently (panel (c)). This is in stark contrast to the model without the threshold or the one with fully-informed rational agents, where inflation fluctuates symmetrically around zero. Interest rate smoothing introduces additional persistence, such that the periods in the high-attention regime last longer and thus, the average level of inflation as well as its volatility increase.

The second main take away is that these losses can be mitigated when monetary policy follows one of the rules (3)-(5). In these cases, inflation is relatively stable and fluctuates almost symmetrically around 0, as the economy very rarely stays in the high-attention regime. In the case of the strict-inflation targeting regime (5) the inflation volatility is exactly 0. However, in that case, the output gap is more volatile (not shown) such that the overall losses are very similar to the ones under the other policy rules from (3) and (4).

Overall, Figure \ref{fig:wf} illustrates that following simple Taylor rules, especially ones with interest-rate smoothing and relatively low inflation-response coefficients, can lead to large central bank losses when the public's attention to inflation increases during times of high inflation. Policies that are more hawkish and induce much smaller inflation fluctuations, in contrast, can mitigate the potentially detrimental effects of sudden increases in attention much more effectively, or even prevent these episodes from happening completely.

\section{Conclusion}\label{sec:conclusion}\vspace{-0.2cm}
The recent inflation surge brought inflation back on people's minds. In this paper, I quantify the inflation attention threshold after which people start to pay more attention to inflation. I estimate this attention threshold to be at an inflation rate of about 4\% and that attention doubles from the low-attention regime to the high-attention regime. Supply shocks become twice as inflationary in the high-attention regime and I find that the change in people's attention in early 2021 likely doubled the inflationary effects of supply shocks.

A New Keynesian model that accounts for the inflation attention threshold can replicate the empirical findings and produces inflation and inflation expectation dynamics that are consistent with the ones recently observed in the U.S. As inflation exceeds the threshold, the increase in people's attention leads to an endogenous amplification of the initial inflation surge. After inflation peaks, inflation comes back down relatively quickly initially, but once attention to inflation decreases again, the
higher prior expectations after the inflation surge may render the `last mile' of inflation back to target more arduous.

Accounting for varying attention levels also matters for the model's normative predictions. I show that that following simple Taylor rules leads to substantially larger central bank losses compared to more hawkish rules. While I do not characterize fully optimal monetary policy when people's attention to inflation is subject to sudden changes, doing so and studying how the conduct of monetary policy feeds back into people's attention is a promising direction for future research.

\singlespacing
\bibliographystyle{ecta}
\bibliography{a_olibib}
\onehalfspacing
\clearpage \newpage \appendix 
\begin{center}
    \title{\Huge{\textbf{Online Appendix}} }
\end{center}
\section{News Coverage of Inflation}\label{app:news}
\begin{figure}[h]
\caption{News coverage of inflation is higher in times of high inflation}
\label{fig:nyt}\vspace{0cm} \centering%
\begin{tabular}{c}
\includegraphics[width=.9\textwidth]{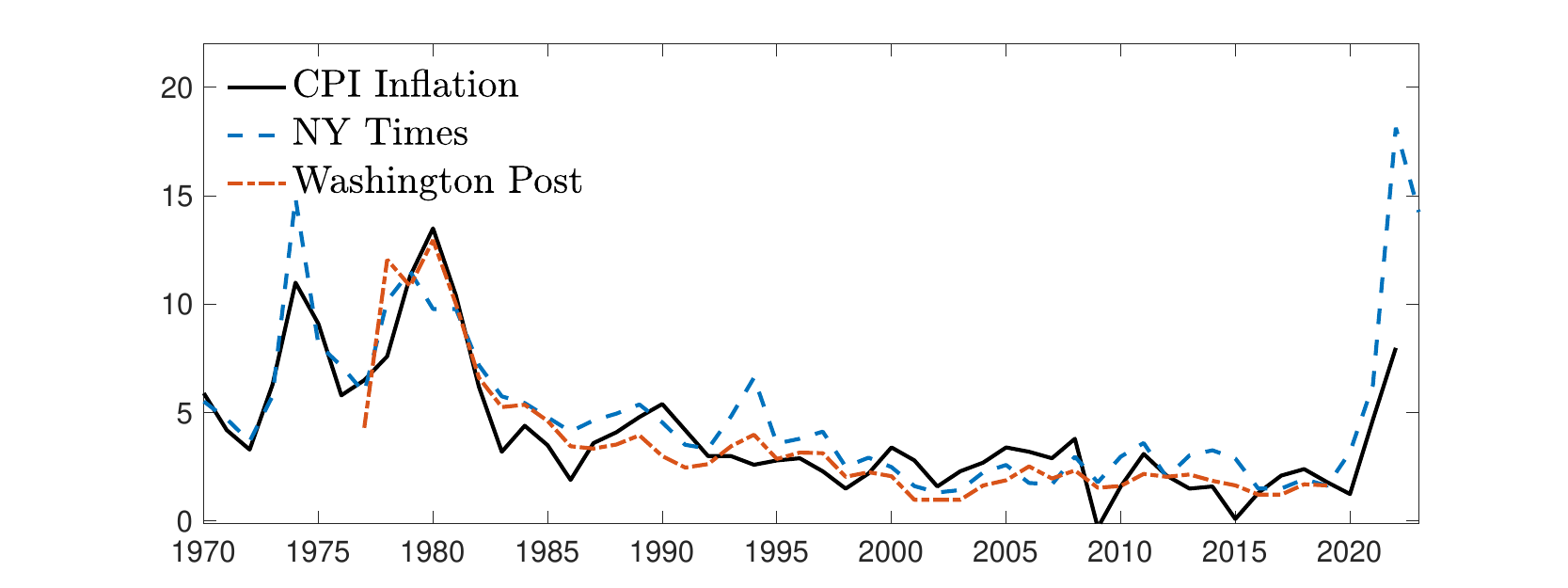}\vspace{-0cm}
\end{tabular}\\
\begin{minipage}{1\textwidth}
\footnotesize{\emph{Notes}: The black-solid line shows the annual CPI inflation rate for the U.S.\ since 1970, the blue-dashed line and the red-dashed-dotted line show the frequency of the word inflation in the New York Times and the Washington post, respectively.
}\vspace{-0.25cm}
 \end{minipage}
\end{figure}

\vspace{-0.2cm}
Figure \ref{fig:nyt} shows that media reporting about inflation is higher in times of higher inflation. The Figure shows the frequency of the word ``inflation'' in the New York Times (1970 to July 2023, blue-dashed line) and the Washington Post (1977-2019, red-dashed-dotted line), together with annual CPI inflation (black-solid line). There is a very strong positive correlation between inflation and news coverage of inflation (the correlation between CPI inflation and the two news-coverage series is 0.86 for the New York Times and 0.90 for the Washington Post). 

News coverage of inflation is also substantially higher in times inflation is above the attention threshold. I find that the average frequency of the word ``inflation'' is 2.7 times as high for the New York Times and 2.9 times as high for the Washington Post when CPI inflation is above 4\% in that year compared to years in which CPI inflation is below 4\%.

Consistent with this, \cite{bracha2023inflation} show that higher inflation rates indeed lead to more media reporting about inflation, and \cite{lamla2014role} find that more intensive news reporting about inflation improves the accuracy of consumers' inflation expectations, consistent with agents being more attentive. \cite{guillochon2023inflation} find that more exposure to inflation news leads to an increase in the probability that households pay attention to inflation. \cite{schmidtinflation} show that during episodes of intensive newspaper coverage of inflation, news reporting has strong effects on inflation expectations but not during other episodes. \cite{larsen2021news} find that news media coverage predicts households’ inflation expectations, and \cite{nimark2019news} show that major events (such as strong inflation increases) lead to a shift in the news focus towards these events.

\begin{figure}[h]
\caption{News coverage of inflation during Covid}
\label{fig:nyt_m}\vspace{0cm} \centering%
\begin{tabular}{c}
\includegraphics[width=.9\textwidth]{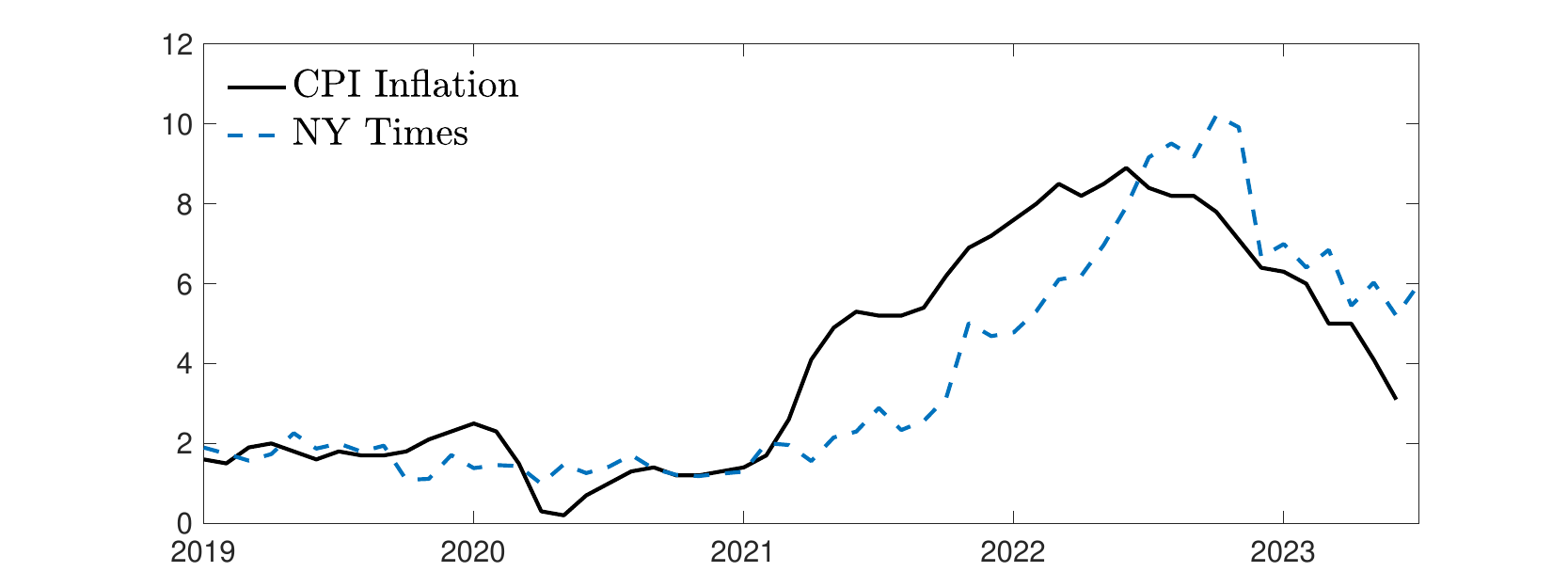}\vspace{-0cm}
\end{tabular}\\
\begin{minipage}{\textwidth}
\footnotesize{\emph{Notes}: The black-solid line shows the monthly year-on-year CPI inflation rate for the U.S.\ and the blue-dashed line the frequency of the word inflation in the New York Times, normalized to have the same standard deviation as CPI inflation.
}\vspace{-0.25cm}
 \end{minipage}
\end{figure}

\paragraph{Covid.} 
Figure \ref{fig:nyt_m} shows news coverage of inflation in the New York Times for the period 2019 until July 2023 at monthly frequency (blue-dashed line) together with monthly year-on-year CPI inflation. The figure shows a strong positive correlation of 0.85 between inflation and news coverage, and that at the peak, news coverage of inflation quintupled from its pre-pandemic level. Consistent with the theory, news coverage seems to lag inflation slightly (the correlation between news coverage and \textit{lagged} inflation is also slightly higher than the one with current inflation, 0.9 instead of 0.85). The correlation of news coverage with Google searches is 0.94 for the period 2017-2023, further supporting the assumption that news coverage is highly correlated with the public's attention to inflation. 

\clearpage\newpage

\section{Appendix to Empirical Section \ref{sec:data}}\label{ap:data}\vspace{-0.2cm}

\paragraph{One-year-ahead expectations.} As my baseline in Section \ref{sec:data}, I focus one one-quarter-ahead expectations and compare them to quarter-on-quarter inflation. When using one-year-ahead expectations and comparing them to year-on-year inflation instead (and using last month's y-o-y inflation), I estimate an attention threshold at 5.03\% and that attention increases from 0.54 to 1.22 from the low-attention to the high-attention regime. The difference between the two is highly statistically significant with an associated $p$-value of 0.000.

\paragraph{Heterogeneity.} In the main text, I focused on average (and median) expectations. I now test whether the attention threshold and the different attention levels depend on people's gender or their age.\footnote{There is a vast literature documenting heterogeneity in inflation expectations, see, e.g., \citet{d2019cognitive,broer2021information,pfauti2022behavioral, d2023data,pedemonte2023aggregate,weber2022subjective,roth2023effects, nord2022cares, meichtry2022sticky}, for recent contributions.} As documented in \cite{d2021gender}, gender and gender roles (e.g., with respect to grocery shopping) play a big role in explaining differences in how men and women form their inflation expectations. I find that men have a higher attention threshold than women (4.4\% vs.\ 3.9\%), indicating that women increase their attention somewhat earlier than men. This might be explained by the fact that women are more likely to go grocery shopping than men \citep{d2021gender} and therefore, experience price changes more directly than men.
I further find that the threshold for younger people (aged 18-34) is lower than for older people (4.44\% vs.\ 6.8\% for people aged between 35 and 54, and 5.7\% for people aged older than 55), but that their attention levels tend to be lower overall (0.21 vs.\ 0.24 below the threshold and 0.41 vs.\ 0.74 above it).

\clearpage\newpage
\section{Additional Results and Robustness to Section \ref{sec:empirical_irfs}}\label{app:robustness_irfs}\vspace{-0.2cm}
\subsection{Other shocks}\vspace{-0.2cm}
\paragraph{Inflation shock.} \cite{angeletos2020business} estimate a VAR including 10 key macroeconomic variables and then identify different shocks by maximizing their contribution to the volatility of a given variable over business-cycle frequency (6-32 quarters). I use their shock that contributes most to the volatility of inflation (using the GDP deflator, as in \cite{angeletos2020business}). These shocks are available at quarterly frequency and span the period 1960 until the end of 2017. I use the previous quarter's CPI inflation as an indicator whether the economy is in the high-attention regime or the low-attention regime (i.e., was the previous annualized CPI inflation rate above or below 4\%.)

\begin{figure}[!ht]
\caption{Price level response to the shock targeting inflation}
\label{fig:emp_irfs_infshock}\vspace{0.15cm} \centering%
\begin{tabular}{cc}
(a) High-Attention Regime  & (b) Low-Attention Regime  \\ 
\includegraphics[width=.44\textwidth]{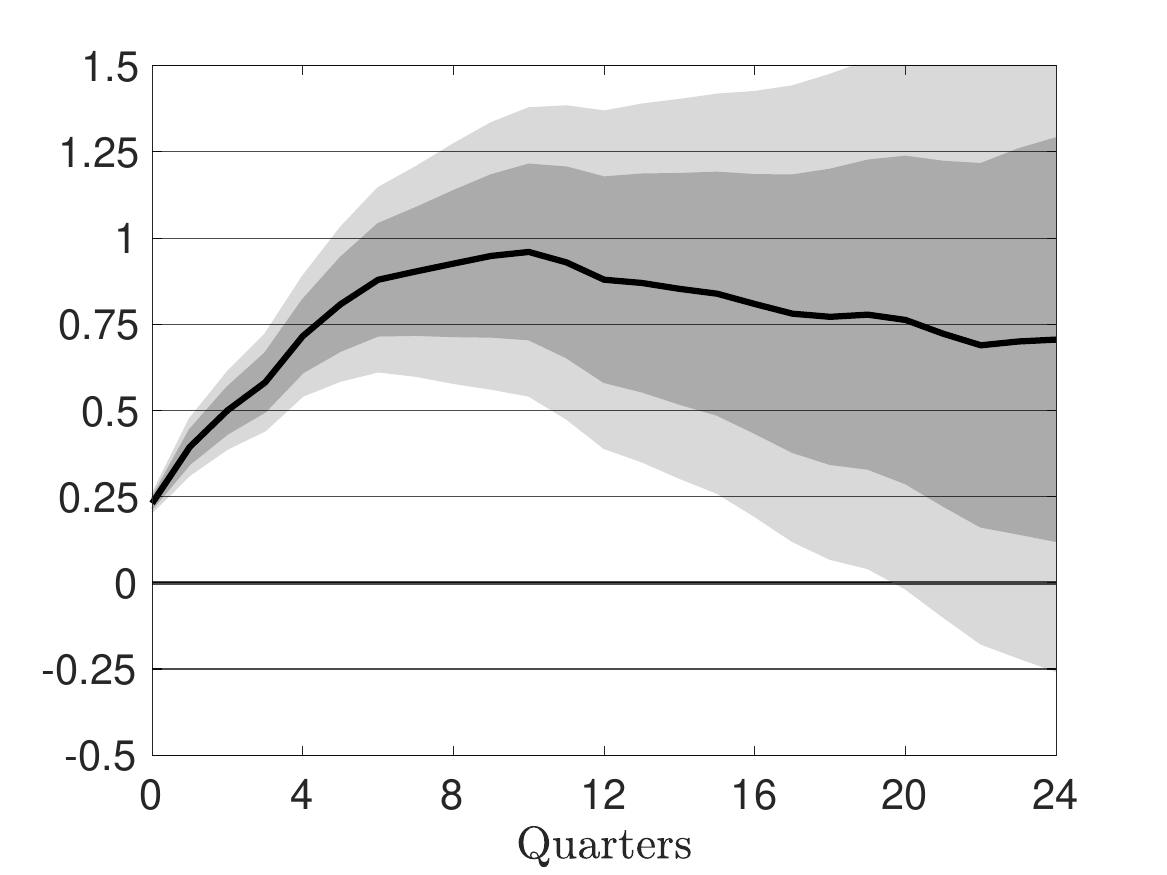} &    \includegraphics[width=.44\textwidth]{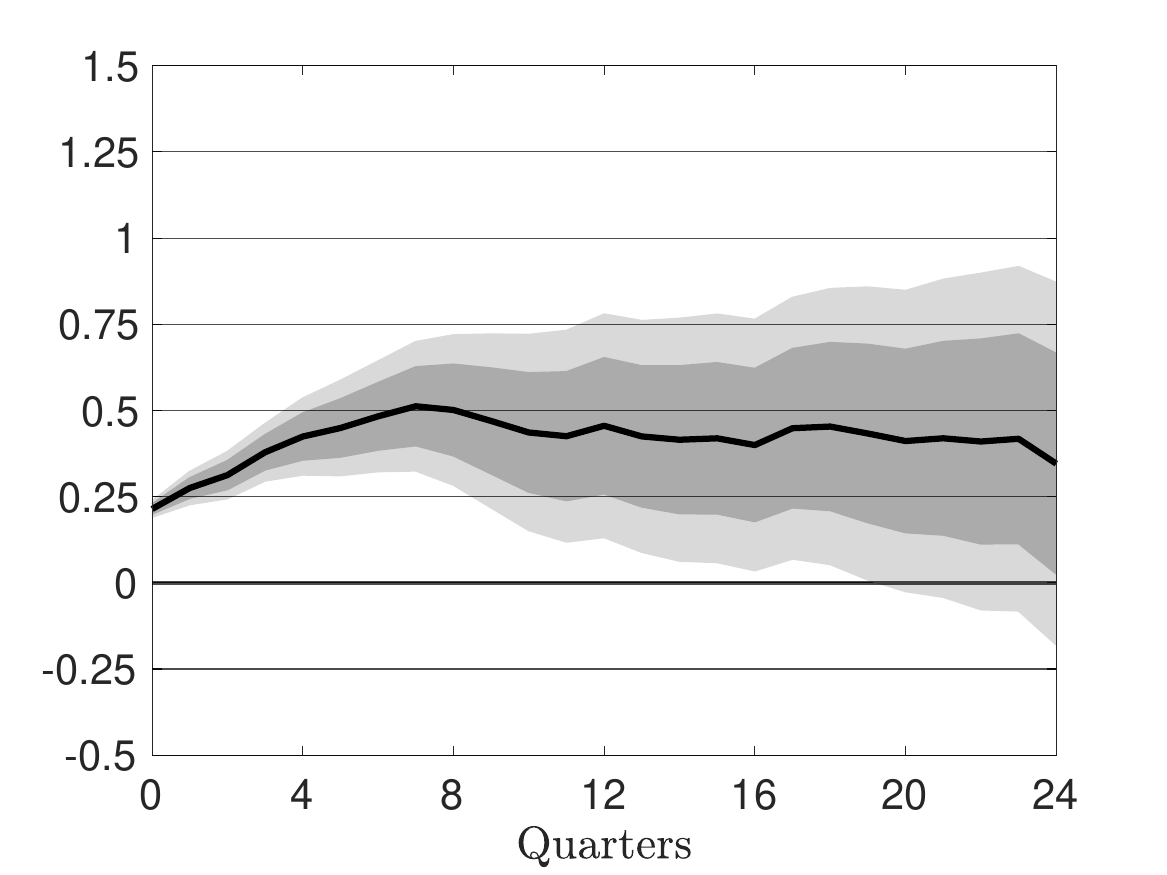} \\
(c) Average effect  & (d) Difference  \\ \includegraphics[width=.44\textwidth]{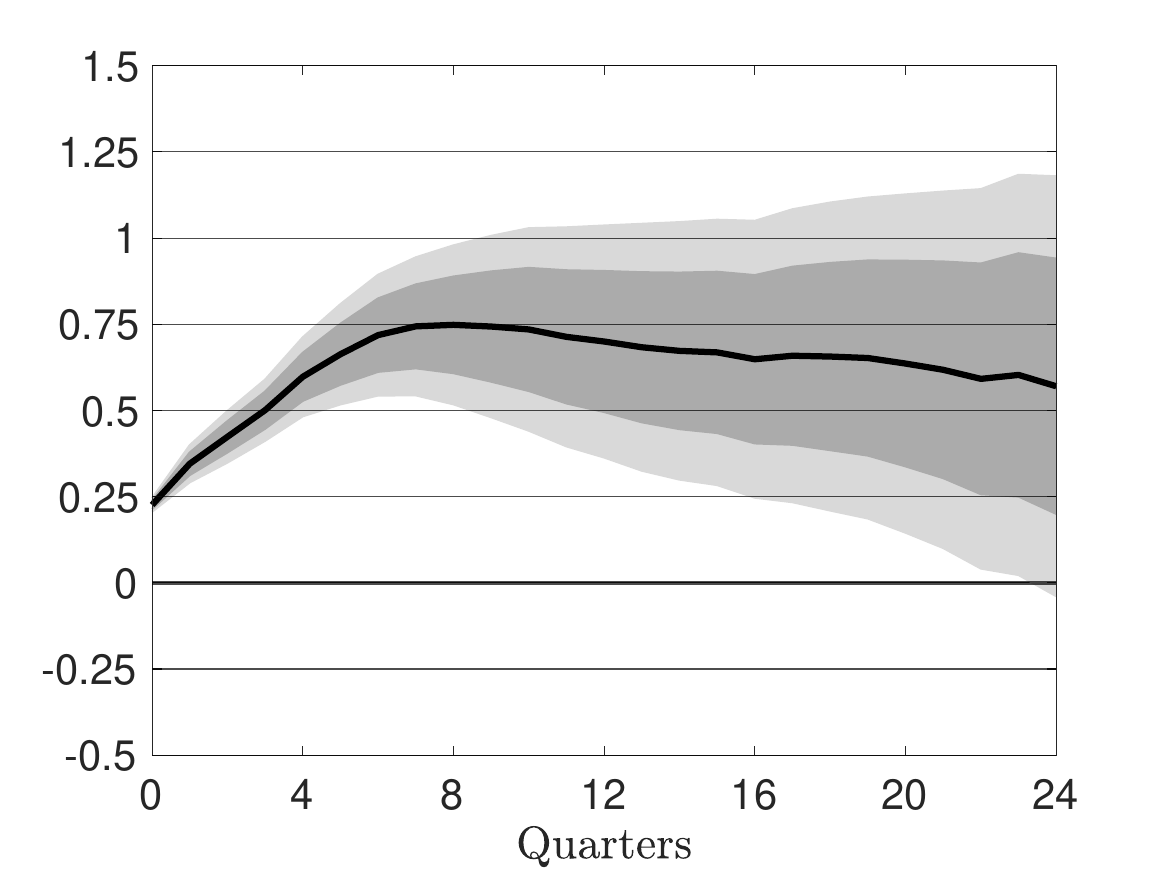} & \includegraphics[width=.44\textwidth]{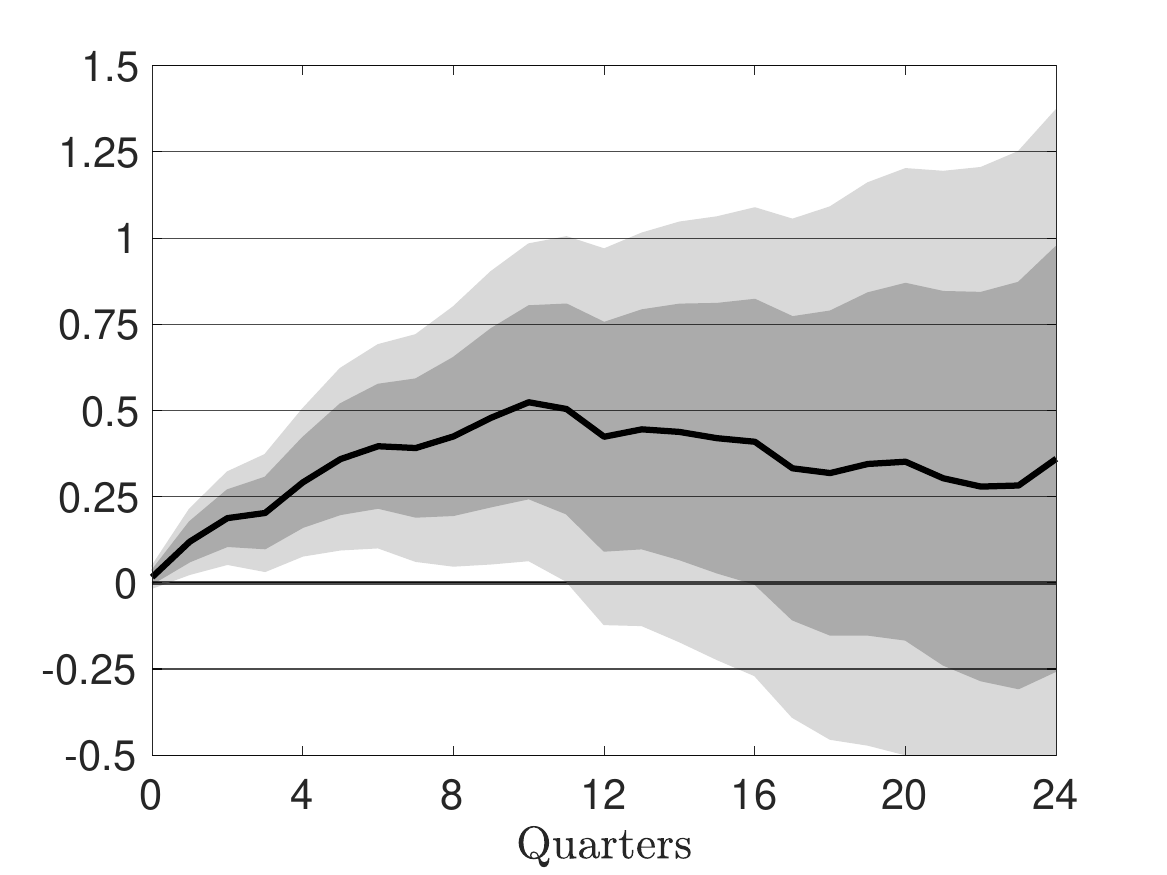}
\end{tabular}%

\begin{minipage}{1\textwidth}
\footnotesize{\emph{Notes}: This figure shows the cumulative price level (using the GDP deflator) response to the shock of \cite{angeletos2020business} that targets inflation in the high-attention regime (panel (a)), the low-attention regime (panel (b)), on average across regimes (panel (c)), and the difference between the two regimes (panel (d)). The dark shaded areas depict the 68\% confidence bands and the light-shaded area the 90\% confidence bands. Standard errors are robust with respect to serial correlation and heteroskedasticity (\cite{newey1987simple} with 4 lags). The attention regimes ares defined based on the previous quarter's CPI inflation rate. }%
 \end{minipage}
\end{figure}

The dependent variable is the change in the log of the GDP deflator (times 100) from quarter $t-1$ to quarter $t+h$.
Figure \ref{fig:emp_irfs_infshock} shows the results. 
As in the specification with oil news shocks, inflation responds about twice as much to the \textit{inflation shock} when hit in the high-attention regime compared to the low-attention regime. As panel (d) shows, these differences are highly statistically significant and quite persistent: the difference is largest 10 quarters after the shock and still highly statistically significant.

\begin{figure}[!ht]
\caption{Inflation response to an monetary policy shock}
\label{fig:emp_irfs_inflation_mp}\vspace{0.15cm} \centering%
\begin{tabular}{cc}
(a) High-Attention Regime  & (b) Low-Attention Regime  \\ 
\includegraphics[width=.44\textwidth]{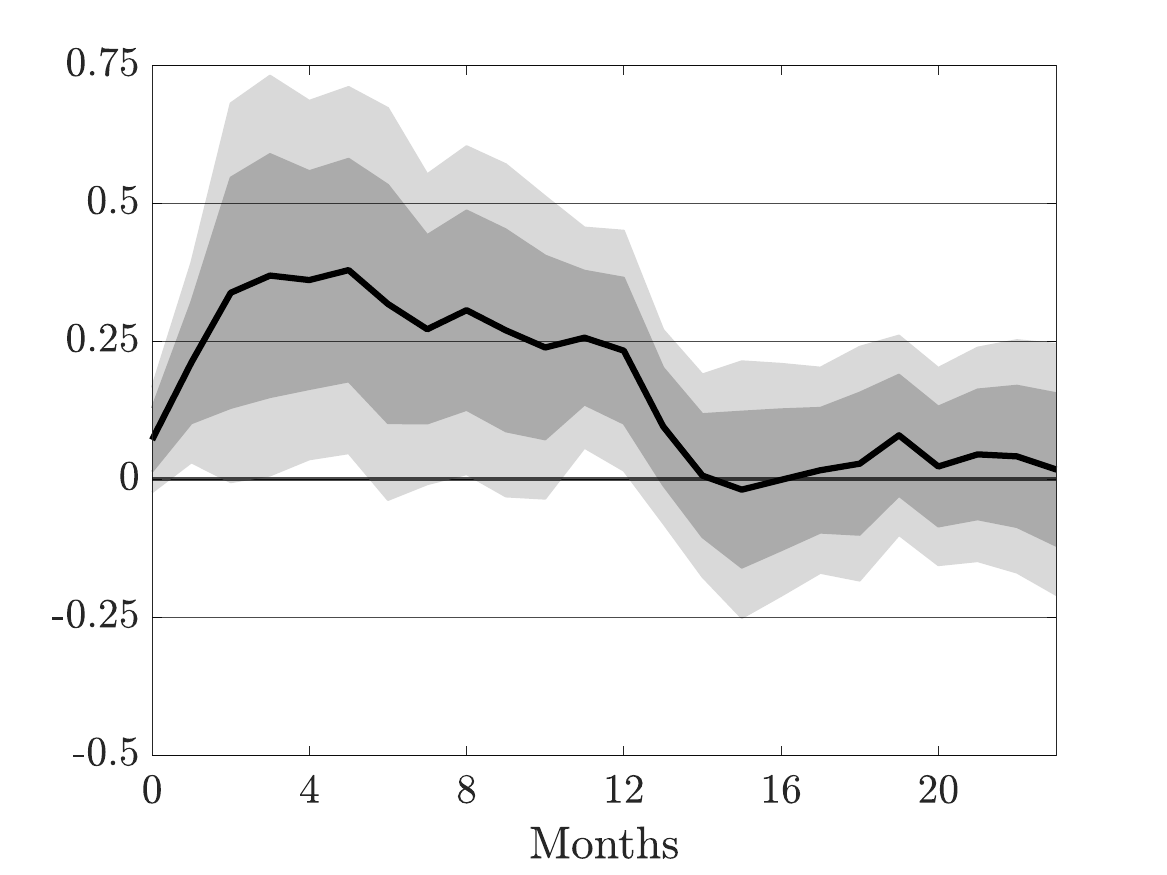} &    \includegraphics[width=.44\textwidth]{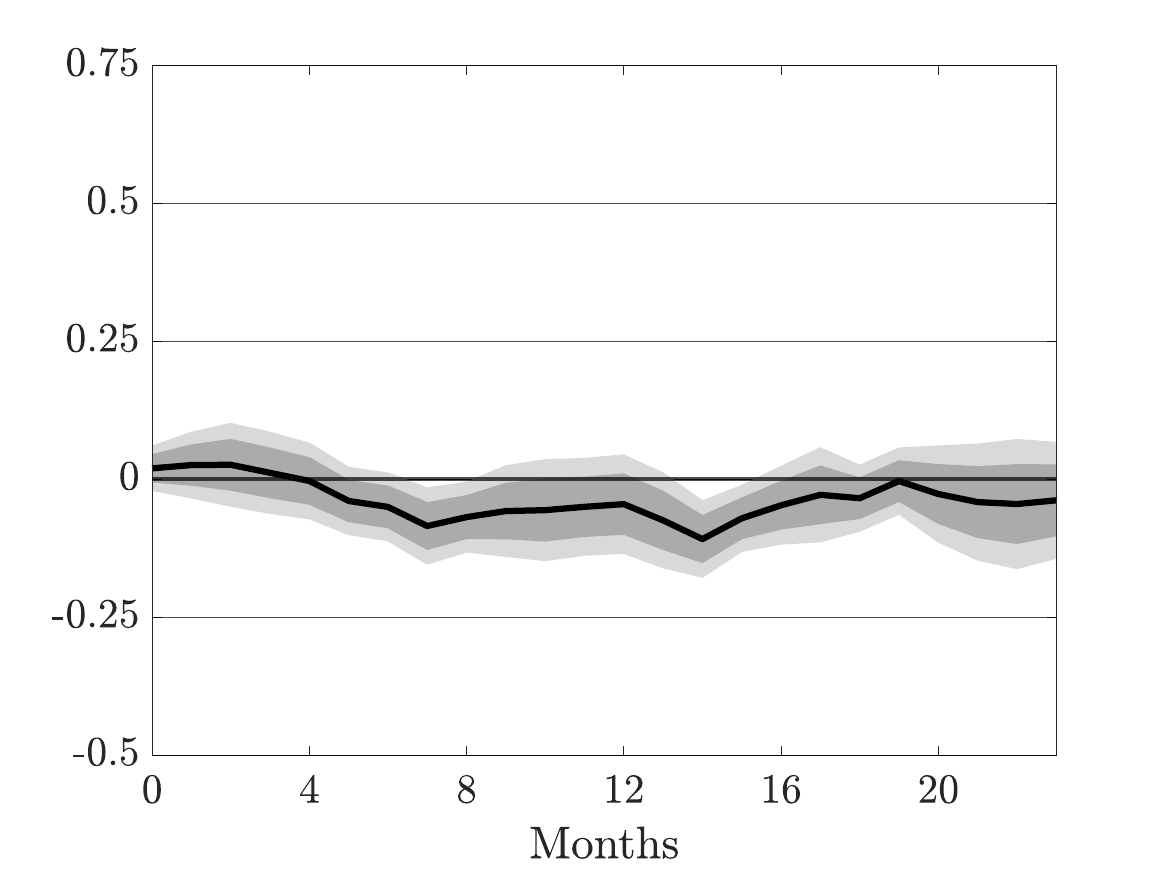} \\
(c) Average effect  & (d) Difference \\ \includegraphics[width=.44\textwidth]{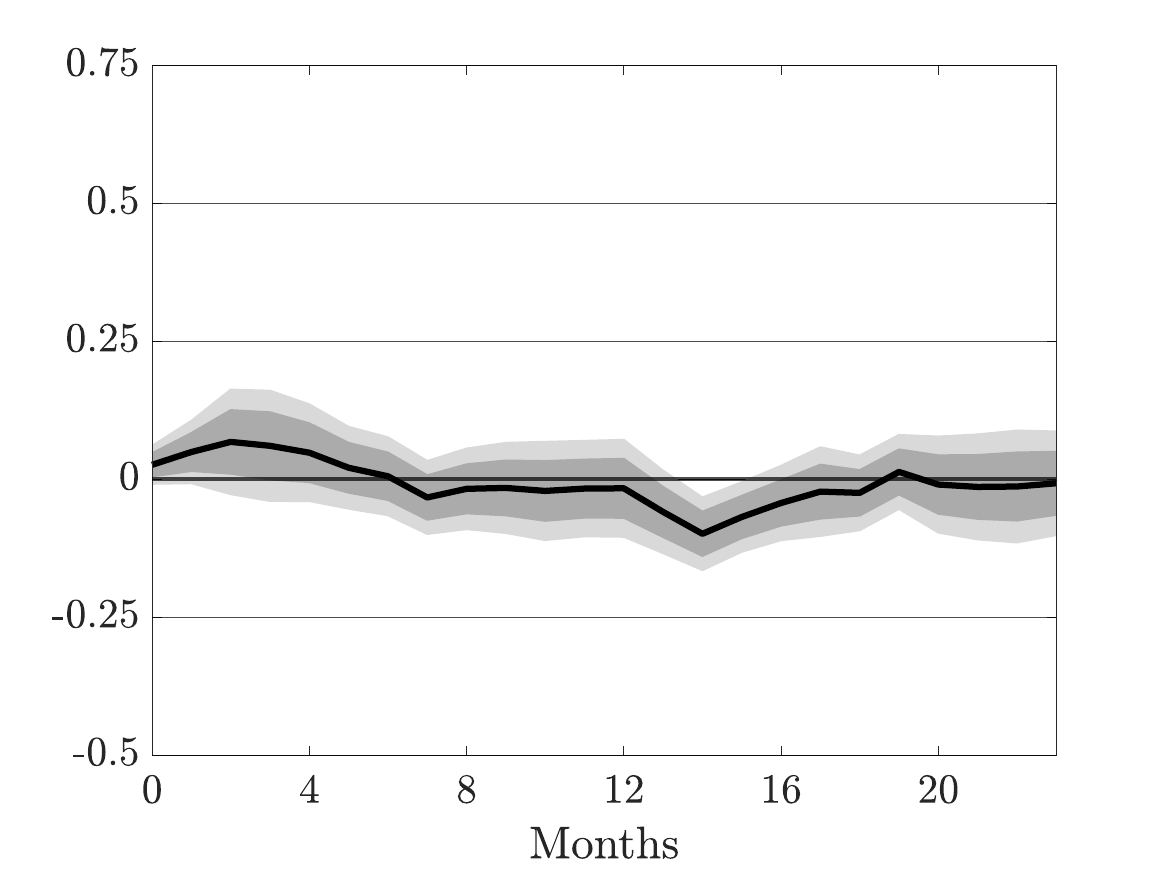}  & \includegraphics[width=.44\textwidth]{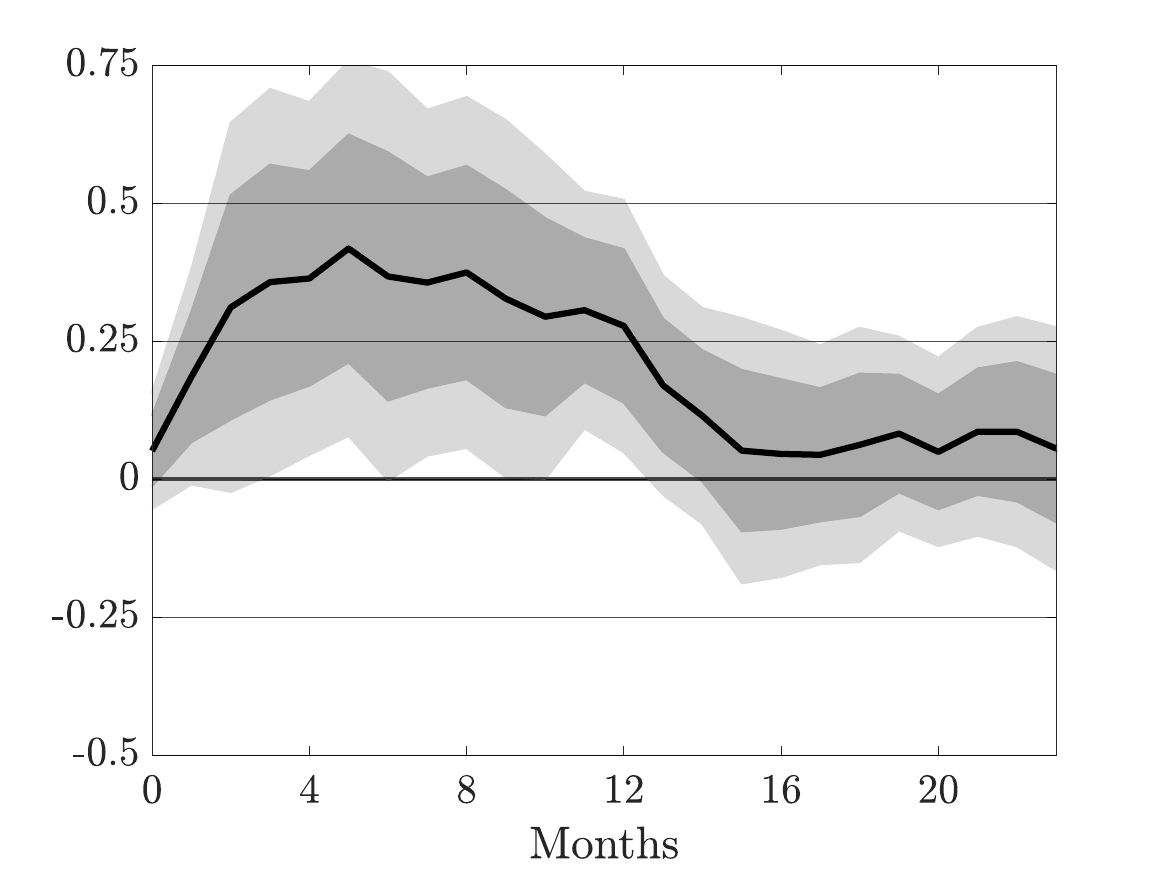} 
\end{tabular}%

\begin{minipage}{1\textwidth}
\footnotesize{\emph{Notes}: This figure shows the inflation response to a monetary policy shock in the high-attention regime (panel (a)), the low-attention regime (panel (b)), on average across regimes (panel (c)), and the difference between the two regimes (panel (d)). The dark shaded areas depict the 68\% confidence bands and the light-shaded area the 90\% confidence bands. Standard errors are robust with respect to serial correlation and heteroskedasticity (\cite{newey1987simple} with 12 lags). The shocks are from \cite{jarocinski2020deconstructing} and are based on a high-frequency identification and cleaned from information effects of monetary policy.
}%
 \end{minipage}
\end{figure}

\paragraph{Monetary policy shocks.} In figure \ref{fig:emp_irfs_inflation_mp}, I show that inflation also responds more strongly in the high-attention regime to monetary policy shocks identified using a high-frequency identification (the shocks are taken from \cite{jarocinski2020deconstructing} and are purged from the information effects of monetary policy statements; I show here the shocks identified using sign restrictions, but the results are practically identical when using the shock series based on the ``poor-man approach''). A drawback of using these monetary policy shocks, however, is that they are only available for the period 1990-2019. In particular, there are 49 months out of 353 in which quarter-on-quarter CPI inflation was above 4\% during that period, so about 10\% of the time, whereas in the full sample the economy spends about 30\% of the time in the high-attention regime. Thus, the sample does not include most of the high-inflation periods. Therefore, while the differences across regimes are substantial in magnitude, the differences are less statistically significant.

\clearpage\newpage

\subsection{Robustness analysis}
\begin{figure}[!ht]
\caption{Inflation response to an oil supply shock with Google data as threshold-defining variable}
\label{fig:emp_irfs_inflation_google}\vspace{0.15cm} \centering%
\begin{tabular}{cc}
(a) High-Attention Regime  & (b) Low-Attention Regime  \\ 
\includegraphics[width=.44\textwidth]{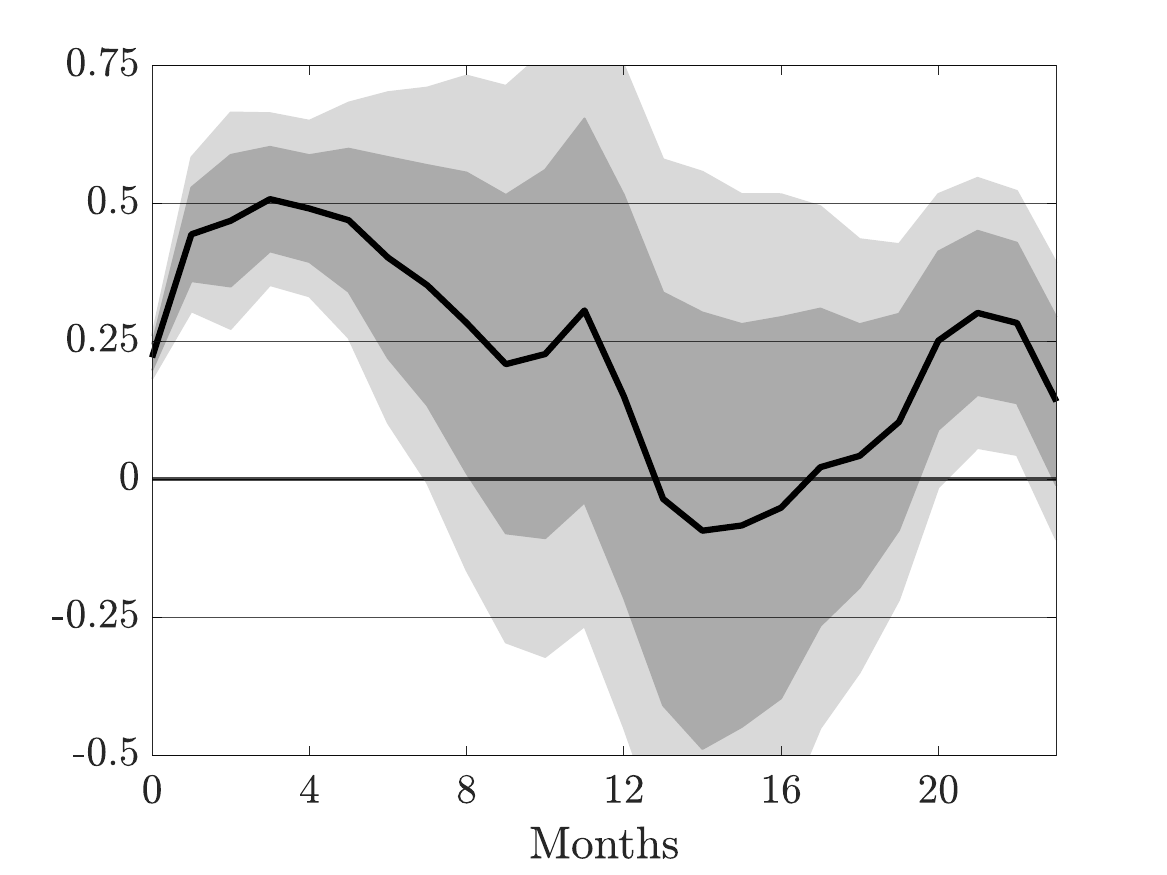} &    \includegraphics[width=.44\textwidth]{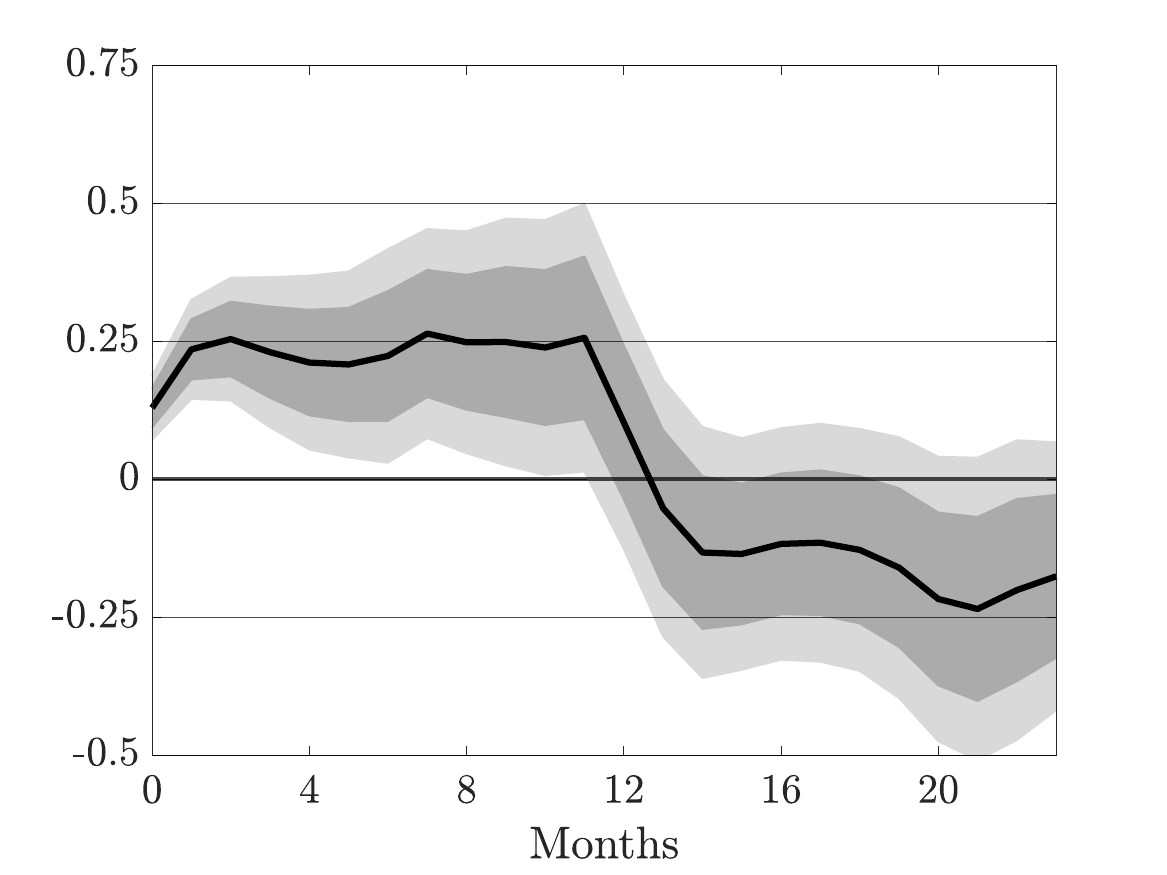} \\
(c) Average effect  & (d) Difference \\ \includegraphics[width=.44\textwidth]{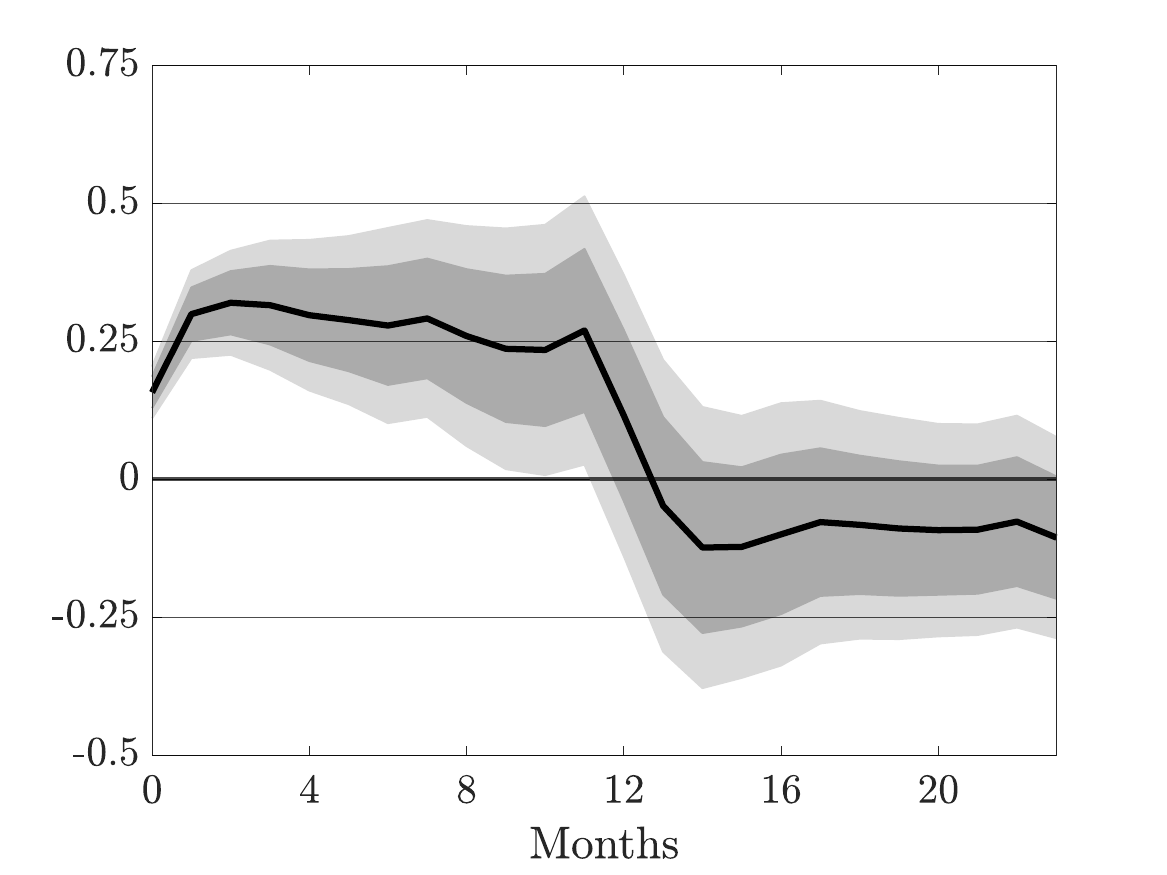} & \includegraphics[width=.44\textwidth]{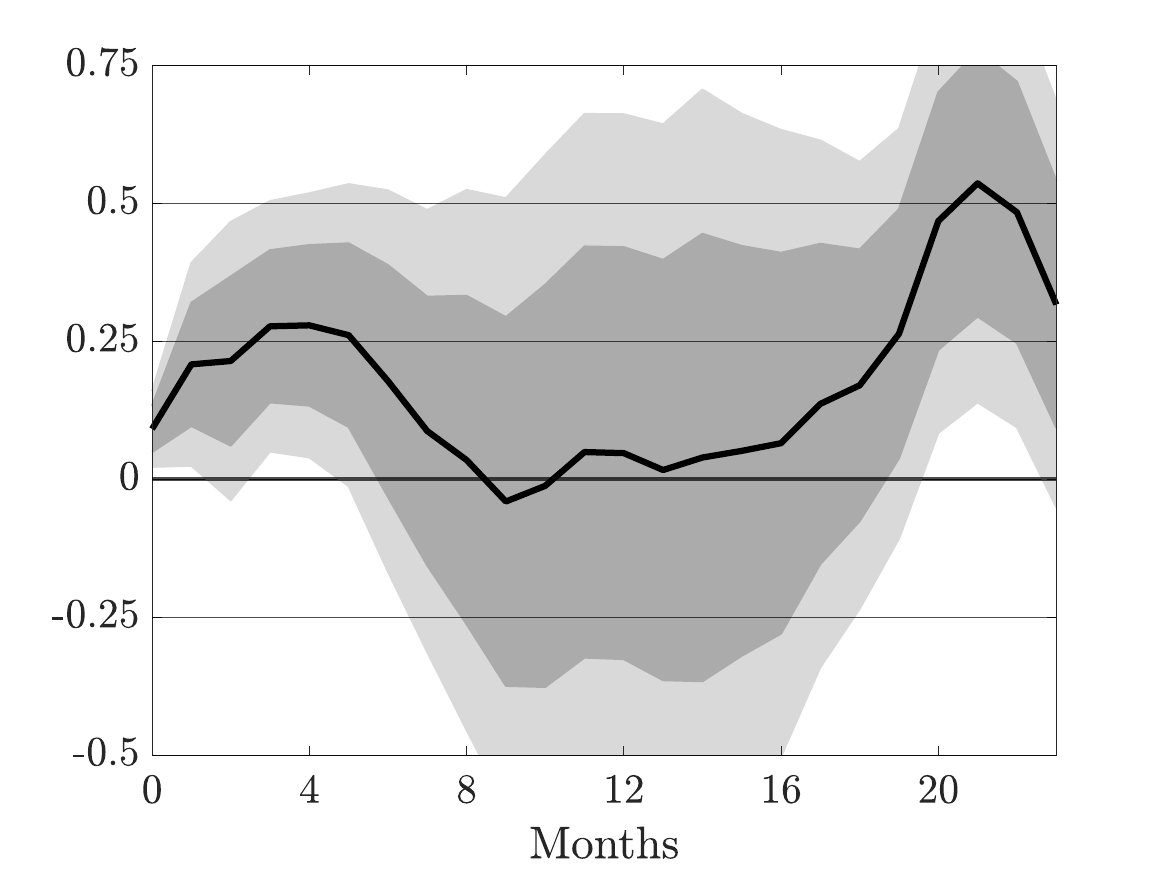}
\end{tabular}%

\begin{minipage}{1\textwidth}
\footnotesize{\emph{Notes}: This figure shows the inflation response to an oil supply news shock in the high-attention regime (panel (a)), the low-attention regime (panel (b)), on average across regimes (panel (c)), and the difference between the two regimes (panel (d)). The attention regimes are defined based on the number of Google searches of the word inflation in the current month. The dark shaded areas depict the 68\% confidence bands and the light-shaded area the 90\% confidence bands. Standard errors are robust with respect to serial correlation and heteroskedasticity (\cite{newey1987simple} with 12 lags). 
}%
 \end{minipage}
\end{figure}

\vspace{-0.2cm}
\paragraph{Using Google Trends as the threshold-defining variable.}
Instead of using the lagged inflation rate as the threshold-defining variable, I now use Google Trends data to define the regimes. As this data is only available since 2004, the high-attention regime largely coincides with the recent inflation surge. 
I use the same control variables as in my baseline specification and I further include four lags of the number of Google searches as control variables. Figure \ref{fig:emp_irfs_inflation_google} shows the results. They confirm the baseline results that inflation responds substantially and significantly more to supply shocks when attention to inflation is high. 

\begin{figure}[!ht]
\caption{Real oil price response to an oil supply news shock}
\label{fig:emp_irfs_realoilp}\vspace{0.15cm} \centering%
\begin{tabular}{cc}
(a) High-Attention Regime  & (b) Low-Attention Regime  \\ 
\includegraphics[width=.44\textwidth]{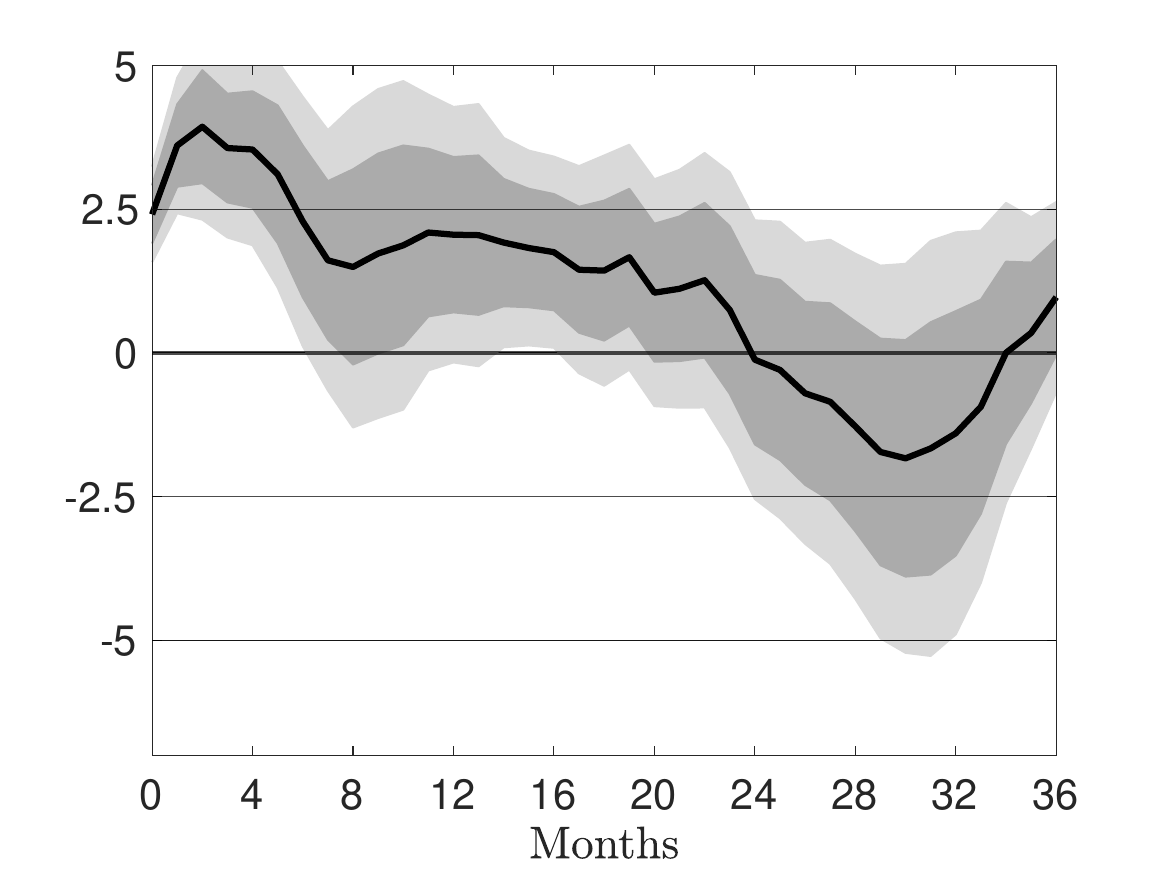} &    \includegraphics[width=.44\textwidth]{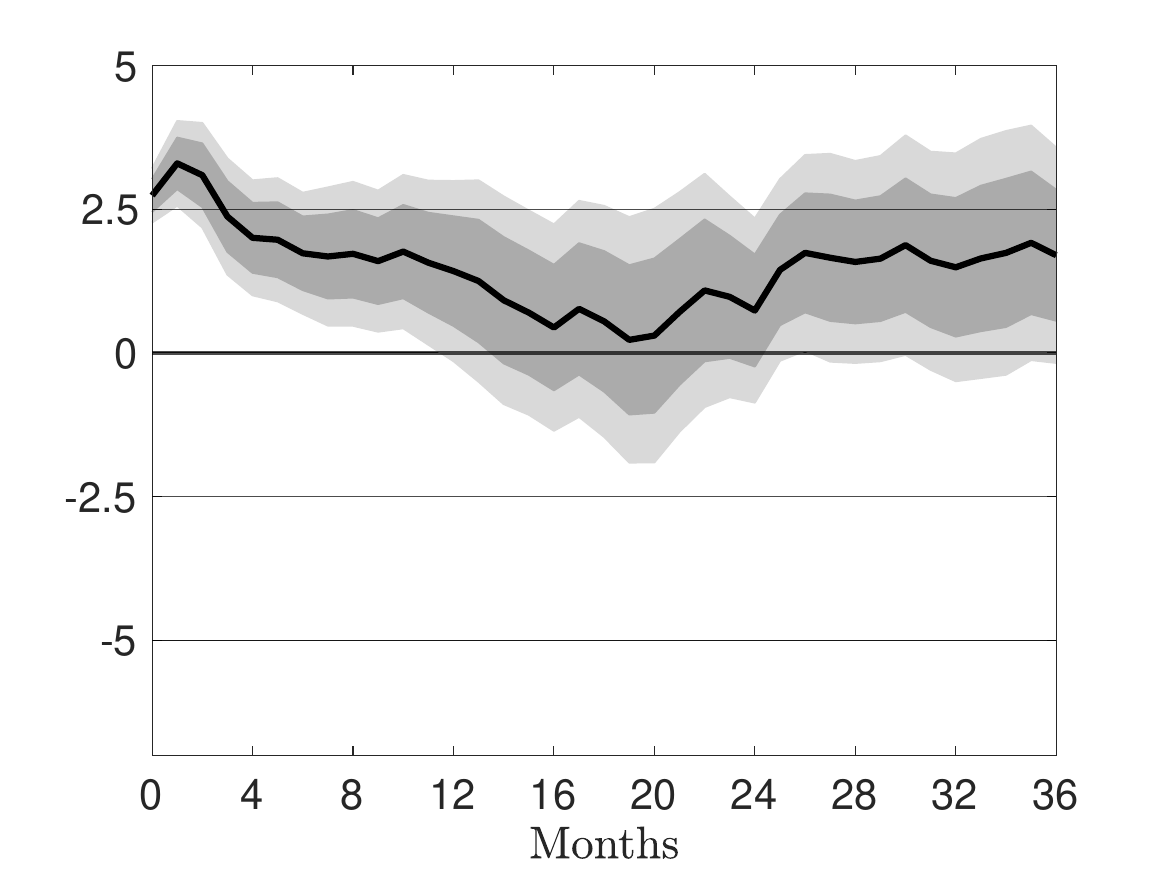} \\
(c) Average effect  & (d) Difference  \\ \includegraphics[width=.44\textwidth]{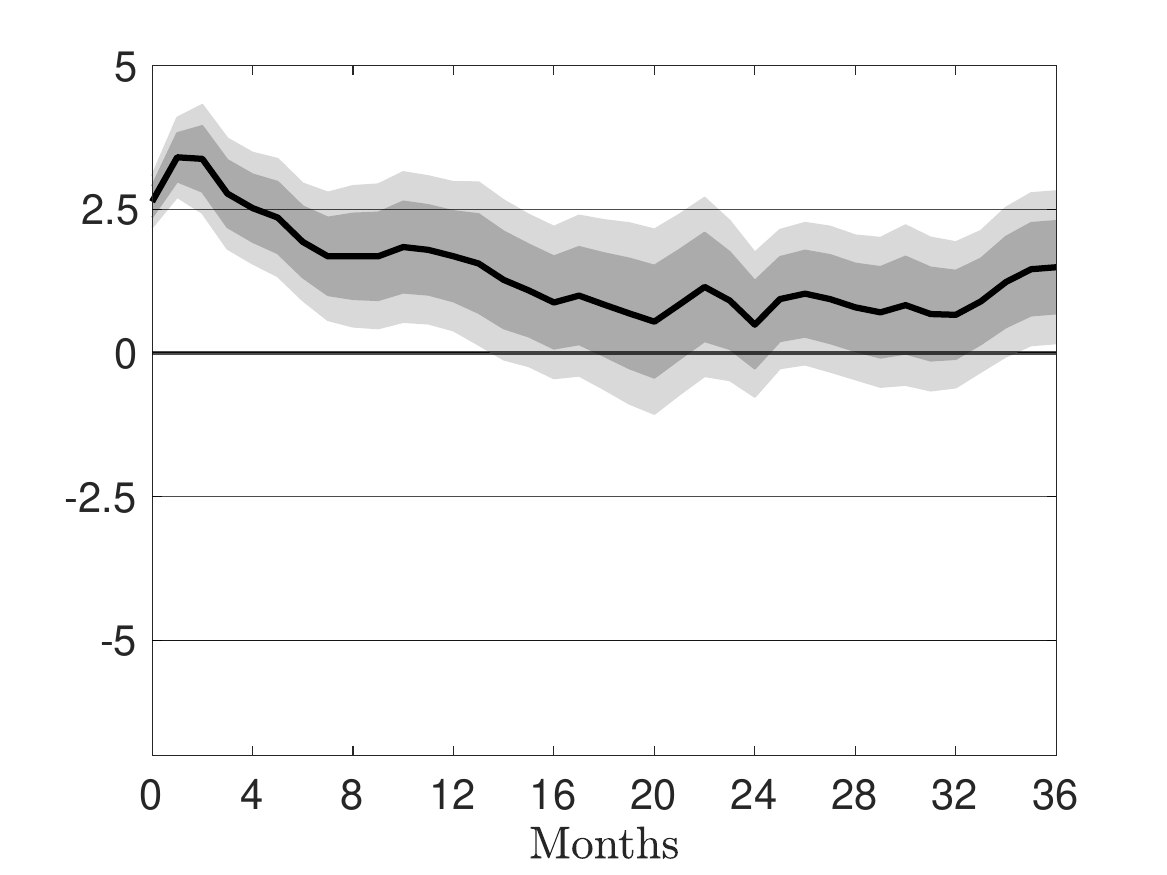} & \includegraphics[width=.44\textwidth]{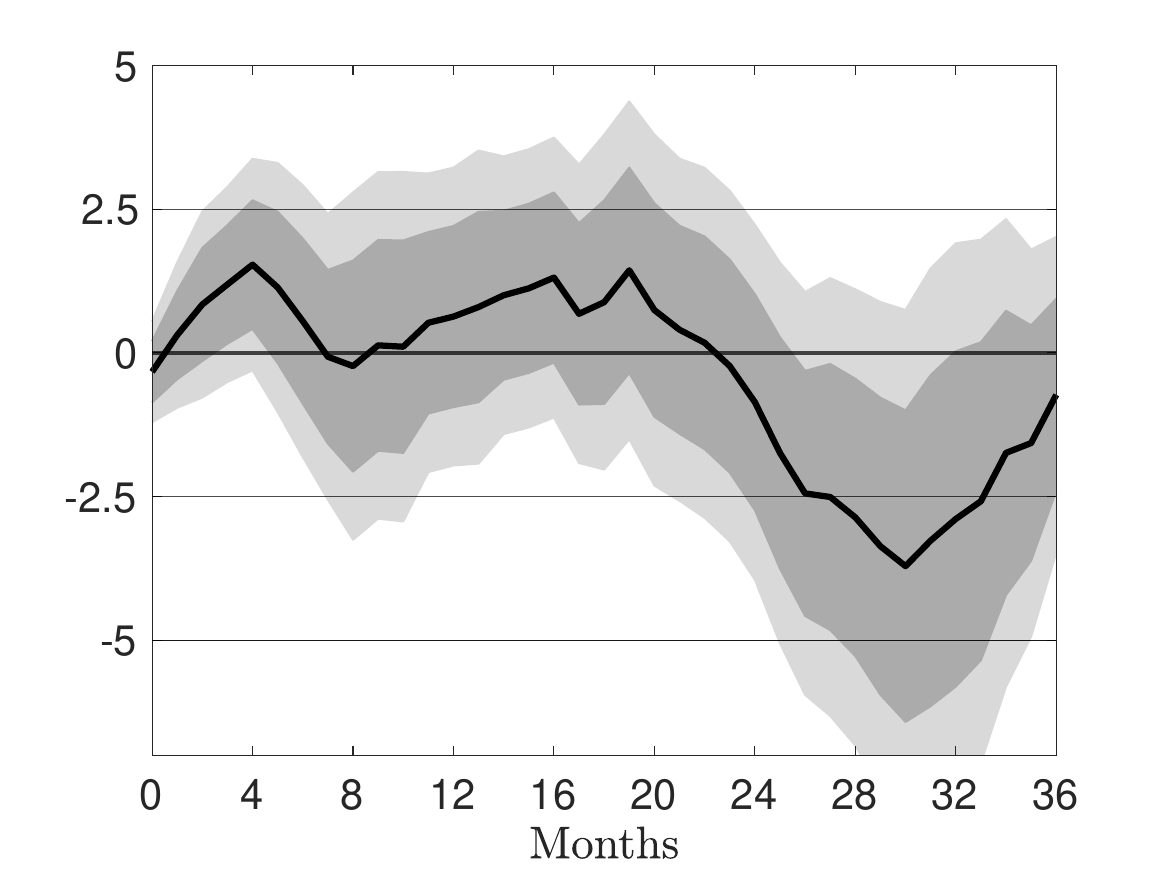}
\end{tabular}%

\begin{minipage}{1\textwidth}
\footnotesize{\emph{Notes}: This figure shows the real oil price response to an oil supply news shock in the high-attention regime (panel (a)), the low-attention regime (panel (b)), on average across regimes (panel (c)), and the difference between the two regimes (panel (d)). The dark shaded areas depict the 68\% confidence bands and the light-shaded area the 90\% confidence bands. Standard errors are robust with respect to serial correlation and heteroskedasticity (\cite{newey1987simple} with 12 lags). The attention regimes ares defined based on the previous month's inflation rate.
}%
 \end{minipage}
\end{figure}

\begin{figure}[!ht]
\caption{Nominal oil price response to an oil supply news shock}
\label{fig:emp_irfs_nominal_oilp}\vspace{0.15cm} \centering%
\begin{tabular}{cc}
(a) High-Attention Regime  & (b) Low-Attention Regime  \\ 
\includegraphics[width=.44\textwidth]{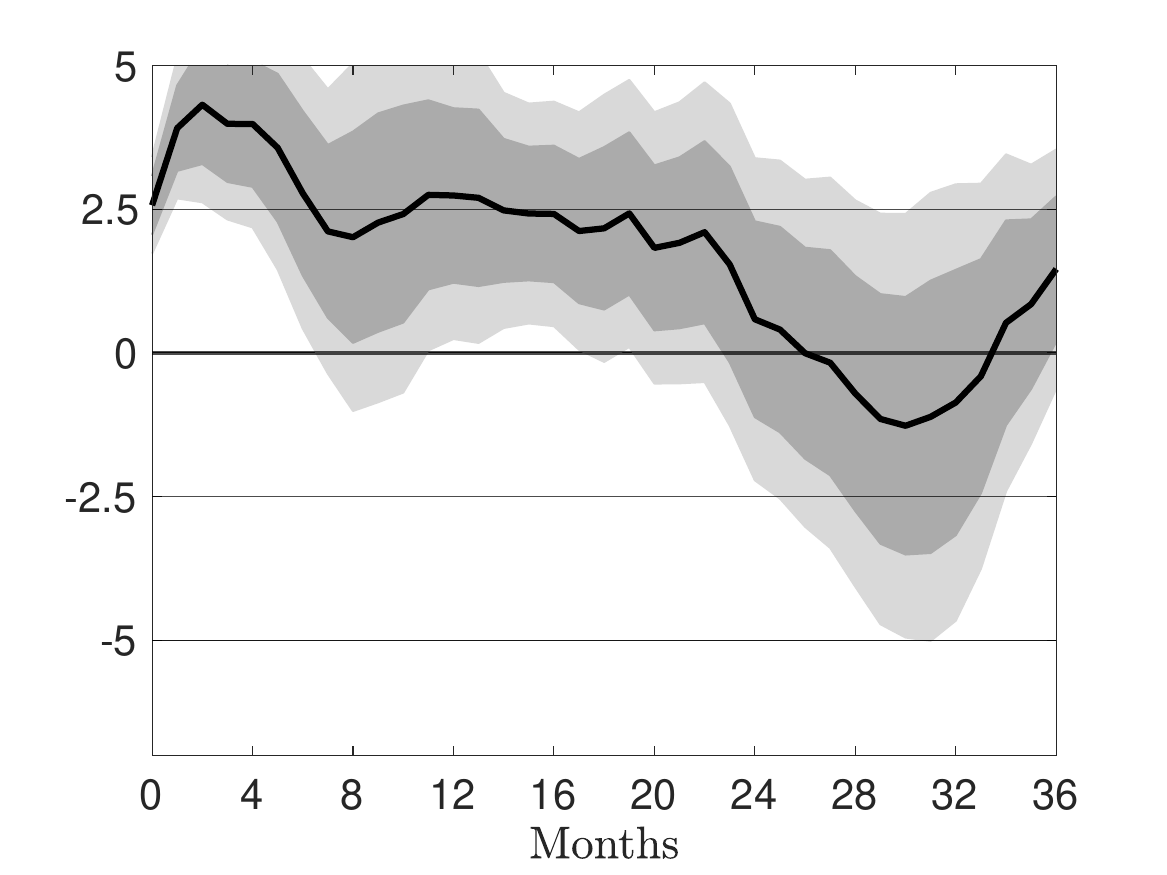} &    \includegraphics[width=.44\textwidth]{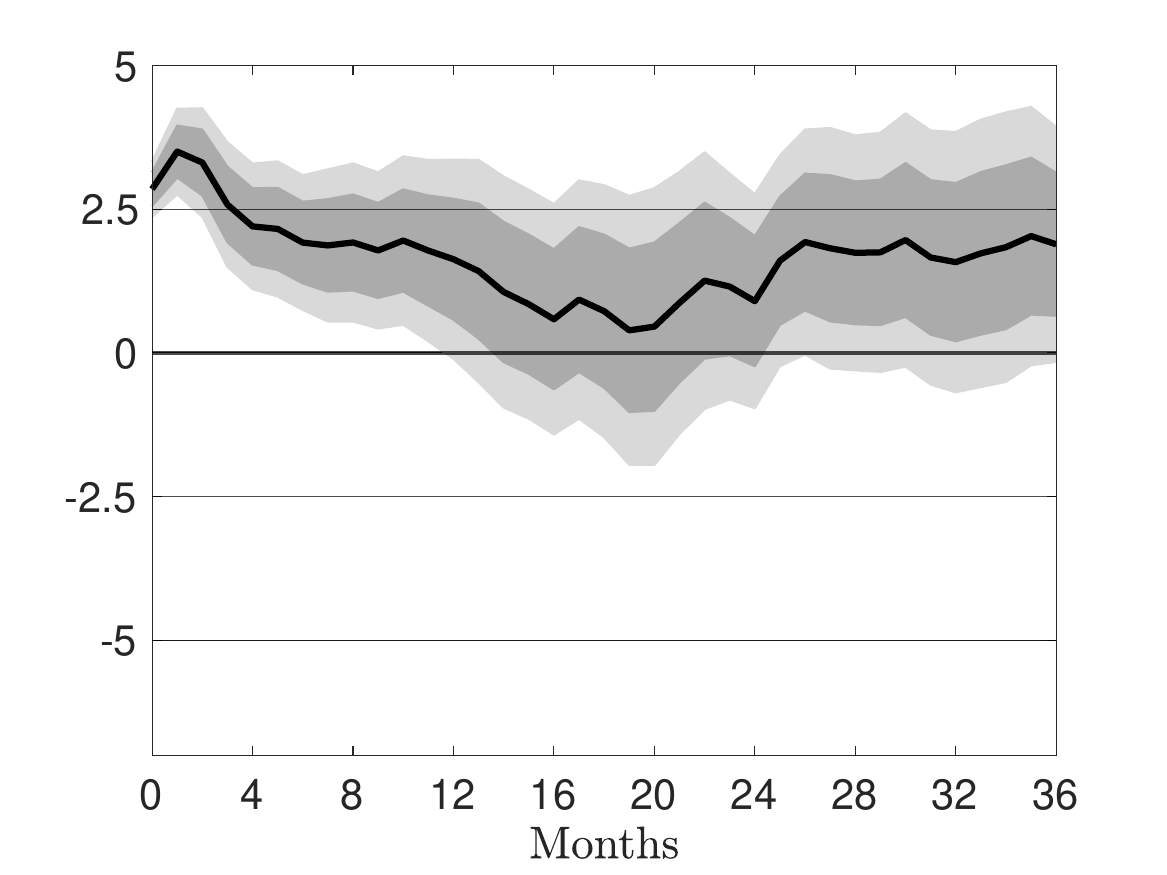} \\
(c) Average effect  & (d) Difference  \\ \includegraphics[width=.44\textwidth]{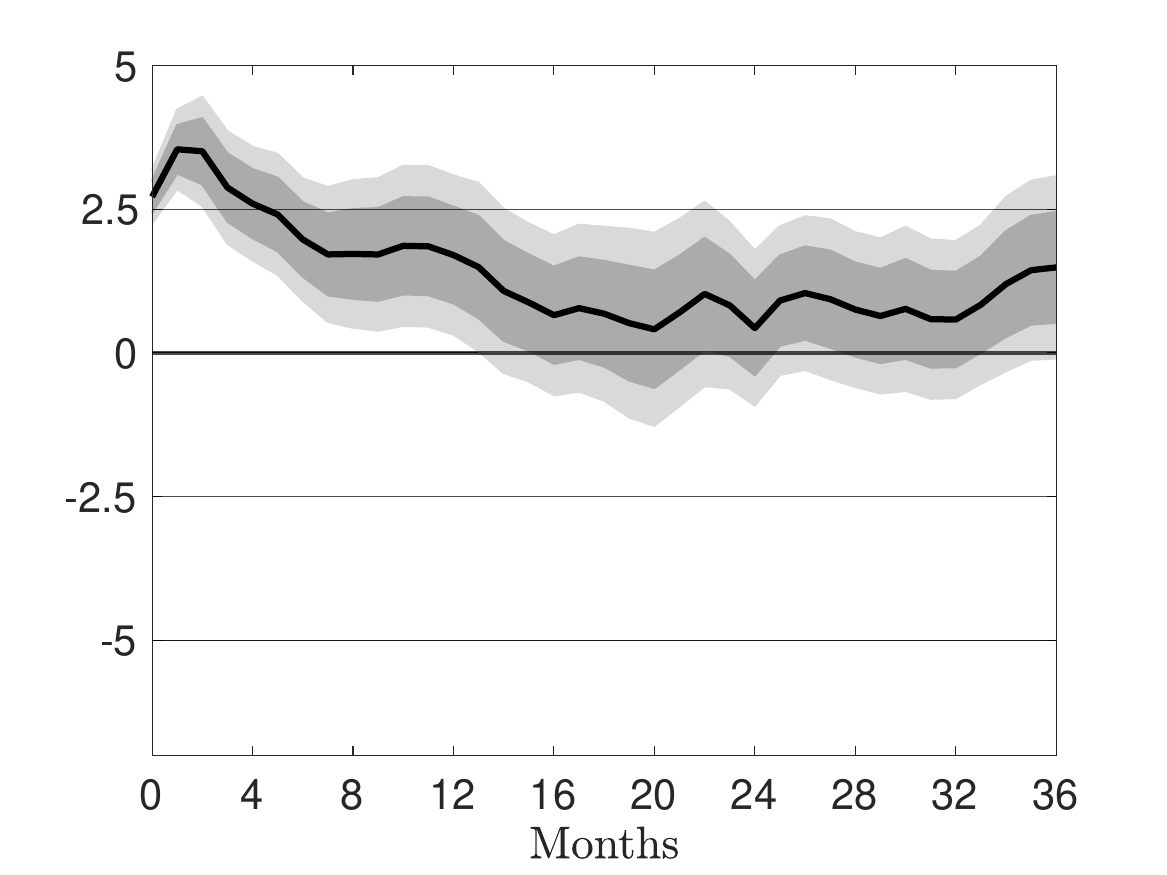} & \includegraphics[width=.44\textwidth]{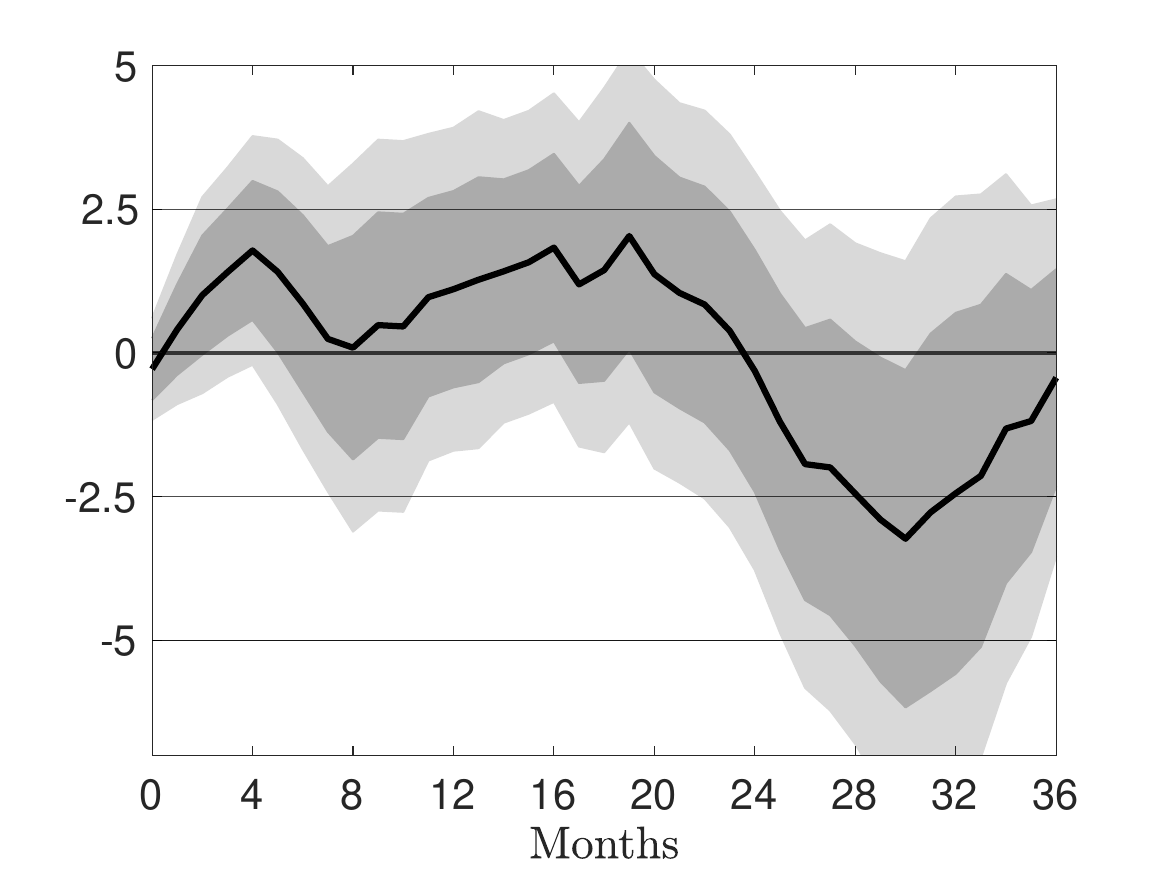}
\end{tabular}%

\begin{minipage}{1\textwidth}
\footnotesize{\emph{Notes}: This figure shows the nominal oil price response to an oil supply news shock in the high-attention regime (panel (a)), the low-attention regime (panel (b)), on average across regimes (panel (c)), and the difference between the two regimes (panel (d)). The dark shaded areas depict the 68\% confidence bands and the light-shaded area the 90\% confidence bands. Standard errors are robust with respect to serial correlation and heteroskedasticity (\cite{newey1987simple} with 12 lags). The attention regimes ares defined based on the previous month's inflation rate.
}%
 \end{minipage}
\end{figure}

\paragraph{Oil price response.} Figures \ref{fig:emp_irfs_realoilp} and \ref{fig:emp_irfs_nominal_oilp} show the responses of the real and nominal oil price, respectively, to a negative oil supply news shock for the two attention regimes (panels (a) and (b)), the average response in panel (c), and the difference across regimes in (d). We see that the differences across regimes are not significant at the 10\% significance level at any horizon, indicating that the differences in the inflation responses discussed in section \ref{sec:empirical_irfs} are unlikely to be driven by different oil price responses.

\clearpage\newpage

\begin{figure}[!ht]
\caption{Unemployment response to an oil supply news shock}
\label{fig:emp_irfs_unemp_nocovid}\vspace{0.15cm} \centering%
\begin{tabular}{cc}
(a) High-Attention Regime   & (b) Low-Attention Regime  \\ 
\includegraphics[width=.44\textwidth]{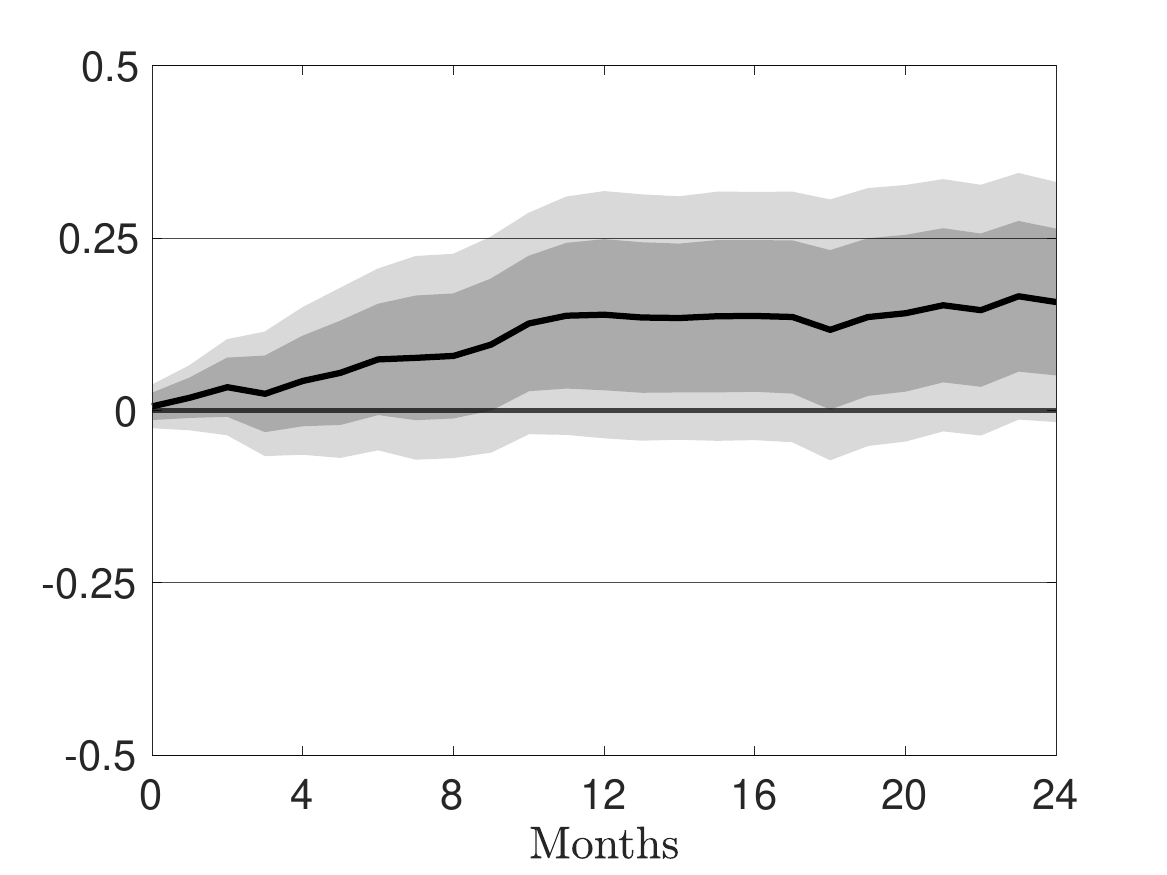} &    \includegraphics[width=.44\textwidth]{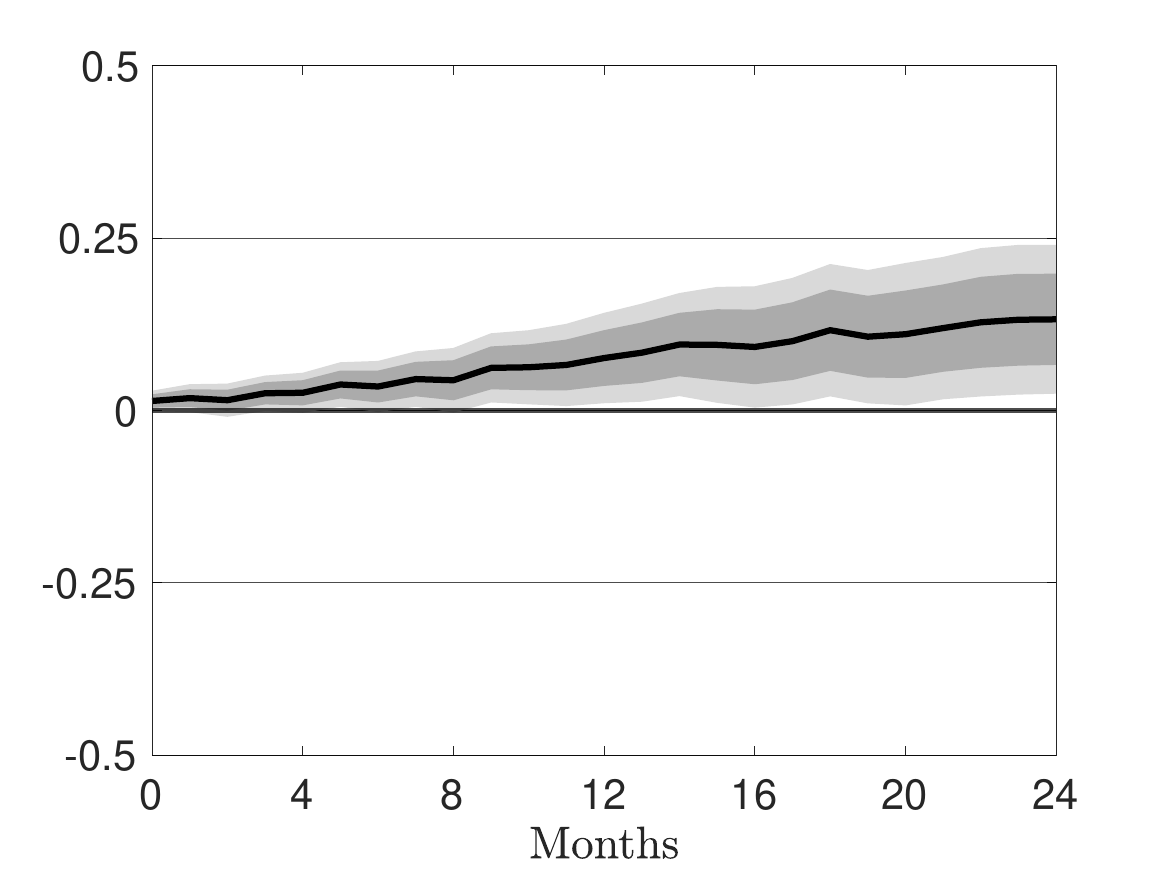} \\
(c) Average effect   & (d) Difference   \\ \includegraphics[width=.44\textwidth]{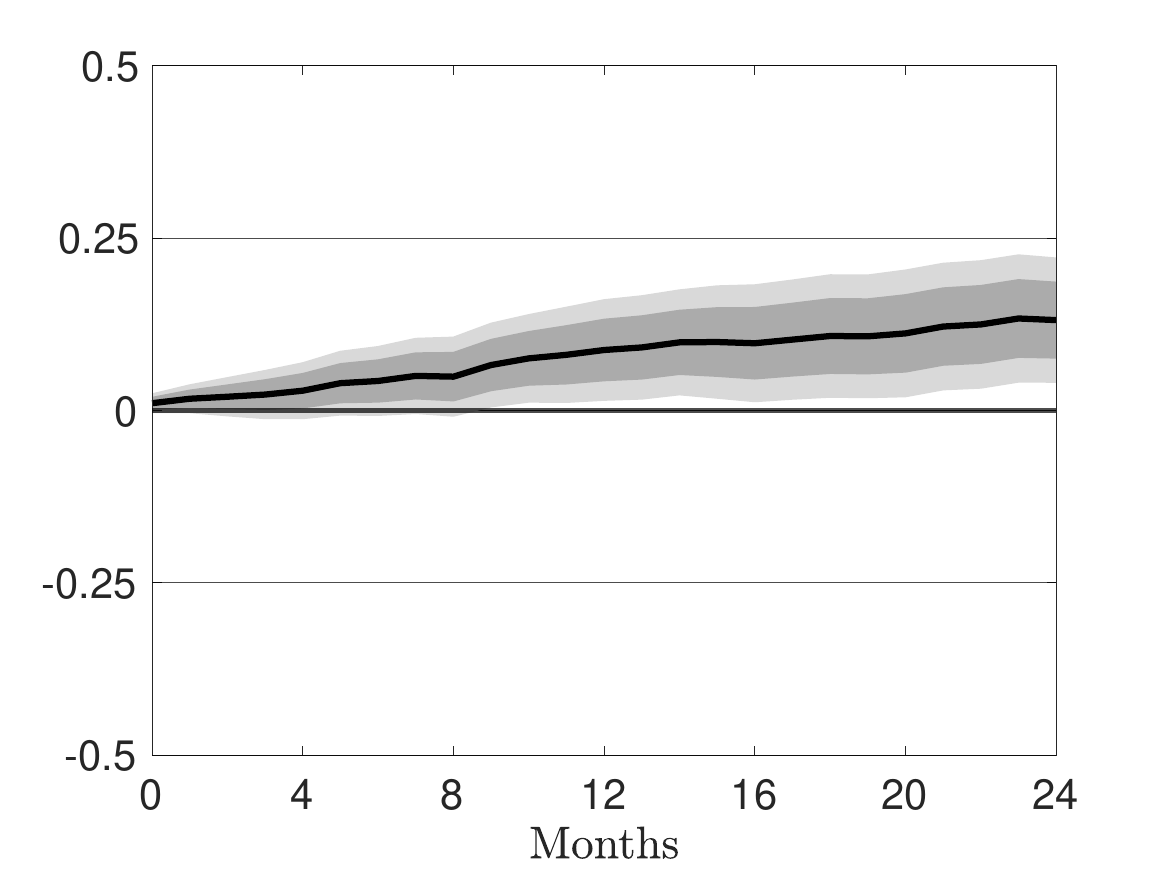} & \includegraphics[width=.44\textwidth]{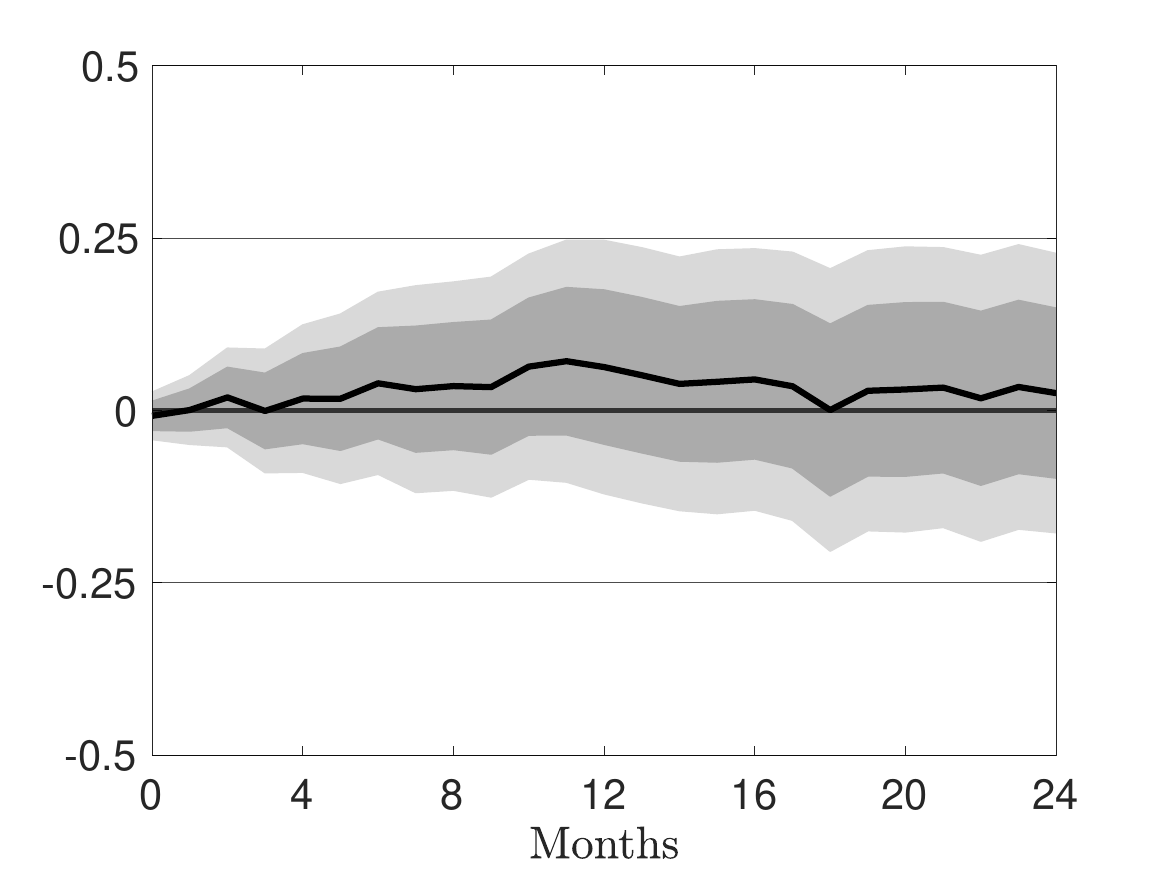}
\end{tabular}%

\begin{minipage}{1\textwidth}
\footnotesize{\emph{Notes}: This figure shows the unemployment rate response to an oil supply news shock when excluding the Covid-19 period (2020-2023) in the high-attention regime (panel (a)), the low-attention regime (panel (b)), on average across regimes (panel (c)), and the difference between the two regimes (panel (d)). The dark shaded areas depict the 68\% confidence bands and the light-shaded area the 90\% confidence bands. Standard errors are robust with respect to serial correlation and heteroskedasticity (\cite{newey1987simple} with 12 lags). 
}%
 \end{minipage}
\end{figure}

\begin{figure}[!ht]
\caption{Unemployment response to an oil supply news shock including Covid}
\label{fig:emp_irfs_unemp}\vspace{0.15cm} \centering%
\begin{tabular}{cc}
(a) High-Attention Regime  & (b) Low-Attention Regime  \\ 
\includegraphics[width=.44\textwidth]{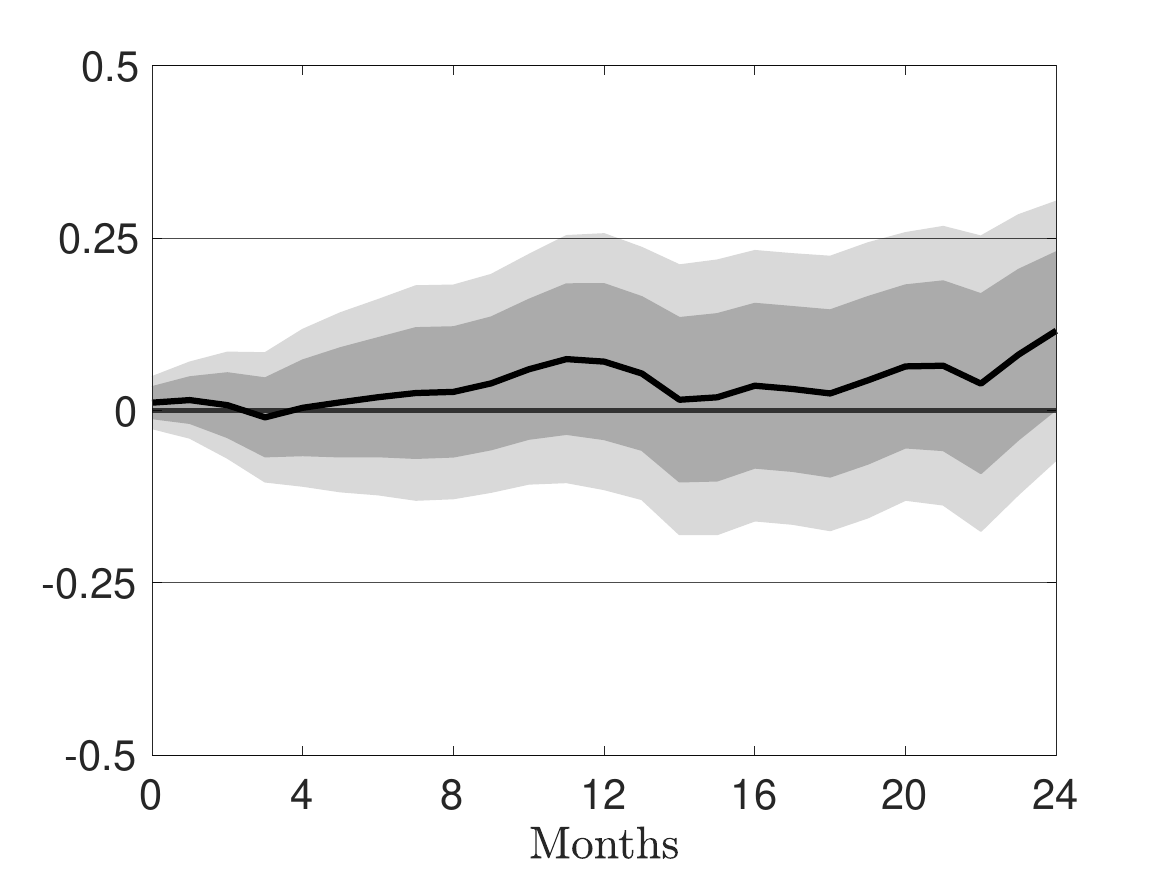} &    \includegraphics[width=.44\textwidth]{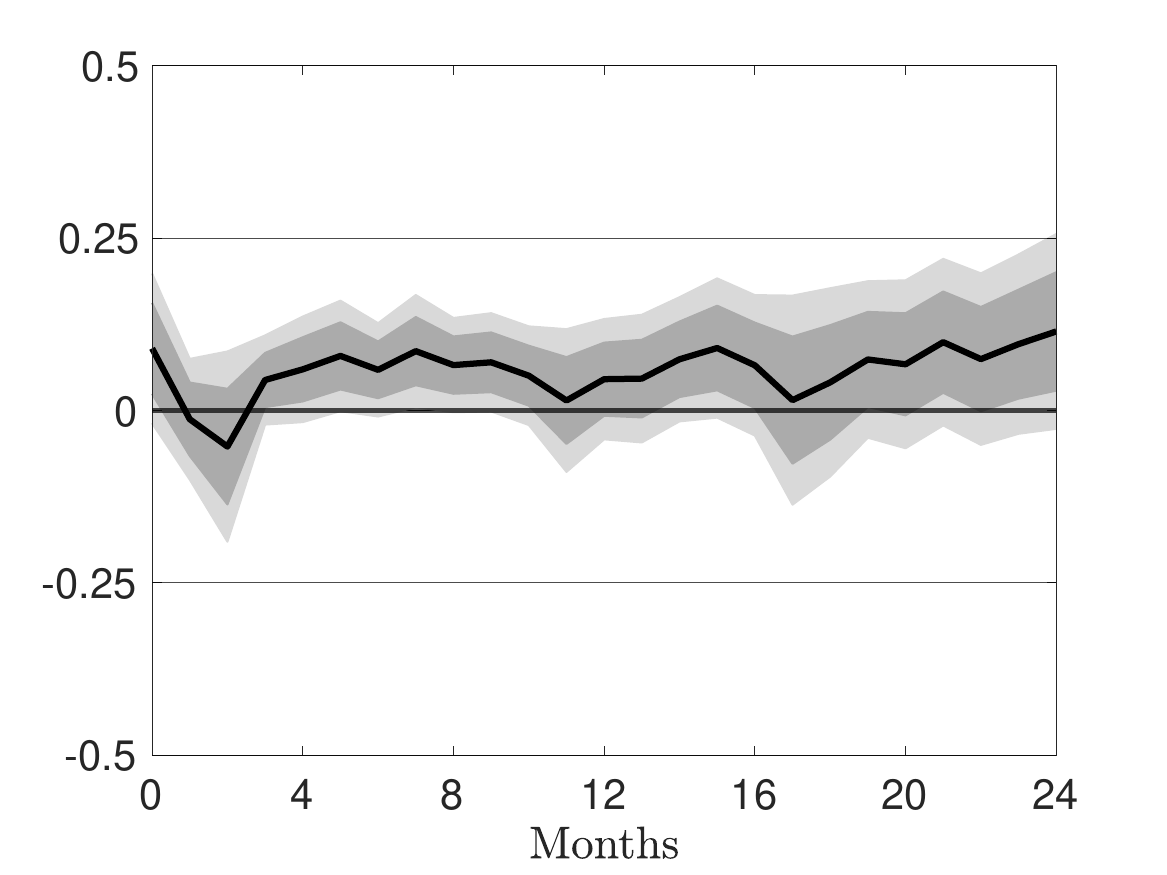} \\
(c) Average effect  & (d) Difference  \\ \includegraphics[width=.44\textwidth]{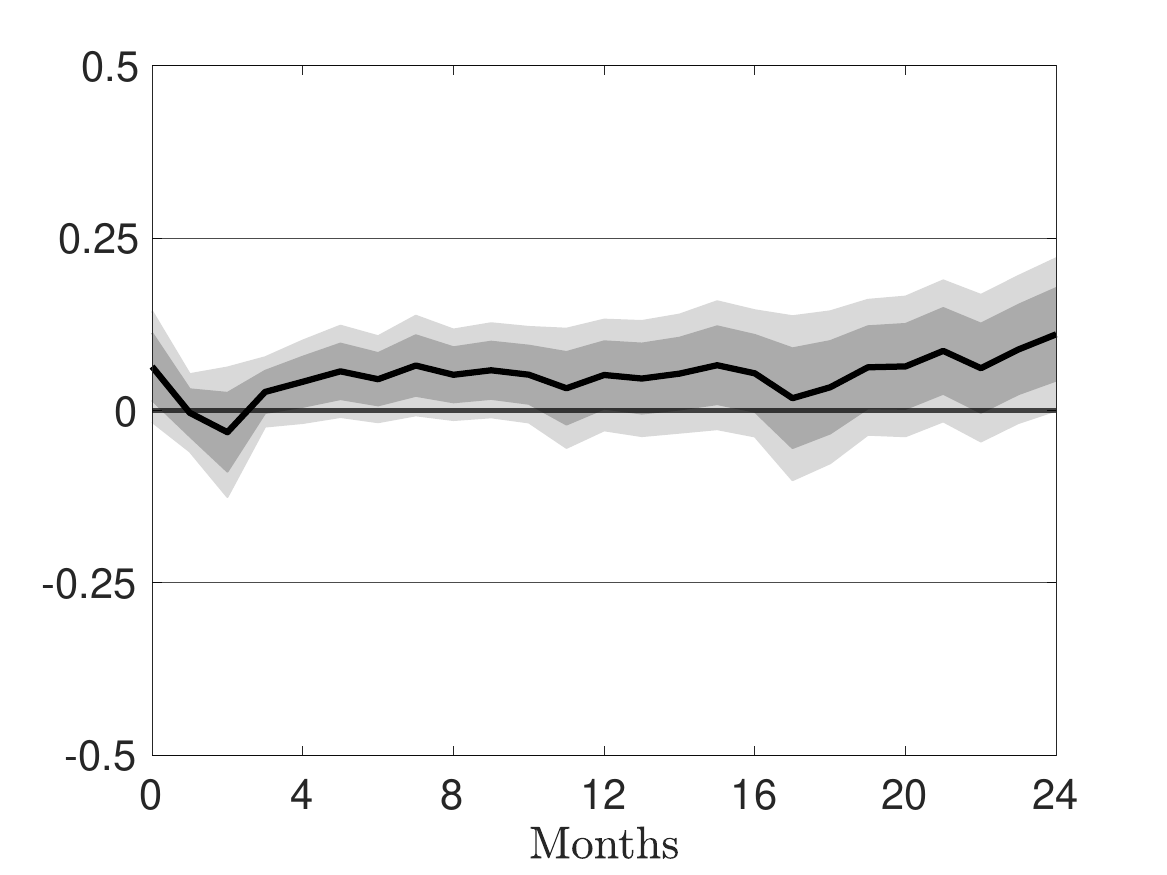} & \includegraphics[width=.44\textwidth]{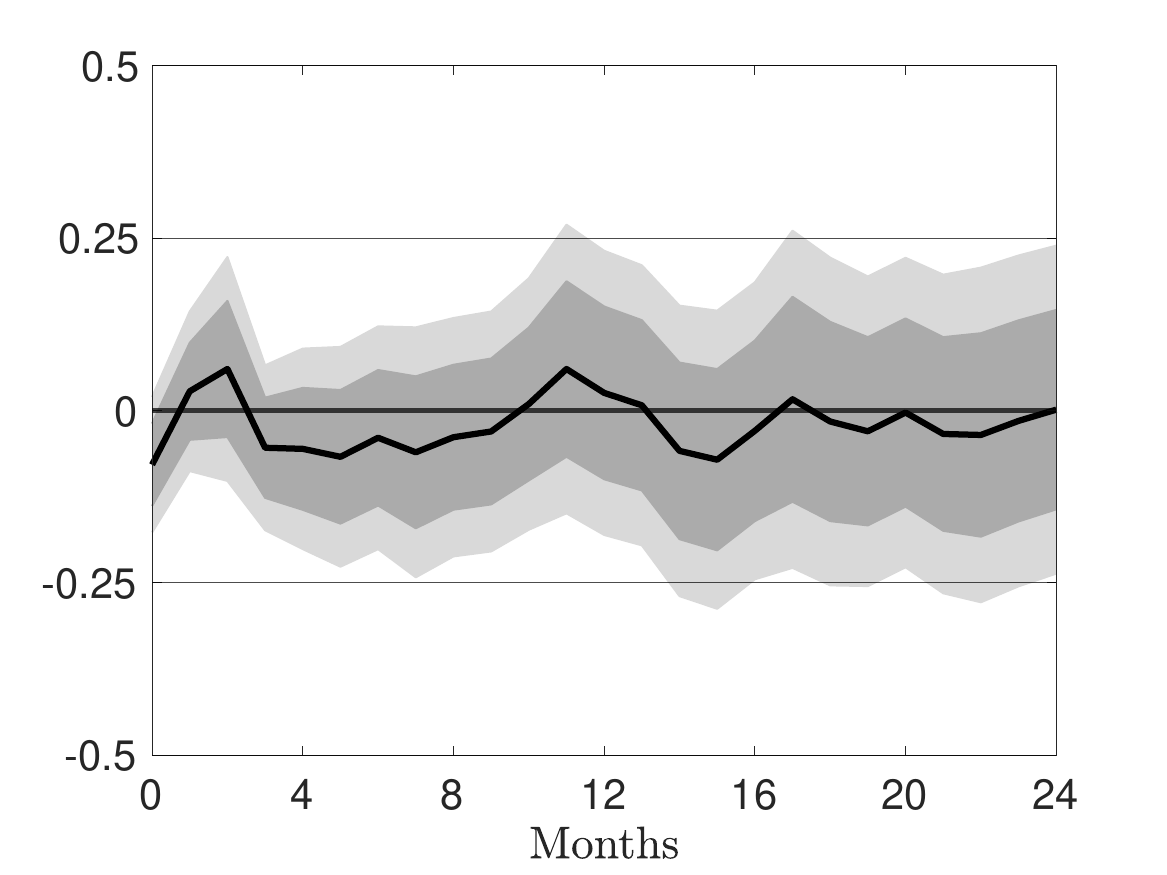}
\end{tabular}%

\begin{minipage}{1\textwidth}
\footnotesize{\emph{Notes}: This figure shows the unemployment rate response to an oil supply news shock when including the Covid-19 period (2020-2023) in the high-attention regime (panel (a)), the low-attention regime (panel (b)), on average across regimes (panel (c)), and the difference between the two regimes (panel (d)). The dark shaded areas depict the 68\% confidence bands and the light-shaded area the 90\% confidence bands. Standard errors are robust with respect to serial correlation and heteroskedasticity (\cite{newey1987simple} with 12 lags). 
}%
 \end{minipage}
\end{figure}

\paragraph{Unemployment response.} Figures \ref{fig:emp_irfs_unemp_nocovid} and \ref{fig:emp_irfs_unemp} show the responses of the unemployment rate (excluding the Covid period 2020-2023 and including it, respectively) to a negative oil supply news shock for the two attention regimes (panels (a) and (b)), the average response in panel (c), and the difference across regimes in (d). We see that the differences across regimes are not significant at any conventional significance level at any horizon, indicating that the differences in the inflation responses discussed in section \ref{sec:empirical_irfs} are unlikely to be driven by generally more responsive macroeconomic variables.

\begin{figure}[!ht]
\caption{Inflation response to an oil supply news shock excluding Covid}
\label{fig:emp_irfs_inflation_noCovid}\vspace{0.15cm} \centering%
\begin{tabular}{cc}
(a) High-Attention Regime  & (b) Low-Attention Regime  \\ 
\includegraphics[width=.44\textwidth]{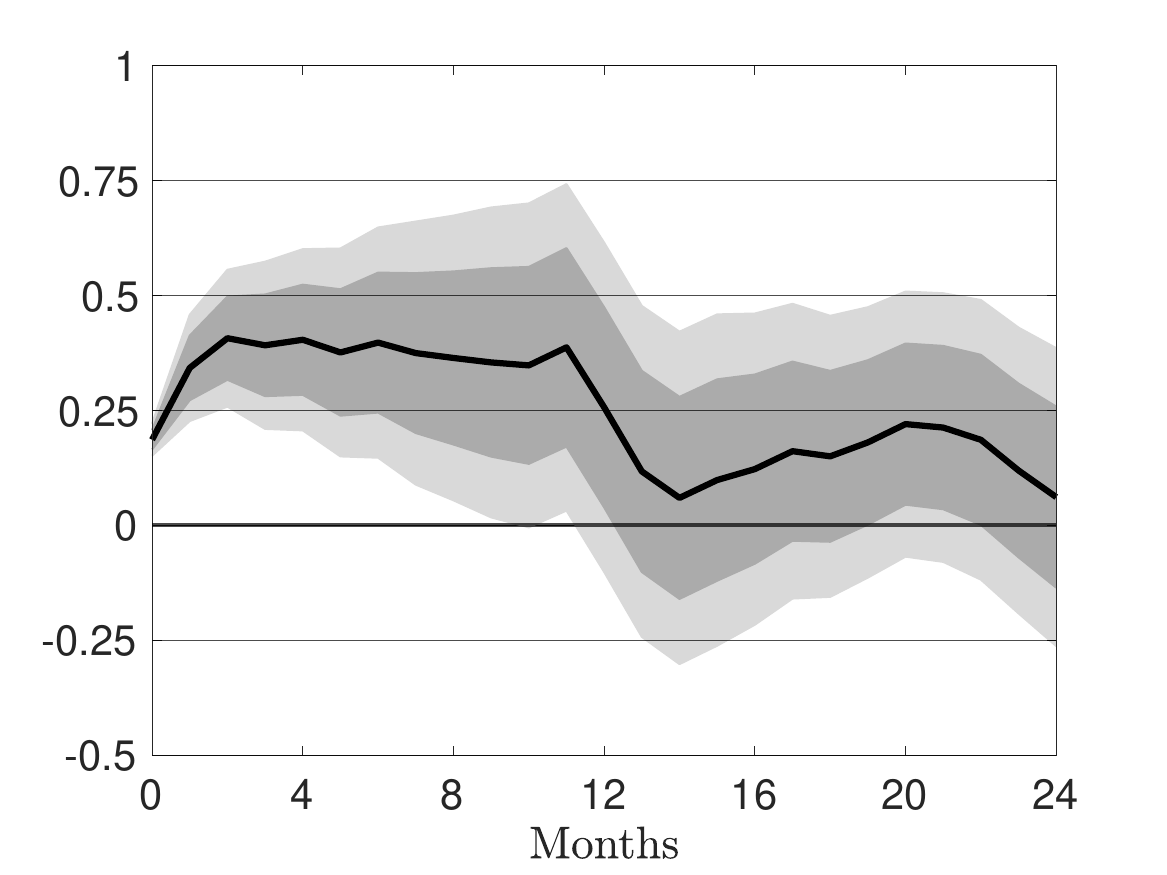} &    \includegraphics[width=.44\textwidth]{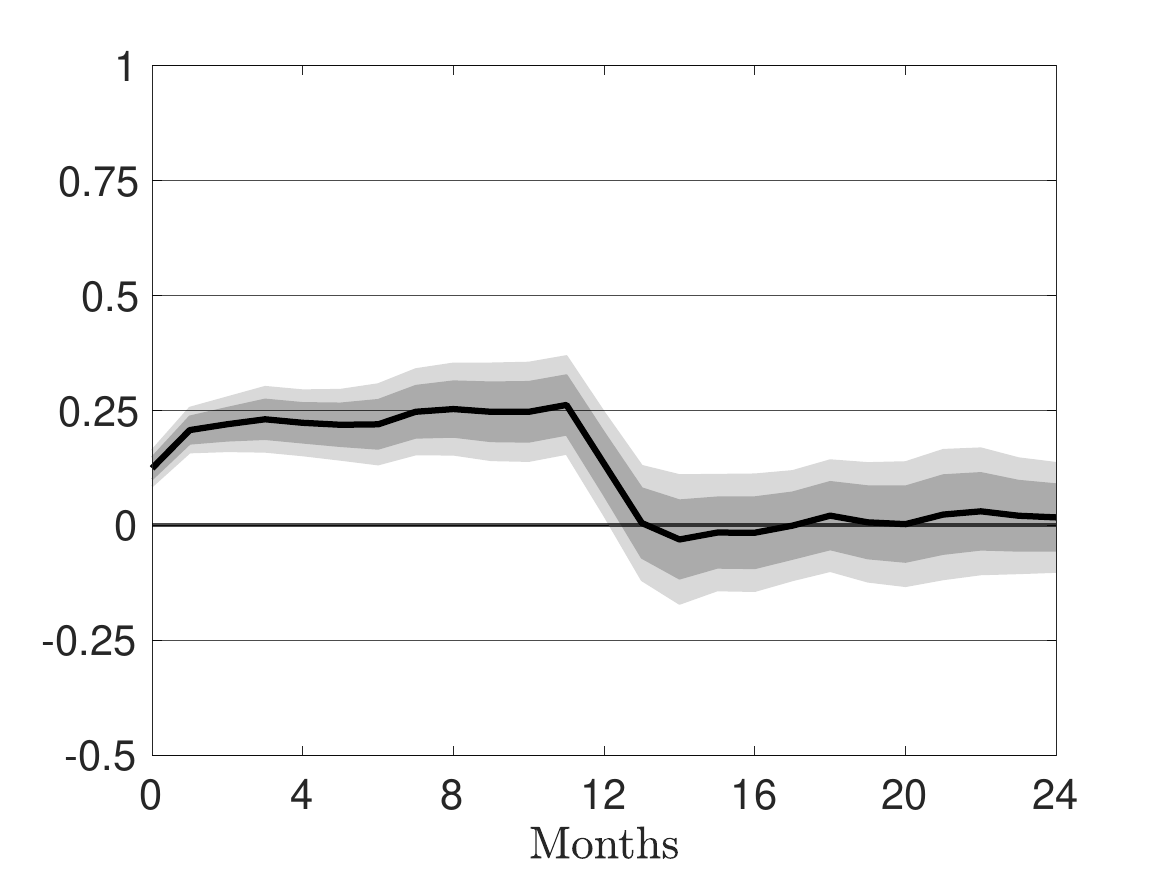} \\
(c) Average effect  & (d) Difference \\\includegraphics[width=.44\textwidth]{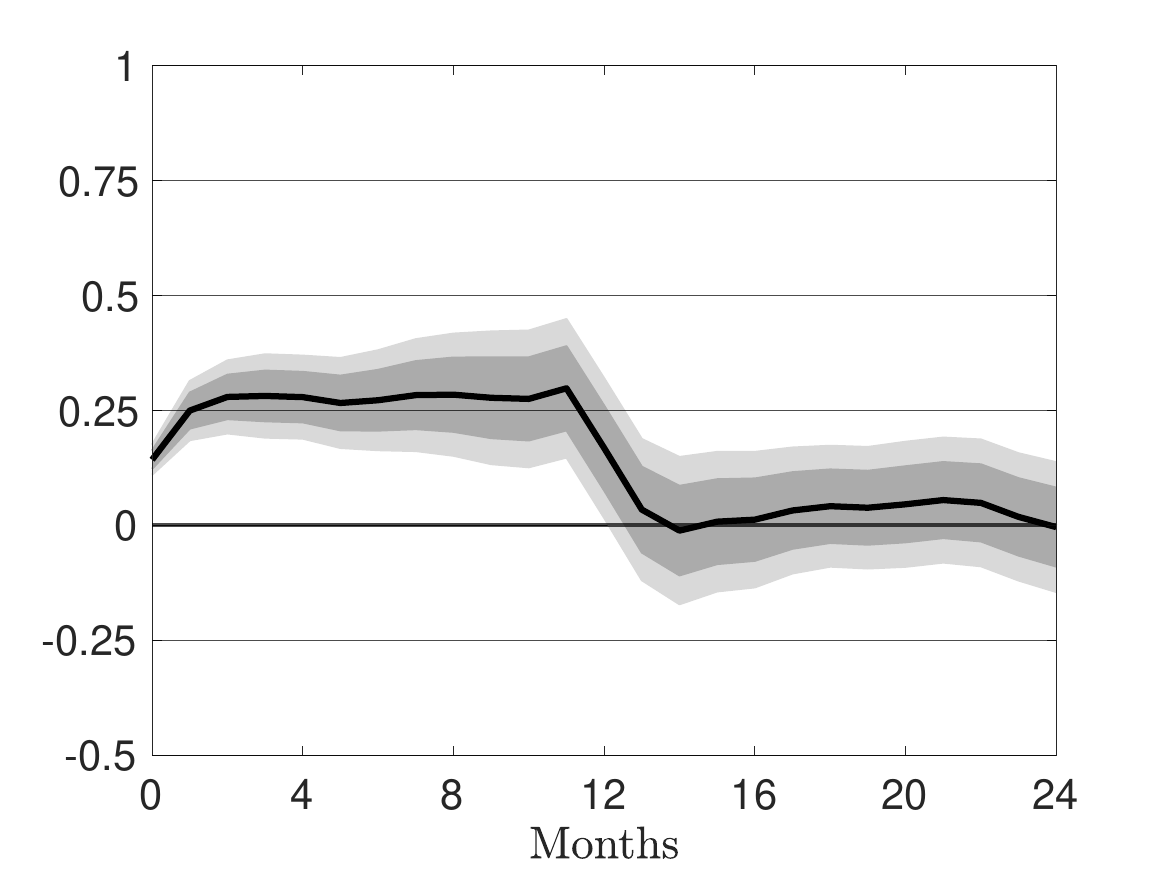} & \includegraphics[width=.44\textwidth]{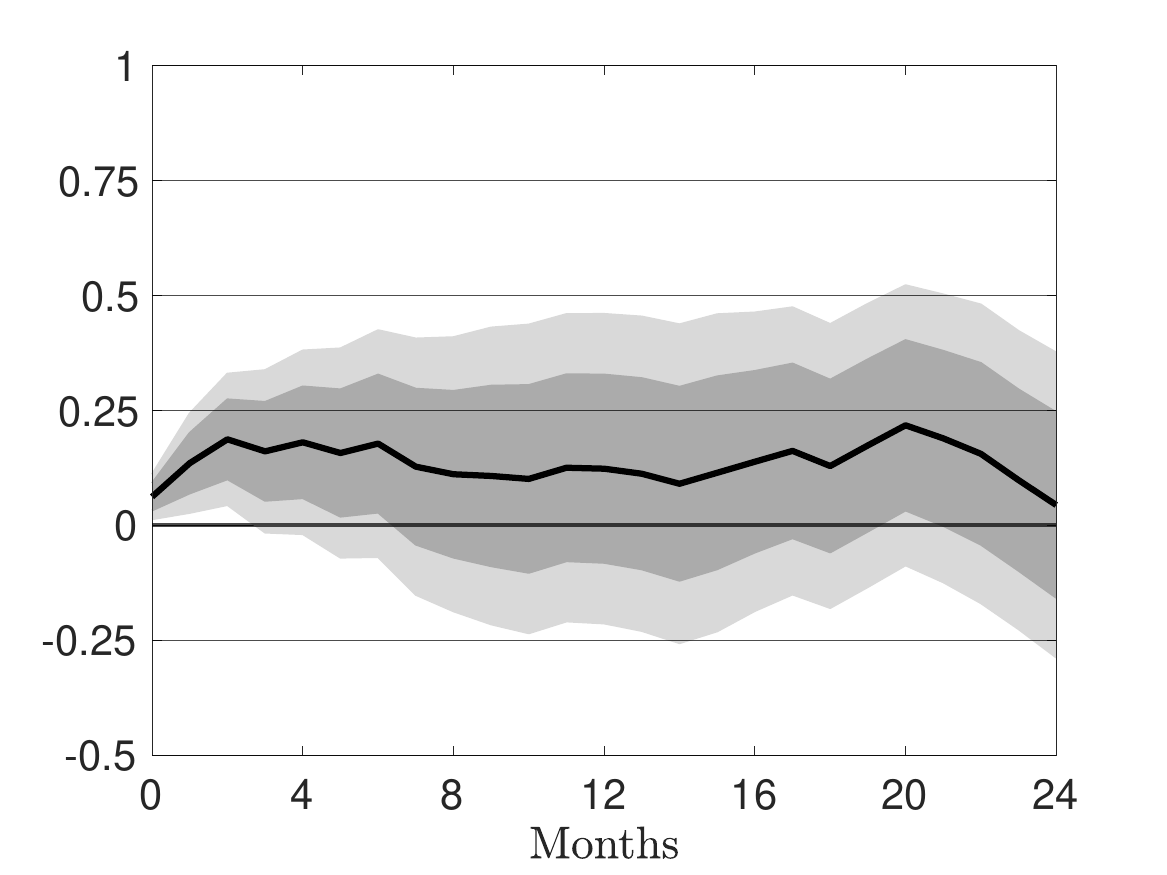}
\end{tabular}%

\begin{minipage}{1\textwidth}
\footnotesize{\emph{Notes}: This figure shows the inflation response to an oil supply news shock when excluding the years 2020, 2021, 2022 and 2023 in the high-attention regime (panel (a)), the low-attention regime (panel (b)), on average across regimes (panel (c)), and the difference between the two regimes (panel (d)). The dark shaded areas depict the 68\% confidence bands and the light-shaded area the 90\% confidence bands. Standard errors are robust with respect to serial correlation and heteroskedasticity (\cite{newey1987simple} with 12 lags). 
}%
 \end{minipage}
\end{figure}

\clearpage\newpage
\paragraph{Covid.} Figure \ref{fig:emp_irfs_inflation_noCovid} shows the inflation response to a negative oil supply news shock for the two attention regimes (panels (a) and (b)), the average response in panel (c), and the difference across regimes in (d) when excluding the Covid period 2020-2023. We see that the differences across regimes are slightly weaker than for the baseline case in Figure \ref{fig:emp_irfs_inflation1} but still substantial.

\begin{figure}[!ht]
\caption{Inflation response to an oil supply news shock after 1990}
\label{fig:emp_irfs_inflation_after1990}\vspace{0.15cm} \centering%
\begin{tabular}{cc}
(a) High-Attention Regime  & (b) Low-Attention Regime  \\ 
\includegraphics[width=.44\textwidth]{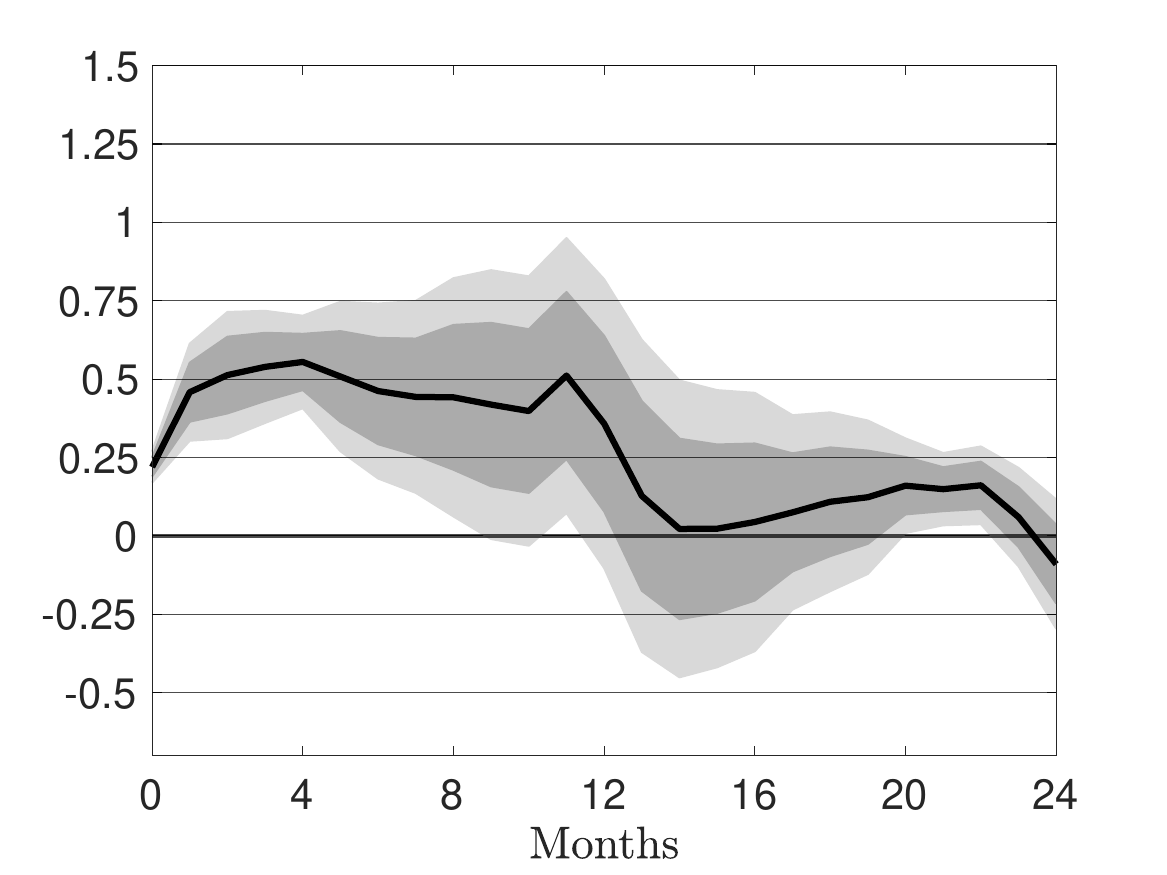} &    \includegraphics[width=.44\textwidth]{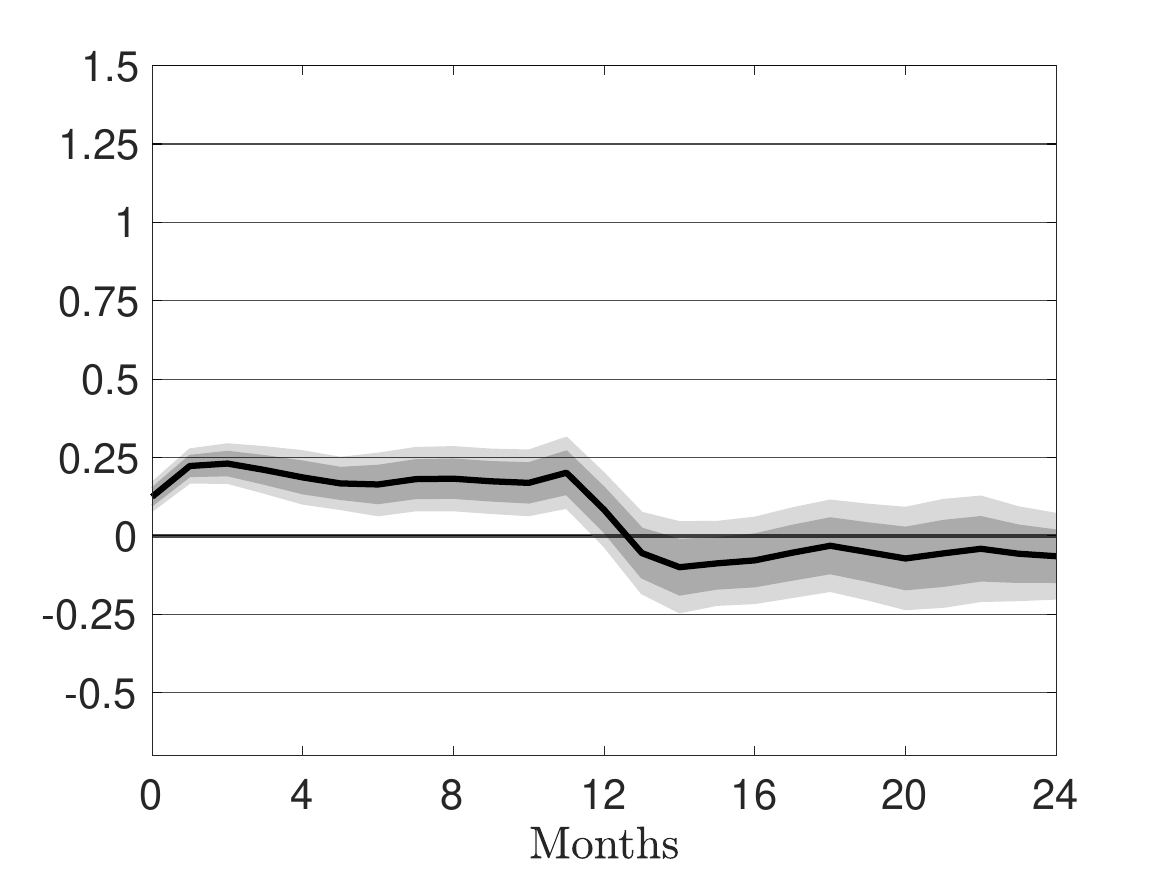} \\
(c) Average effect  & (d) Difference \\\includegraphics[width=.44\textwidth]{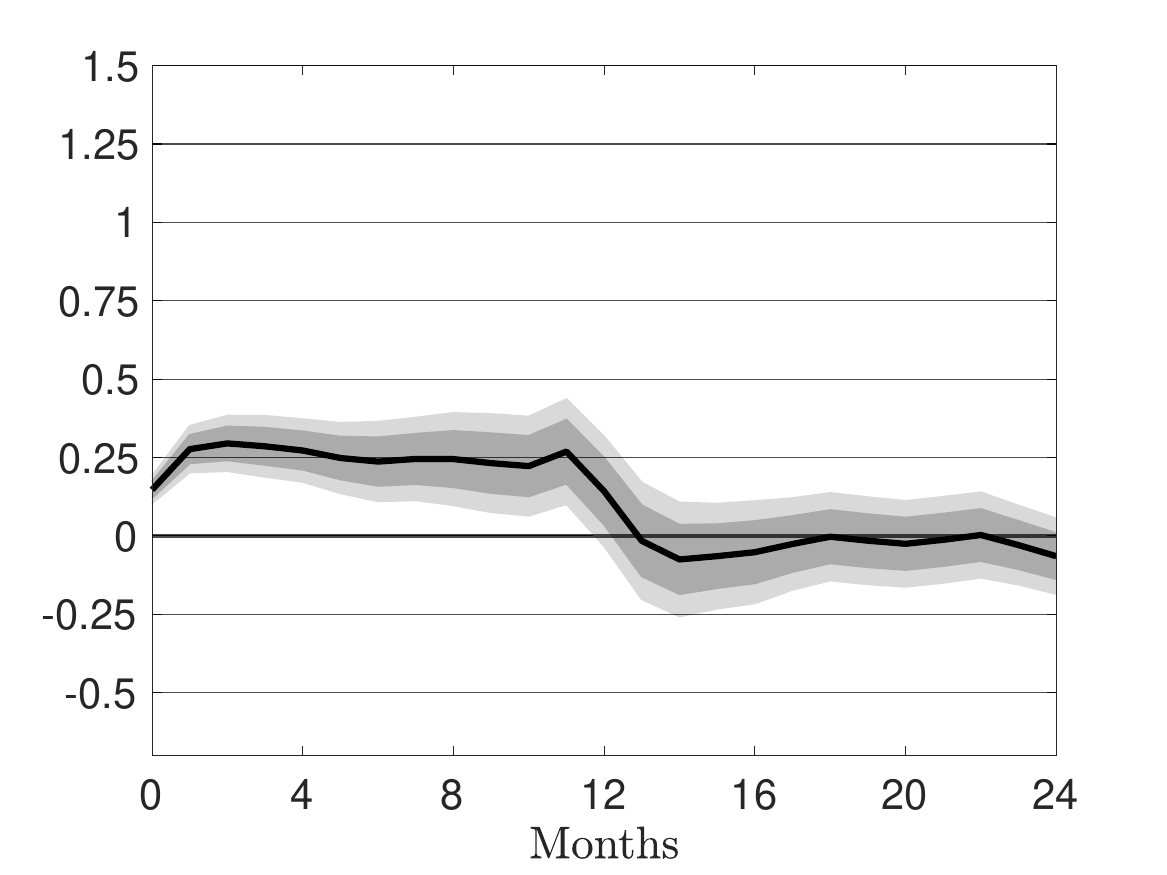} & \includegraphics[width=.44\textwidth]{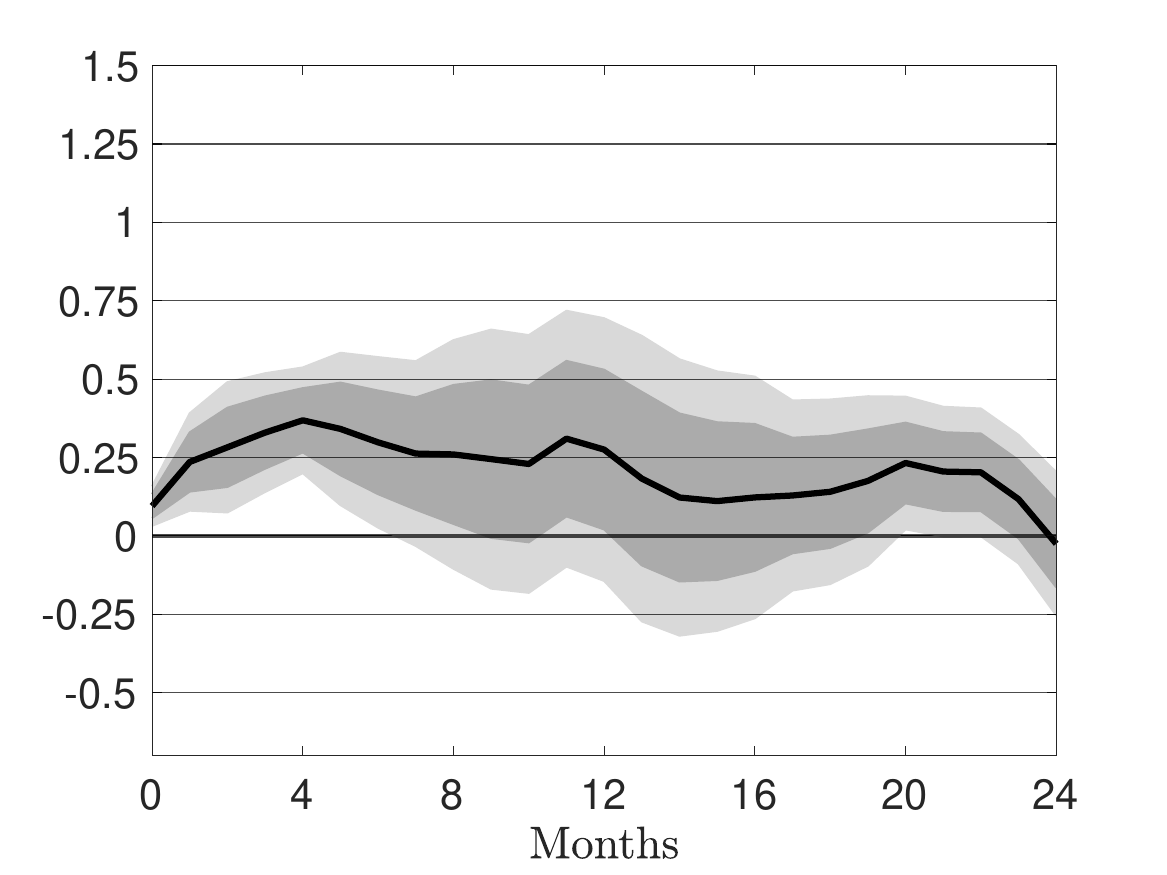}
\end{tabular}%

\begin{minipage}{1\textwidth}
\footnotesize{\emph{Notes}: This figure shows the inflation response to an oil supply news shock when excluding the years prior to 1990, in the high-attention regime (panel (a)), the low-attention regime (panel (b)), on average across regimes (panel (c)), and the difference between the two regimes (panel (d)). The dark shaded areas depict the 68\% confidence bands and the light-shaded area the 90\% confidence bands. Standard errors are robust with respect to serial correlation and heteroskedasticity (\cite{newey1987simple} with 12 lags). 
}%
 \end{minipage}
\end{figure}

\clearpage\newpage
\paragraph{Exluding the Great Inflation period.} Figure \ref{fig:emp_irfs_inflation_after1990} shows the inflation response to a negative oil supply news shock for the two attention regimes (panels (a) and (b)), the average response in panel (c), and the difference across regimes in (d) when excluding the Great inflation period, i.e., when excluding observations before 1990. We see that the differences across regimes are even slightly stronger than in the baseline specification.

\begin{figure}[!ht]
\caption{Inflation response to an oil supply news shock (no controls)}
\label{fig:emp_irfs_inflation_noctrls}\vspace{0.15cm} \centering%
\begin{tabular}{cc}
(a) High-Attention Regime  & (b) Low-Attention Regime  \\ 
\includegraphics[width=.44\textwidth]{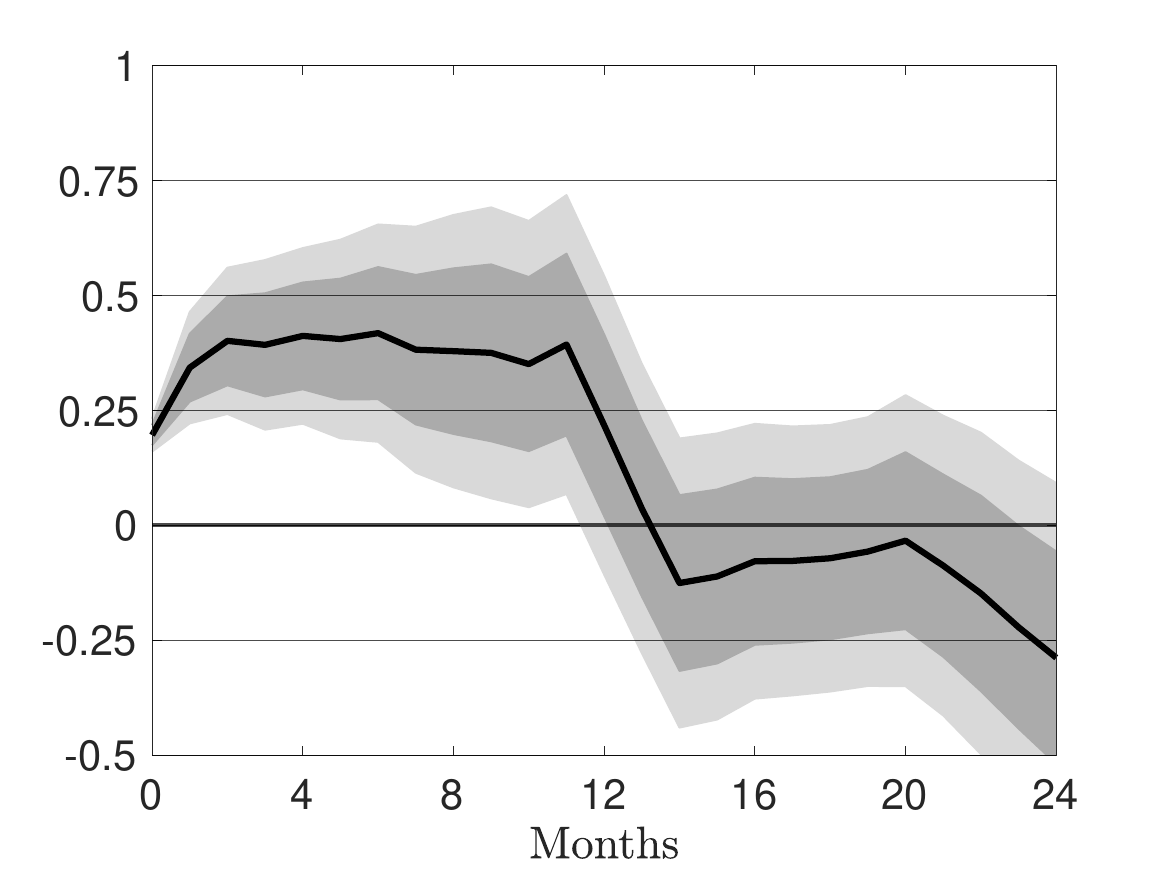} &    \includegraphics[width=.44\textwidth]{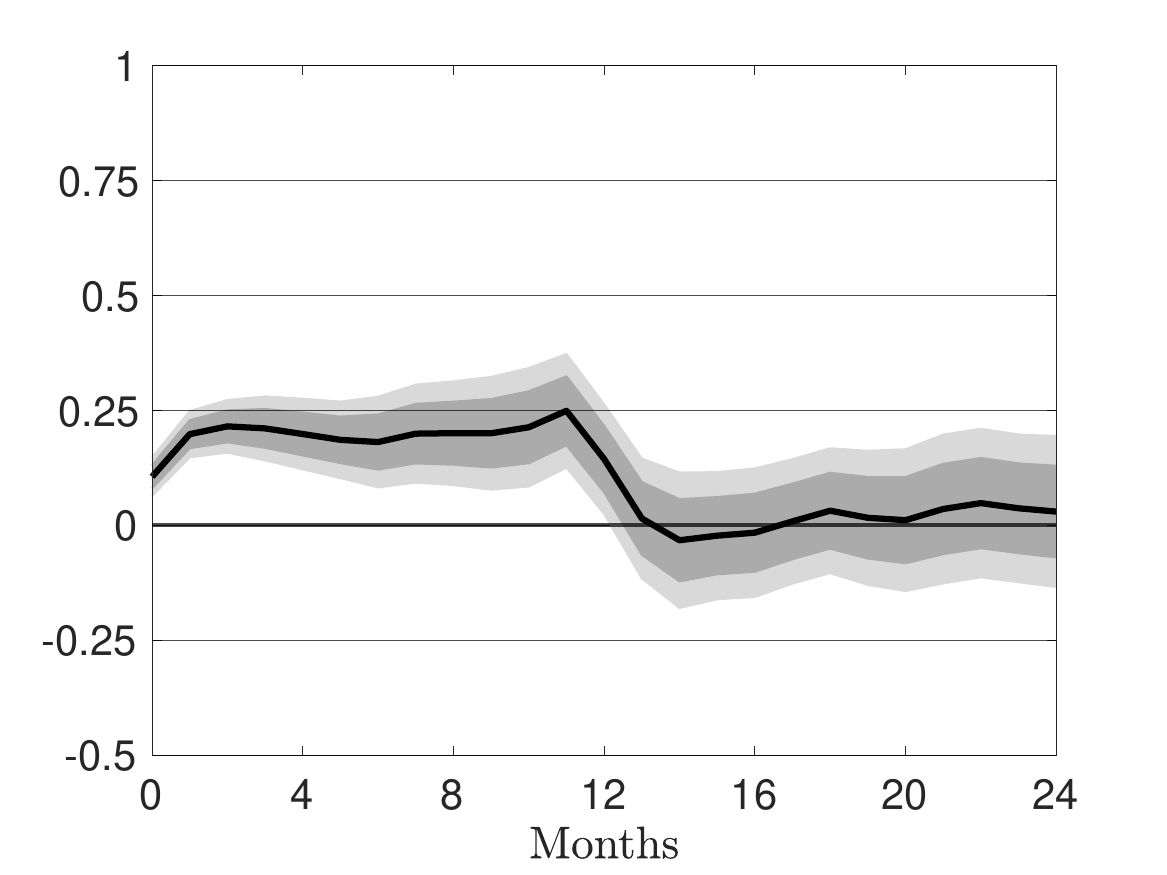} \\
(c) Average effect  & (d) Difference   \\\includegraphics[width=.44\textwidth]{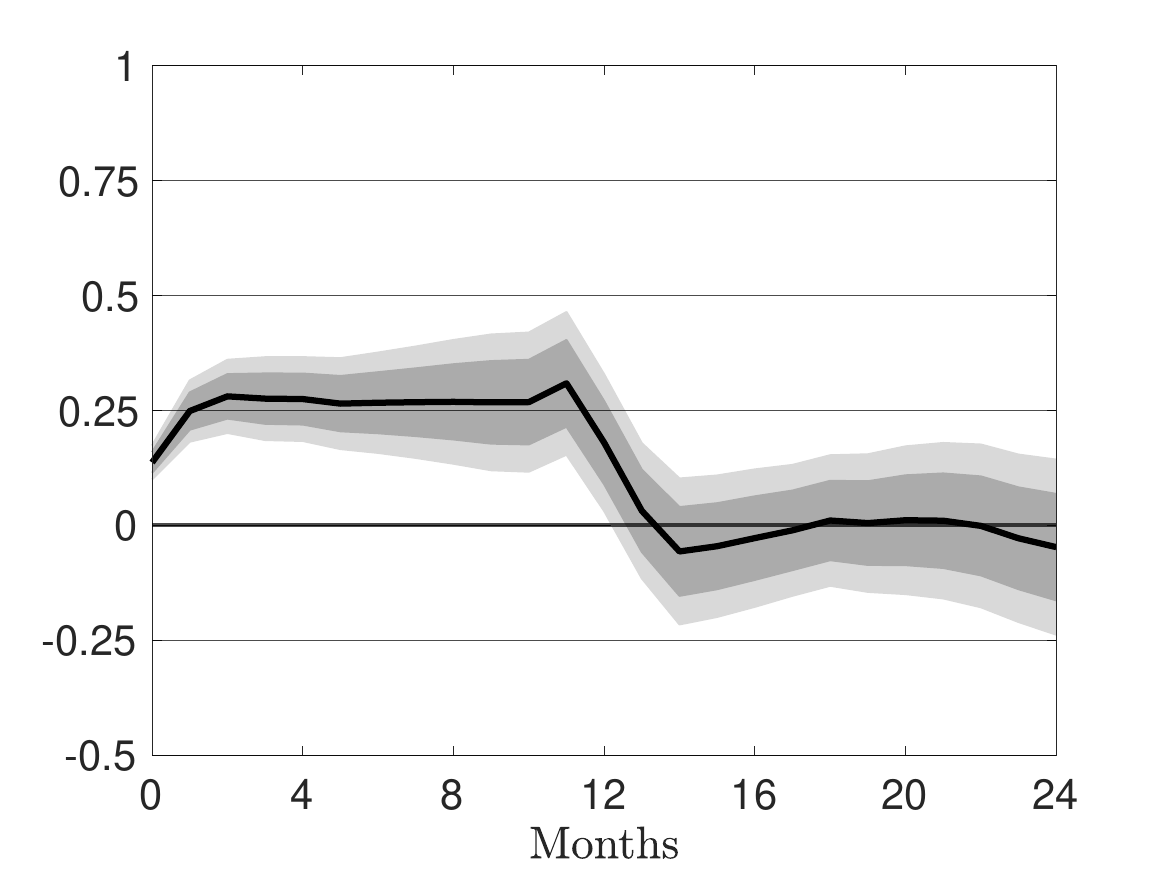} & \includegraphics[width=.44\textwidth]{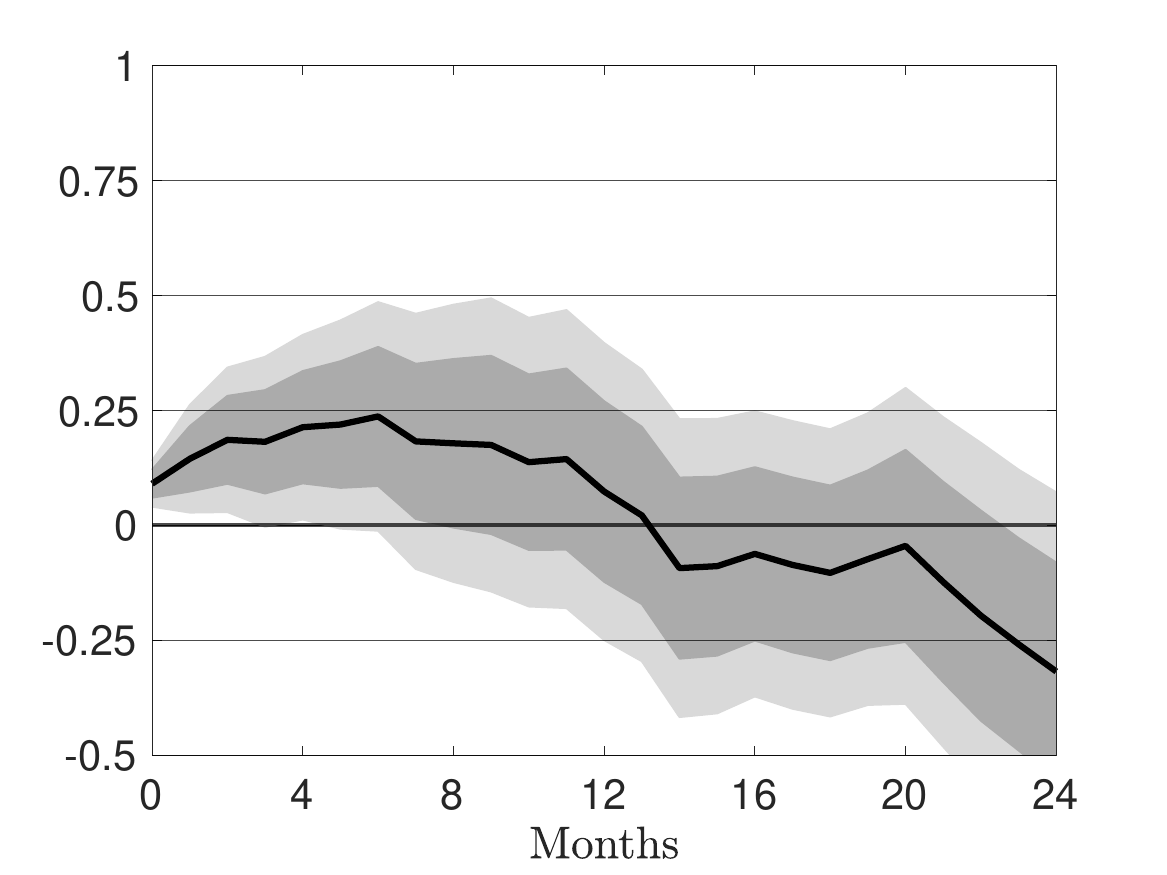}
\end{tabular}%

\begin{minipage}{1\textwidth}
\footnotesize{\emph{Notes}: This figure shows the inflation response to an oil supply news shock when using no control variables in the high-attention regime (panel (a)), the low-attention regime (panel (b)), on average across regimes (panel (c)), and the difference between the two regimes (panel (d)). The dark shaded areas depict the 68\% confidence bands and the light-shaded area the 90\% confidence bands. Standard errors are robust with respect to serial correlation and heteroskedasticity (\cite{newey1987simple} with 12 lags). 
}%
 \end{minipage}
\end{figure}

\clearpage\newpage
\paragraph{Role of control variables.} Figure \ref{fig:emp_irfs_inflation_noctrls} shows the inflation response to a negative oil supply news shock for the two attention regimes (panels (a) and (b)), the average response in panel (c), and the difference across regimes in (d) when not including any control variables. We see that the differences across regimes are slightly weaker than for the baseline case in Figure \ref{fig:emp_irfs_inflation1} but still substantial.

\begin{figure}[!ht]
\caption{Price level response to an oil supply news shock}
\label{fig:emp_irfs_oilprice}\vspace{0.15cm} \centering%
\begin{tabular}{cc}
(a) High-Attention Regime  & (b) Low-Attention Regime  \\ 
\includegraphics[width=.44\textwidth]{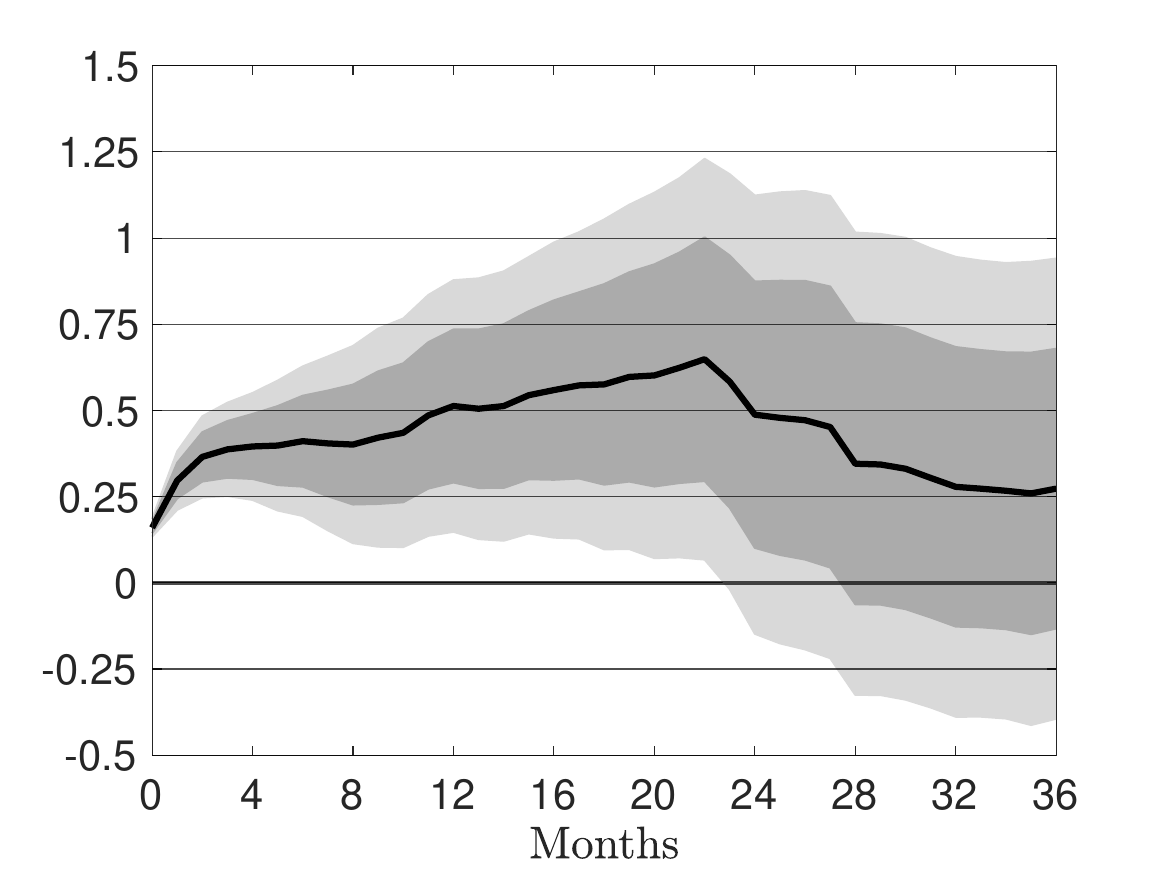} &    \includegraphics[width=.44\textwidth]{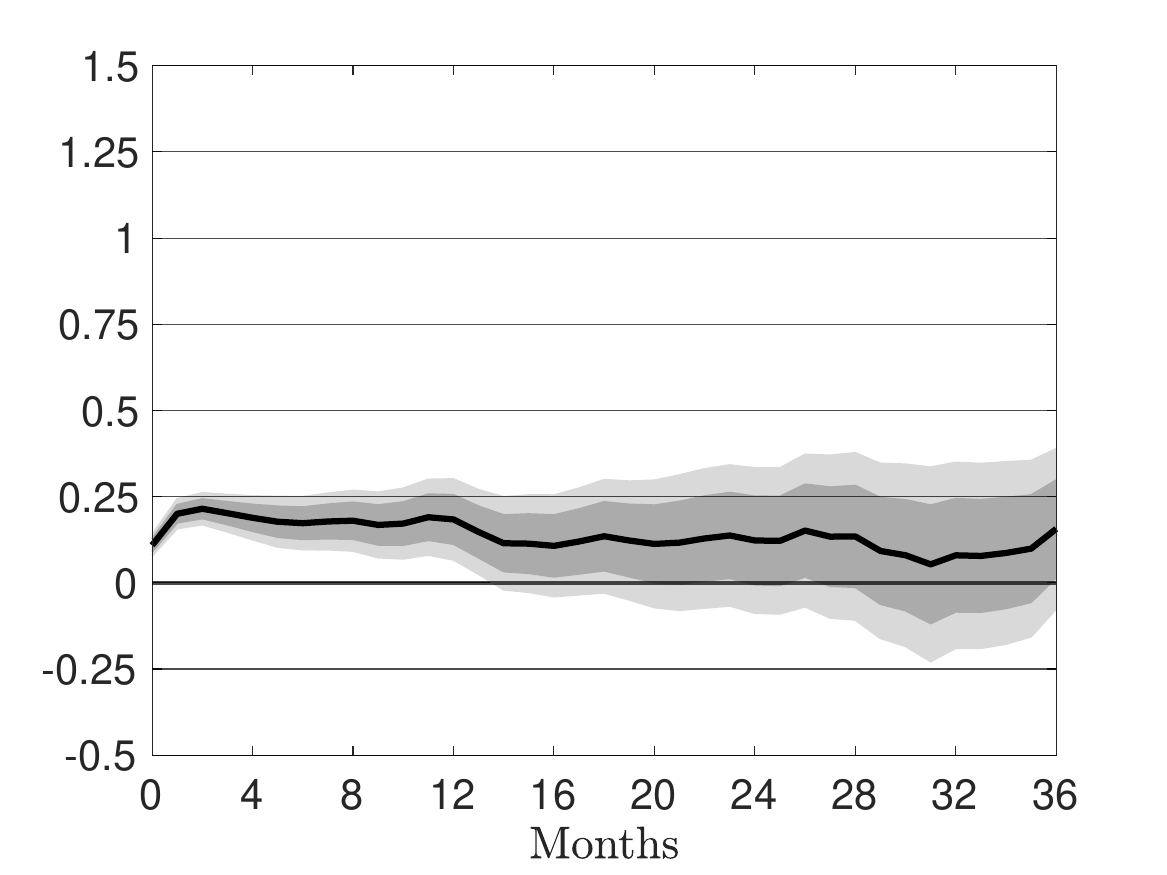} \\
(c) Average effect  & (d) Difference  \\ \includegraphics[width=.44\textwidth]{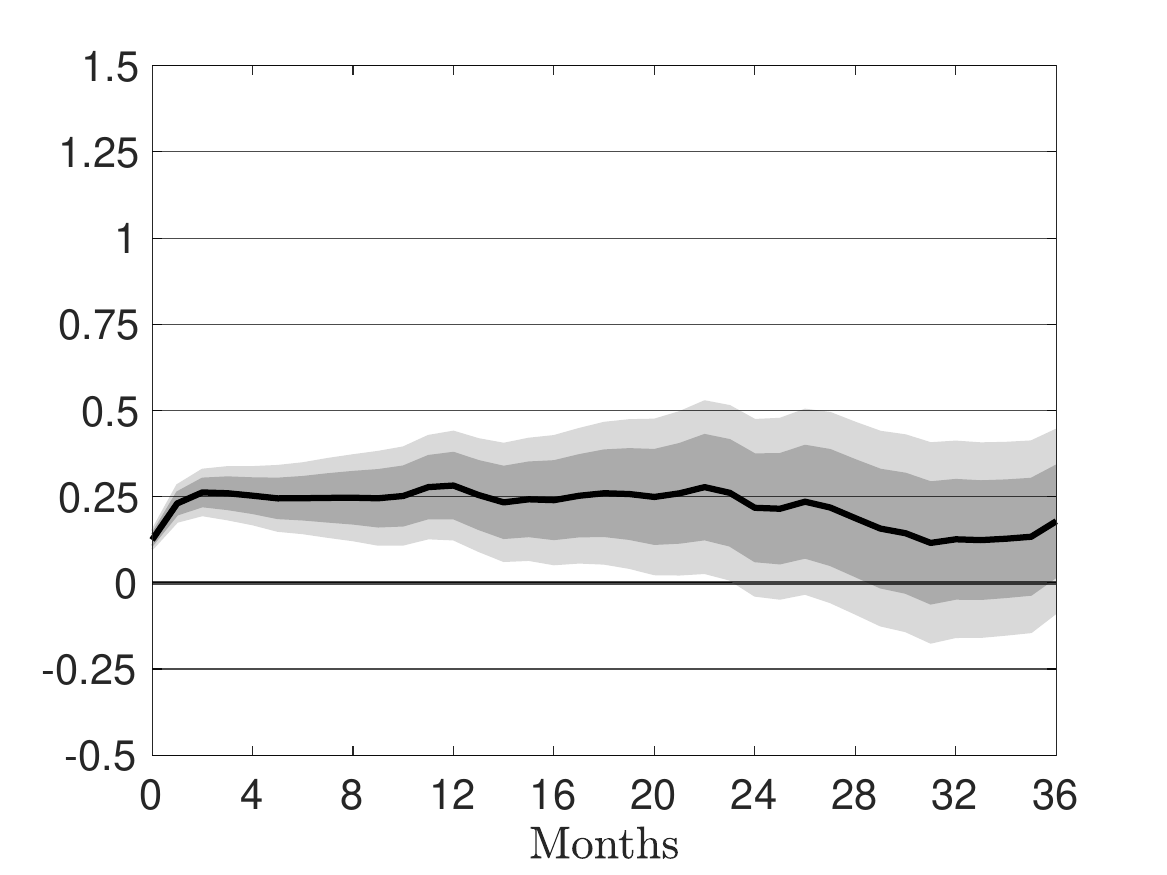} & \includegraphics[width=.44\textwidth]{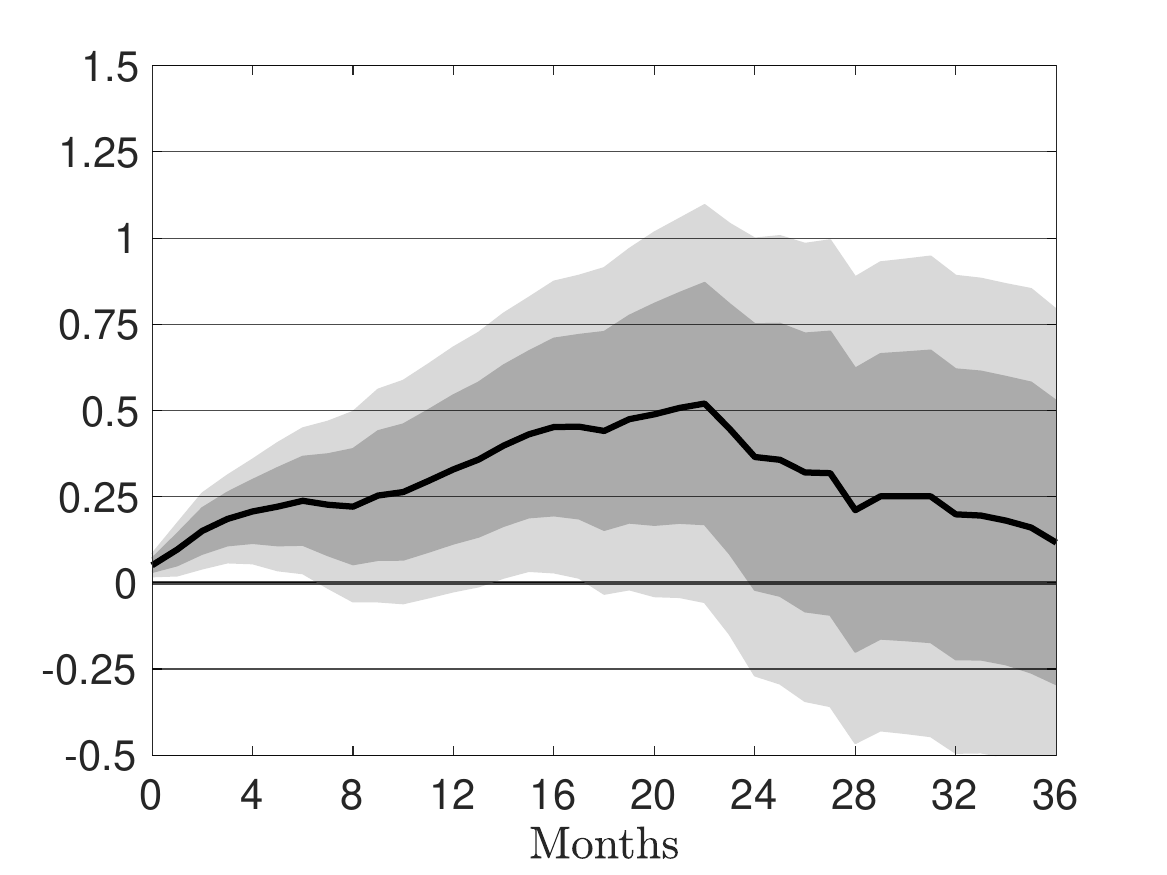}
\end{tabular}%

\begin{minipage}{1\textwidth}
\footnotesize{\emph{Notes}: This figure shows the cumulative price level (using the CPI) response to an oil supply news shock in the high-attention regime (panel (a)), the low-attention regime (panel (b)), on average across regimes (panel (c)), and the difference between the two regimes (panel (d)). The dark shaded areas depict the 68\% confidence bands and the light-shaded area the 90\% confidence bands. Standard errors are robust with respect to serial correlation and heteroskedasticity (\cite{newey1987simple} with 12 lags). The attention regimes ares defined based on the previous month's inflation rate. }%
 \end{minipage}
\end{figure}

\clearpage\newpage
\paragraph{Price-level response.} Figure \ref{fig:emp_irfs_oilprice} shows the case where the dependent variable is the cumulative change in the price level: $P_{t+h} - P_{t-1}$, where $P_t$ is the natural logarithm (times 100) of the price level using the CPI. We see from Figure \ref{fig:emp_irfs_oilprice} that the results are similar to the baseline case in figure \ref{fig:emp_irfs_inflation1}: inflation responds 2-3 times more strongly in the high-attention regime and the differences are highly statistically significant, at least in the first 1-2 years.

\begin{figure}[!ht]
\caption{Inflation and expected inflation}
\label{fig:emp_irfs_expinf}\vspace{0.15cm} \centering%
\begin{tabular}{c}
 \includegraphics[width=.44\textwidth]{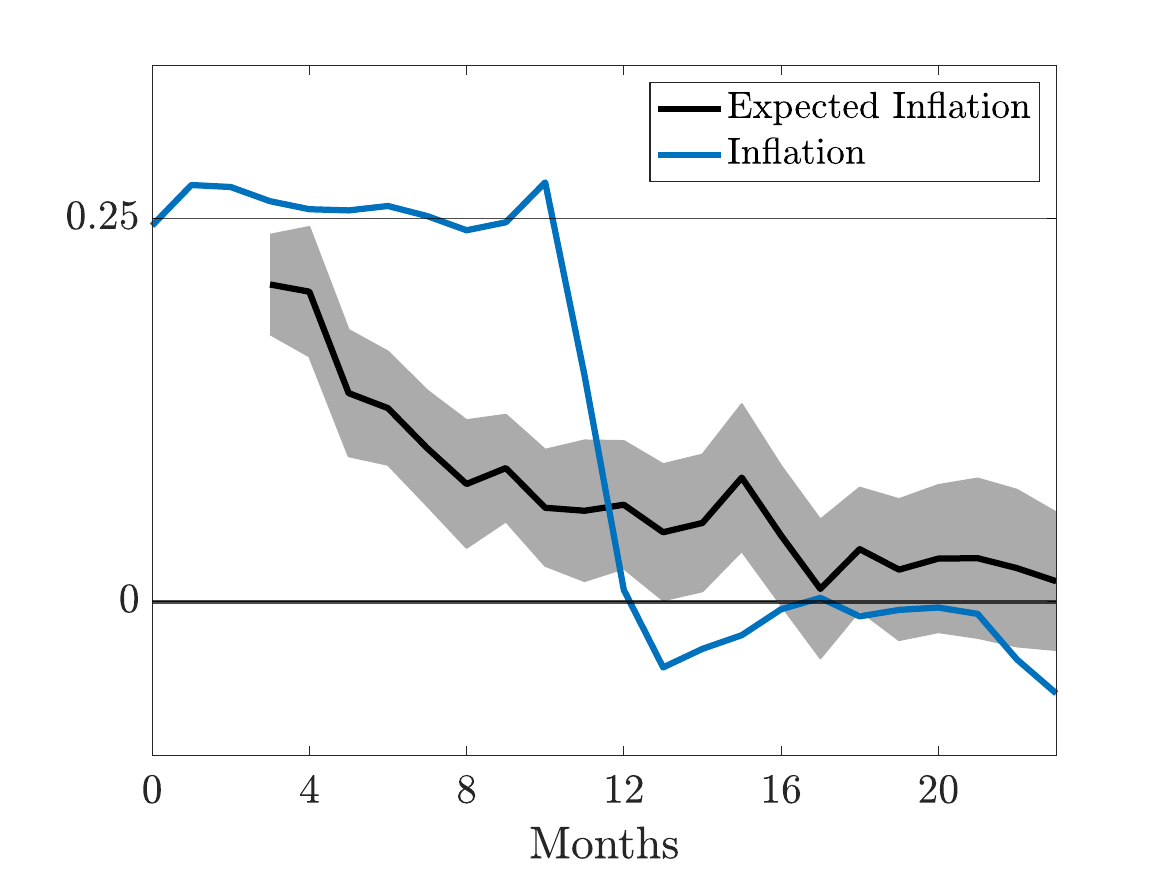}
\end{tabular}%

\begin{minipage}{1\textwidth}
\footnotesize{\emph{Notes}: This figure shows the response of inflation forecasts using average expectations from the Survey of Consumers (black-solid line) together with the response of actual inflation (blue-solid line). The response of expectations is shifted such that the vertical distance between the two lines captures forecast errors. The shaded areas depict the 68\% confidence bands, standard errors are robust with respect to serial correlation and heteroskedasticity (\cite{newey1987simple} with 12 lags). }%
 \end{minipage}
\end{figure}

\clearpage\newpage
\paragraph{Expectations and inflation.} Figure \ref{fig:emp_irfs_expinf} shows the response of inflation forecasts to an adverse oil supply news shock using average expectations from the Survey of Consumers (black-solid line) together with the response of actual inflation (blue-solid line). The response of expectations is shifted such that the vertical distance between the two lines captures forecast errors. Consistent with the model, we see that expectations initially undershoot, followed by a delayed overreaction.

\section{Model Derivations and Calibration of $\gamma_x$}\label{app:model}\vspace{-0.2cm}
\paragraph{Derivation of equation \eqref{eq:euler}.} Linearizing the Euler equation \eqref{eq:euler_model} yields
\begin{equation}
    \widehat{c}_t = \tilde{E}_t\widehat{c}_{t+1} - \left(\Tilde{i}_t - \Tilde{E}_t\pi_{t+1} - (\widehat{z_t} - \Tilde{E}_tz_{t+1})\right). \label{eq:euler_lin}
\end{equation}
The representative household therefore needs to form expectations about future consumption $\tilde{E}_t\widehat{c}_{t+1}$, inflation $\Tilde{E}_t\pi_{t+1}$ and the exogenous preference shock $\Tilde{E}_t\widehat{z}_{t+1}$.  The household holds rational expectations about the shock, so $\Tilde{E}_t\widehat{z}_{t+1} = E_t\widehat{z}_{t+1}$, where $E_t(\cdot)$ denotes the rational expectations operator, and since $\widehat{z}_t$ follows an AR(1) process with persistence $\rho_{z}$, it follows that $\Tilde{E}_t\widehat{z}_{t+1} = E_t\widehat{z}_{t+1} = \rho_z \widehat{z}_t$.

In order to express equation \eqref{eq:euler_lin} in terms of the output gap, we need to derive the efficient allocation, i.e., the one that prevails in the economy with fully flexible prices. From the production function, we have $Y^*_t = H^*_t$. The real wage is constant $w^*_t = 1$. From the labor-leisure equation \eqref{eq:ll}, we get that potential output is therefore also constant and equal to
\begin{equation}
    Y^*_t = \Xi^{-1}.
\end{equation}
Thus, potential output in log-deviations is 0. The Euler equation in the flexible-price economy is therefore given by
\begin{equation}
    0 =  -\left(r_t - (\widehat{z_t} - \Tilde{E}_tz_{t+1})\right). \notag
\end{equation}
Since the natural rate is defined as the real rate that prevails under flexible prices, $r_t$, it follows directly that
\begin{equation}
    r^*_t = \widehat{z}_t - \Tilde{E}_tz_{t+1}.
\end{equation}
Substituting $\widehat{z}_t - \Tilde{E}_tz_{t+1}$ with $r^*_t$ in \eqref{eq:euler_lin} and using that $\widehat{c}_t = \widehat{y}_t = \widehat{x}_t$, since potential output is 0, yields the IS equation \eqref{eq:euler}:
\begin{equation}
    \widehat{x}_t = \tilde{E}_t\widehat{c}_{t+1} - \left(\tilde{i}_t-\tilde{E}_t\pi_{t+1}-r^*_t\right). \label{eq:ISintermediate}
\end{equation} 
Note, however, that if we assume initial values $\tilde{E}_{-1}\widehat{c}_{0} = \tilde{E}_{-1}\widehat{x}_{0}$, it follows that $\tilde{E}_t\widehat{c}_{t+1} = \tilde{E}_t\widehat{x}_{t+1}$ because $\widehat{c}_t = \widehat{x}_t$ in equilibrium. This holds true, even if the household does not know that they are equal in equilibrium. We can thus write equation \eqref{eq:ISintermediate} as 
\begin{equation}
    \widehat{x}_t = \tilde{E}_t\widehat{x}_{t+1} - \left(\tilde{i}_t-\tilde{E}_t\pi_{t+1}-r^*_t\right),
\end{equation} 
which is equation \eqref{eq:euler} in the main text.

\paragraph{Derivation of equation \eqref{eq:nkpc}.}
To derive the New Keynesian Phillips Curve \eqref{eq:nkpc}, we start by maximizing
\begin{equation}
    \Omega_0(j) = \Tilde{E}_0\sum\limits_{t=0}^{\infty}\beta^t\left[(1-\tau_t)P_t(j)\left( \frac{P_t(j)}{P_t}\right)^{-\epsilon} \frac{Y_t}{P_t} - w_t H_t(j) - \frac{\psi}{2}\left(\frac{P_t(j)}{P_{t-1}(j)} - 1\right)^2Y_t + t^F_t(j)\right], \notag
\end{equation}
which yields the first-order condition
\begin{align*}
        (1-\tau_t) (\epsilon-1)P_t(j)^{-\epsilon}\frac{Y_t}{P_t^{1-\epsilon}}  &= \epsilon mc_t \left(\frac{P_t(j)}{P_t} \right)^{-\epsilon -1} \frac{Y_t}{P_t} - \psi\left(\frac{P_t(j)}{P_{t-1}(j)} -1\right)\frac{Y_t}{P_{t-1}(j)}\\
&+ \beta \psi \tilde{E}_t\left[\left(\frac{P_{t+1}(j)}{P_t(j)}-1\right)\frac{P_{t+1}(j)}{P_t(j)}\frac{Y_{t+1}}{P_t(j)} \right].
\end{align*}

I define $T_t \equiv 1-\tau_t$, and set it such that marginal costs in the steady state are zero, i.e., $T = \frac{\epsilon}{\epsilon -1}$. Linearizing around the zero-inflation steady state in which $P(j) = P$ for all $j$, I obtain
\begin{align}
    &T(\epsilon-1)\frac{Y}{P}\left[\widehat{T}_t - \epsilon \widehat{p}_t(j) + \widehat{y}_t - (1-\epsilon)\widehat{p}_t \right] \\  = & \epsilon mc \frac{Y}{P}\left[ (-\epsilon -1) \widehat{p}_t(j) - (-\epsilon - 1) \widehat{p}_t + \widehat{mc}_t + \widehat{y}_t - \widehat{p}_t\right] \\
    -& \psi\frac{Y}{P}\left(\widehat{y}_t - \widehat{p}_{t-1}(j) - \pi_t(j) - \widehat{y}_t + \widehat{p}_{t-1}(j) \right) \\
    +  & \beta\psi\frac{Y}{P}\tilde{E}^j_t\pi^j_{t+1}.
\end{align}
Grouping terms, using $mc = 1$, $T = \frac{\epsilon}{\epsilon -1}$ and dividing by $\frac{Y}{P}$ yields
\begin{equation}
    \widehat{p}_t(j) = \frac{1}{\psi+\epsilon}\left[\psi \widehat{p}_{t-1}(j) + \epsilon\left( \widehat{mc}_t - \widehat{T}_t + \widehat{p}_t\right) + \beta\psi \tilde{E}^j_t\pi^j_{t+1} \right].
\end{equation}
Given the assumptions discussed in Section \ref{sec:model}, it follows that $\tilde{E}^j_t \pi ^j_{t+1} = \Tilde{E}_t\pi_{t+1}$ and $\widehat{p}_{t-1}(j) = \widehat{p}_{t-1}$ such that
\begin{equation}
    \widehat{p}_t(j) = \frac{1}{\psi+\epsilon}\left[\psi \widehat{p}_{t-1} + \epsilon\left( \widehat{mc}_t - \widehat{T}_t + \widehat{p}_t\right) + \beta\psi \tilde{E}_t\pi_{t+1} \right].
\end{equation}
Therefore, the optimal price $\widehat{p}_t(j)$ is the same for all firms $j$, $\widehat{p}_t(j) = \widehat{p}_t$, such that we get
\begin{equation}
    \psi\underbrace{\left(\widehat{p}_t - \widehat{p}_{t-1}  \right)}_{= \pi_t} = \epsilon\left( \widehat{mc}_t - \widehat{T}_t\right) + \beta\psi \tilde{E}_t\pi_{t+1}.
\end{equation}
Dividing by $\psi$ yields 
\begin{equation}
    \pi_t = \frac{\epsilon}{\psi}\left(\widehat{mc}_t - \widehat{T}_t \right) + \beta \tilde{E}_t \pi_{t+1}.
\end{equation}
Using $\widehat{mc}_t = \widehat{y}_t$ and $\widehat{y}_t = \widehat{x}_t$ and defining cost-push shocks as $u_t \equiv -\frac{\epsilon}{\psi}\widehat{T}_t$ and the slope parameter $\kappa \equiv \frac{\epsilon}{\psi}$, I arrive at the linearized New Keynesian Phillips Curve under subjective expectations:
\begin{equation}
    \pi_t = \kappa \widehat{x}_t + \beta \Tilde{E}_t\pi_{t+1} + u_t.
\end{equation}

\subsection{Unemployment expectations}\label{app:unemp}\vspace{-0.2cm}
In order to calibrate the attention parameter with respect to the output gap, $\gamma_x$, I use expectations about unemployment changes from the Survey of Consumers.
The survey asks households whether they expect unemployment to increase, decrease or to remain about the same over the next twelve months. I follow \citet{carlson1975inflation}, \citet{bhandari2019survey} and \cite{pfauti2022behavioral} to translate these categorical unemployment expectations into numerical expectations.

Let $q^{D}_t$, $q^{S}_t$ and $q^{U}_t$ denote the shares reported at time $t$ that think unemployment will go down, stay roughly the same, or go up over the next year, respectively. 
I assume that these shares are drawn from a cross-sectional distribution of responses that are normally distributed according to $\mathcal{N}\left(\mu_t,(\sigma_t)^2\right)$ and a threshold $a$ such that when a household expects unemployment to remain within the range $[-a, a]$ over the next year, she responds that unemployment will remain "about the same". We thus have
\begin{align*}
    q^{D}_t = \Phi\left(\frac{-a-\mu_t}{\sigma_t}\right) \qquad q^{U}_t = 1 - \Phi\left( \frac{a-\mu_t}{\sigma_t}\right),
\end{align*}
which after some rearranging yields
\begin{align*}
    \sigma_t &= \frac{2a}{\Phi^{-1}\left(1-q^{U}_t\right)-\Phi^{-1}\left(q^{D}_t\right)}\\
    \mu_t &=  a- \sigma_t\Phi^{-1}\left(1-q^{U}_t\right).
\end{align*}

This leaves us with one degree of freedom, namely $a$. I follow \cite{pfauti2022behavioral} and set $a = 0.5$ which means that if a household expects the change in unemployment to be less than half a percentage point (in absolute terms), she reports that she expects unemployment to be about the same as it is at the time of the survey.
I use data from FRED for the actual unemployment changes and restrict the sample to end in 2019Q4, due to the extreme behavior of unemployment changes with the outbreak of the Covid-19 pandemic. 

I then estimate $\gamma_x$ separately for whether lagged inflation is above or below the estimated threshold of 4\%. This results in estimates $\widehat{\gamma}_{x,L} = 0.25$ and $\widehat{\gamma}_{x,H} = 0.25$. Thus, there are no differences in attention to unemployment changes across regimes and hence, I impose $\gamma_x$ to be the same across regimes in the model.

\clearpage\newpage
\section{Additional Figures and Model Results}\label{app:model_extensions}\vspace{-0.2cm}
\subsection{An Illustrative Example}\label{sec:example}\vspace{-0.2cm}
To provide intuition how the attention threshold can trigger self-reinforcing inflation surges, I start with a slightly stylized version of the model. In particular, I assume that agents are completely inattentive to the output gap, i.e., $\gamma_x =0$, and that the Taylor rule is given by $\Tilde{i}_t = \phi_{\pi}\pi_t$ with $\phi_{\pi} > 1$. 

\begin{figure}[!ht]
\caption{An illustrative example}
\label{fig:asad}\vspace{0.15cm} \centering%
\begin{tabular}{ccc}
(a) Steady state & (b) Cost-push shock: AS shifts up \\
\includegraphics[width=.45\textwidth]{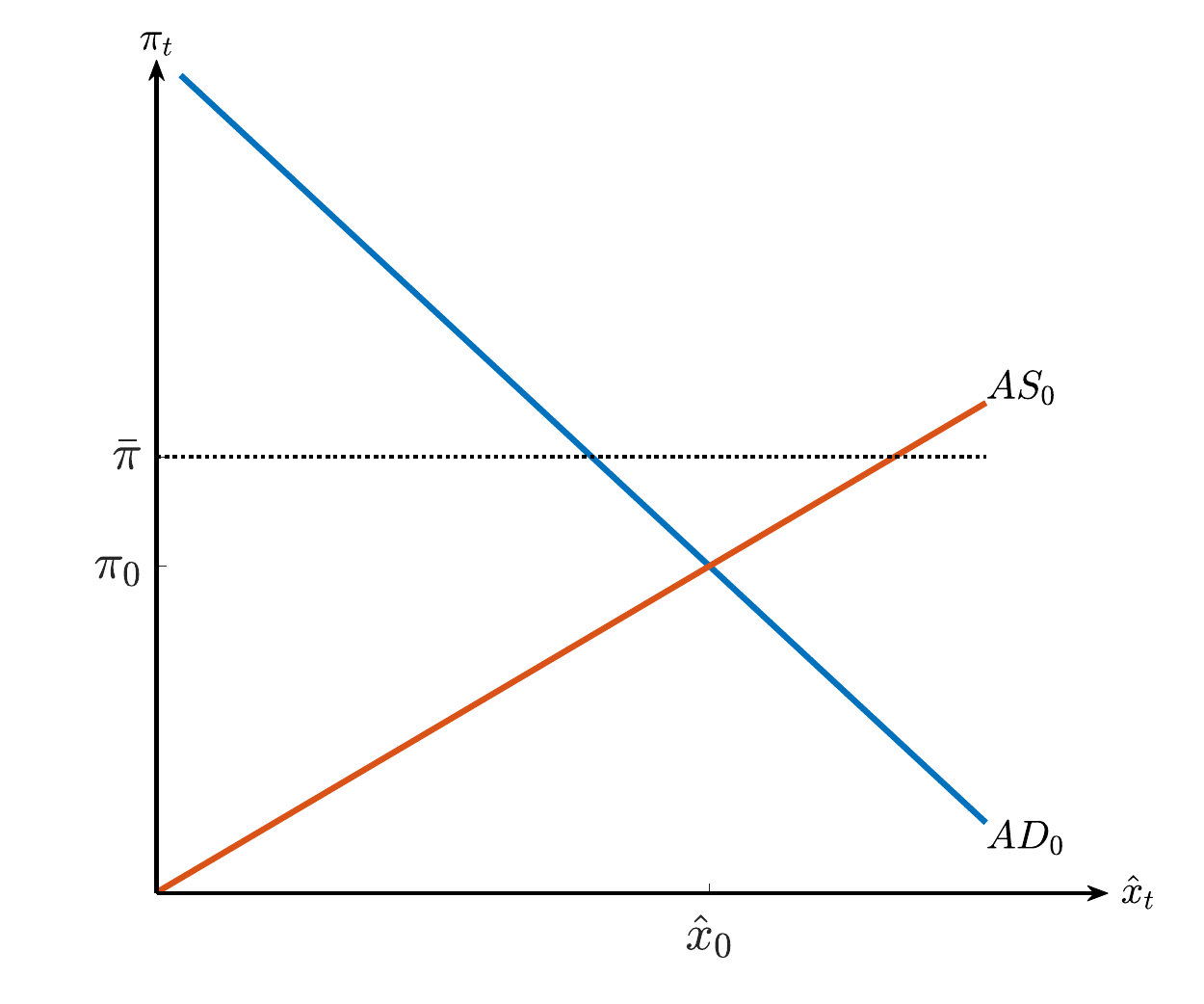} &    \includegraphics[width=.45\textwidth]{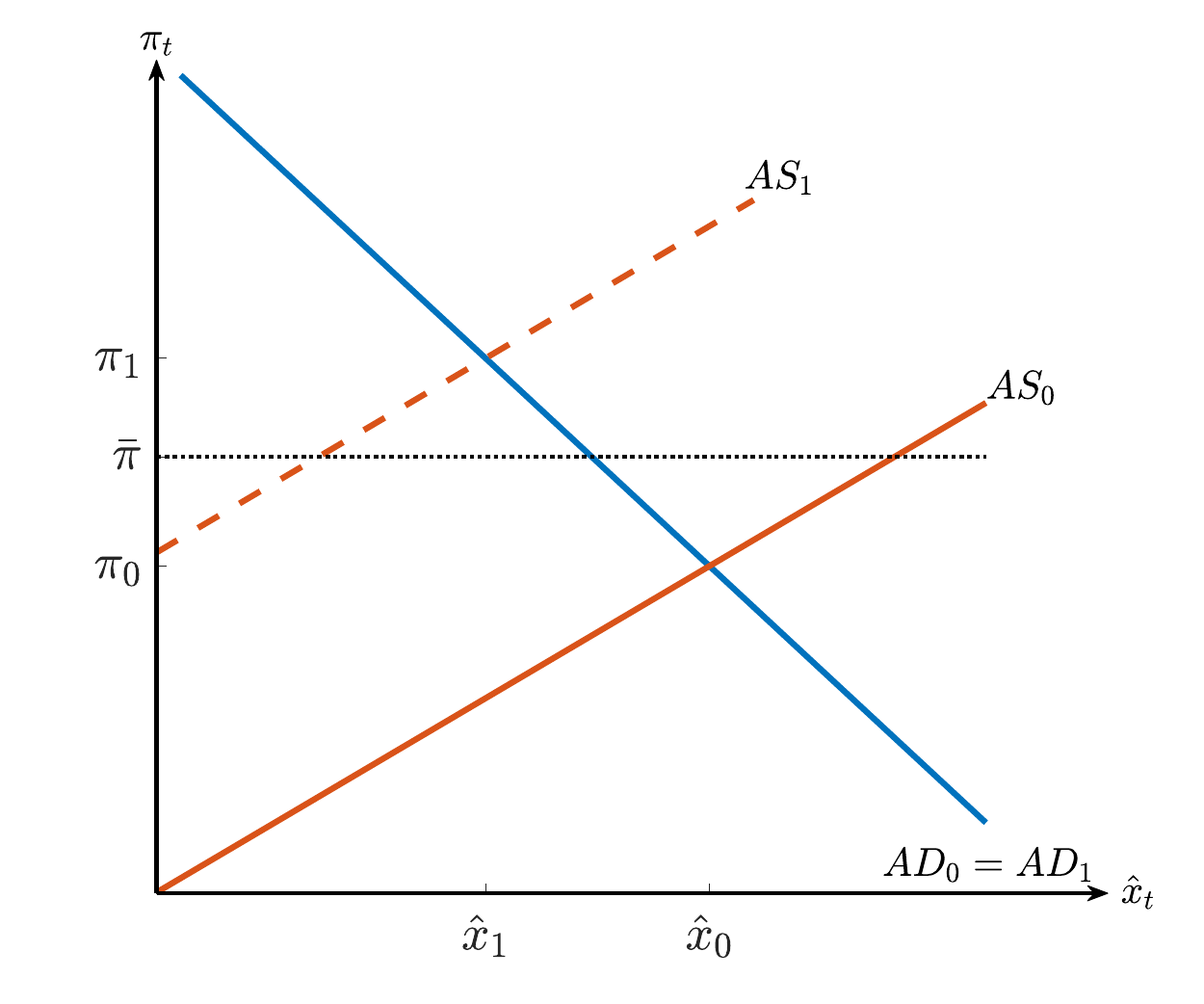} \\
(c) AS: further up and steeper & (d) AD: out and steeper \\
\includegraphics[width=.45\textwidth]{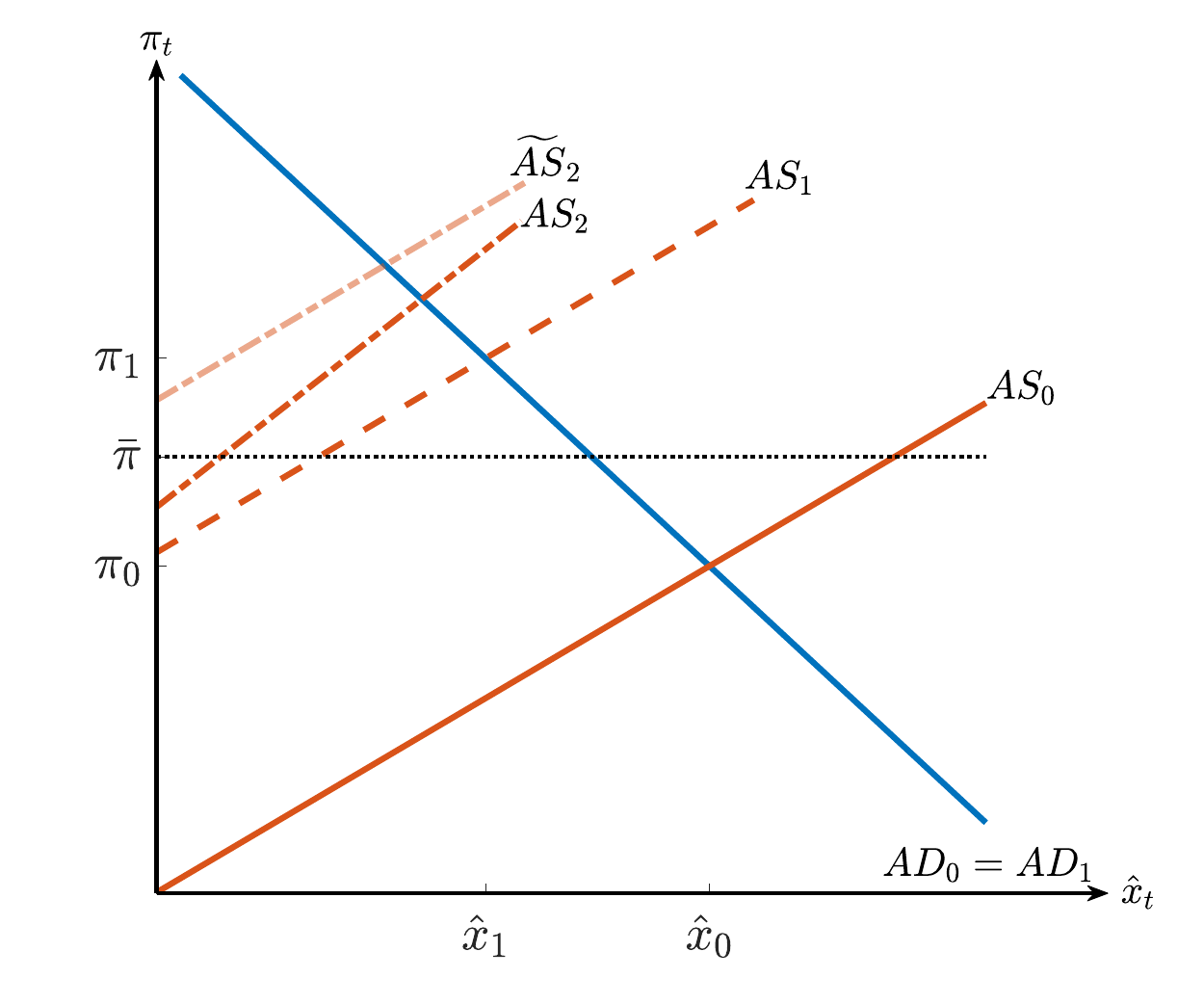} &    \includegraphics[width=.45\textwidth]{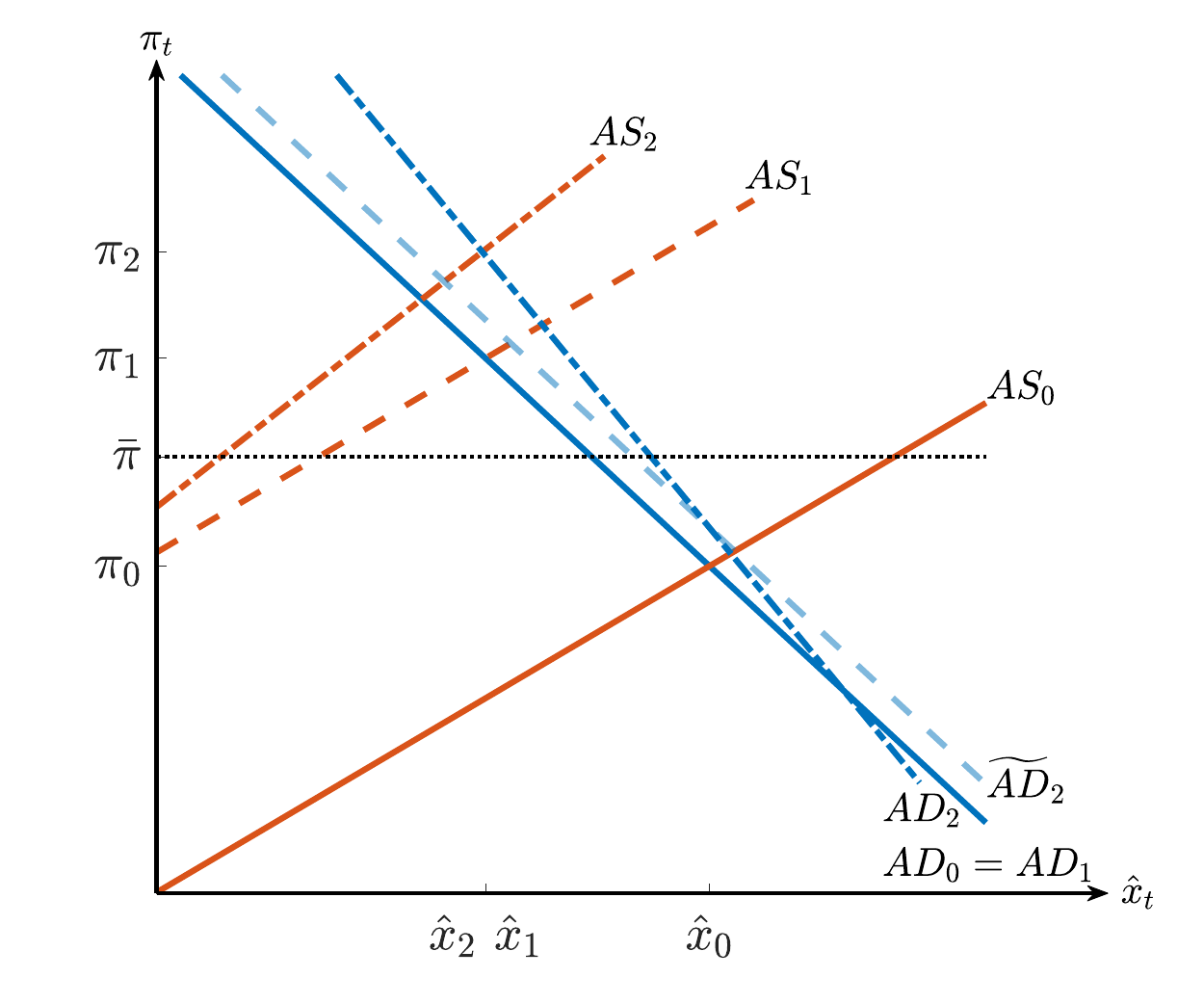} 
\end{tabular}%

\begin{minipage}{1\textwidth}
\footnotesize{\emph{Notes}: Panel (a) shows the initial situation in period 0 when the economy is in the steady state. In period 1, a positive cost-push shock hits the economy, leading to an upward shift of the AS curve (panel (b)). In period 2, the AS curve shifts further up (shift from $AS_1$ to $\widetilde{AS}_2$) and the curve becomes steeper (rotation from $\widetilde{AS}_2$ to $AS_2)$, shown in panel (c). Simultaneously, the AD curve shifts out (shift from $AD_1$ to $\widetilde{AD}_2$) and becomes stepper (rotation from $\widetilde{AD}_2$ to $AD_2$), as shown in panel (d). The black-dotted line at $\Bar{\pi}$ depicts the inflation attention threshold.
}%
 \end{minipage}
\end{figure}

The economy starts in the steady state, with $u_0 = 0$, $r^*_0 = 0$ and prior expectations at their long-run averages of 0: $\Tilde{E}_{-1}\pi_0 = 0$ and $\Tilde{E}_{-1}\widehat{x}_0 = 0.$ The aggregate supply (AS) curve---derived by plugging in the inflation expectations in the Phillips Curve---in the initial period is given by
\begin{equation}
    AS_0: \quad \pi_0 = \frac{\kappa}{1-\beta\gamma_{\pi,L}}\widehat{x}_0,
\end{equation}
and aggregate demand (AD), which follows from combining the Taylor rule with the aggregate Euler (or IS) equation, is given by
\begin{equation}
    AD_0: \quad \pi_0 = -\frac{1}{\phi_{\pi}-\gamma_{\pi,L}}\widehat{x}_0.
\end{equation}
Panel (a) in Figure \ref{fig:asad} depicts this initial situation graphically.\footnote{The values I use in this stylized example are: $\Bar{\pi} = 4\%$, $\gamma_{\pi,L} = 0.2$, $\gamma_{\pi,H} = 0.4$, $\beta = 0.99$, $\kappa = 0.6$, $\phi_{\pi} = 1.05$ and $u_1 = u_2 = 10$.}  Inflation and the output gap are both at their steady state values of 0 and therefore, below the inflation attention threshold $\Bar{\pi}$.

In period 1, a positive cost-push shock hits and I assume that it persists for two periods:  $u_1 = u_2 > 0$ and returns to zero afterwards, $u_t = 0$ for $t\geq 3$. The AS and AD equations are now given by
\begin{align*}
    AS_1: \quad& \pi_1 = \frac{\kappa}{1-\beta\gamma_{\pi,L}}\widehat{x}_1 + \frac{1}{1-\beta\gamma_{\pi,L}}u_1 \\
    AD_1: \quad& \pi_1 = -\frac{1}{\phi_{\pi}- \gamma_{\pi,L}}\widehat{x}_1.
\end{align*}
This situation is shown in panel (b) of Figure \ref{fig:asad}. The cost-push shock shifts the AS curve up along the AD curve. The resulting equilibrium is characterized by output below potential, i.e., a negative output gap, and positive inflation. The shock is assumed to be large enough, such that inflation exceeds the threshold.

Due to the increase in inflation in period 1, firms enter period 2 with positive prior inflation expectations: $\Tilde{E}_1\pi_2 = \gamma_{\pi,L}\pi_1 > 0$. These higher prior expectations together with the still ongoing cost-push shock shift the AS curve further up. This shift is illustrated in panel (c) of Figure \ref{fig:asad} by the $\widetilde{AS}_2$ curve, which is given by
\begin{equation*}
    \widetilde{AS}_2: \quad \pi_2 = \frac{\kappa}{1-\beta\gamma_{\pi,L}}\widehat{x}_2 + \underbrace{\frac{1}{1-\beta\gamma_{\pi,H}}u_2 + \frac{\beta(1-\gamma_{\pi,H})\gamma_{\pi,L}}{1-\beta\gamma_{\pi,H}}\pi_1}_{\text{Intercept} > 0}.
    \end{equation*}
The terms denoted ``Intercept'' capture this shift in the AS curve. Since inflation in the previous period exceeded the attention threshold, attention is now higher. This increase in attention leads to an unambiguously stronger effect of the cost-push shock compared to the case in which attention would have remained constant:
\begin{equation*}
    \frac{1}{1-\beta\gamma_{\pi,H}}u_2 > \frac{1}{1-\beta\gamma_{\pi,L}}u_2.
\end{equation*}
The cost-push shock leads to an increase in inflation, and this inflation increase now leads to a larger increase in firms' inflation expectations due to their higher attention. These higher expectations then feed back into higher prices and thus, higher inflation. 

The effect of the increase in firm managers' prior expectations on inflation in the second period, captured by $\tfrac{\beta(1-\gamma_{\pi,H})\gamma_{\pi,L}}{1-\beta\gamma_{\pi,H}}\pi_1$, however, is smaller at the higher attention level. There are two counteracting forces. First, the increase in attention means that firm managers now update their expectations more strongly and put less weight on their prior expectations. This per se leads to a smaller effect. Second, the higher prior increases overall inflation expectations which, ceteris paribus, increases current inflation. But because inflation expectations are discounted by $\beta \leq 1$, the first effect dominates. Therefore, the shift in the AS curve due to the higher prior expectations is smaller at higher attention levels. Quantitatively, however, these differences are very small as $\beta \approx 1$. In fact, if $\beta = 1$, the change in attention has no effect on the shift due to higher prior expectations, and therefore, the total shift of the AS curve is unambiguously higher when attention is higher, due to stronger effect of the cost-push shock.

The shift of the AS curve to $\widetilde{AS}_2$, however, is only part of the story because the increase in attention also steepens the slope of the AS curve. That is, the aggregate supply curve becomes steeper in periods of high attention---a \textit{dynamic} non-linearity.
Taking this into account, the AS curve in the second period is given by
\begin{equation*}
    AS_2: \quad \pi_2 = \underbrace{\frac{\kappa}{1-\beta\gamma_{\pi,H}}}_{\text{Slope}}\widehat{x}_2 + \frac{1}{1-\beta\gamma_{\pi,H}}u_2 + \frac{\beta(1-\gamma_{\pi,H})\gamma_{\pi,L}}{1-\beta\gamma_{\pi,H}}\pi_1.
\end{equation*}
This steepening of the AS curve is illustrated in panel (c) of Figure \ref{fig:asad} by the rotation of the AS curve from $\widetilde{AS}_2$ to $AS_2$. This steepening of the AS curve eases the inflationary pressures due to the negative output gap. Nevertheless, the steeper AS curve implies that if the AD curve would now shift out, the inflationary effects of this increase in demand would become larger. 

It turns out that the higher prior expectations endogenously lead to such a demand increase.
This is illustrated by $\widetilde{AD}_2$ in panel (d) of Figure \ref{fig:asad}, which is given by
\begin{equation*}
    \widetilde{AD}_2: \quad \pi_2 = -\frac{1}{\phi_{\pi} - \gamma_{\pi,L}}\widehat{x}_2 + \frac{(1-\gamma_{\pi,H})\gamma_{\pi,L}}{\phi_{\pi} - \gamma_{\pi,H}}\pi_1.
\end{equation*}
The higher prior expectations, ceteris paribus, decrease the real rate which leads to the outward shift of the AD curve. These effects, however, are smaller at higher levels of attention as long as the Taylor principle, $\phi_{\pi} > 1$, is satisfied, because in that case the higher inflation rates due to the higher prior expectations are counteracted by a more than one-for-one increase in the nominal rate. If monetary policy is relatively dovish, i.e., $\phi_{\pi}$ is close to 1, these differences are small. Furthermore, since the AD curve is now shifted along a steeper AS curve due to the heightened attention, the inflationary effects of a given shift are larger at the higher level of attention. This also implies that additional demand stimulus---for example, due to loose monetary policy or a fiscal stimulus---would have relatively large inflationary effects.

Additionally, the AD curve also becomes steeper, as illustrated by the rotation from $\widetilde{AD}_2$ to $AD_2$ in panel (d), where $AD_2$ is given by
\begin{equation*}
    AD_2: \quad \pi_2 = -\frac{1}{\phi_{\pi} - \gamma_{\pi,H}}\widehat{x}_2 + \frac{(1-\gamma_{\pi,H})\gamma_{\pi,L}}{\phi_{\pi} - \gamma_{\pi,H}}\pi_1.
\end{equation*}
This leads to a further increase in inflation, especially now because the AS curve is steep. In this stylized example, inflation increases substantially from period 1 to period 2 through the change in attention, whereas the output gap remains practically constant. Thus, the attention threshold offers a potential mechanism for how inflation surges may occur and exhibit self-reinforcing dynamics without large changes in output.

\paragraph{After the shock.}
In the third period, when the shock has died out, the AS curve shifts back down. The AS curve is given by
\begin{equation*}
    AS_3: \quad \pi_3 = \frac{\kappa}{1-\beta\gamma_{\pi,H}}\widehat{x}_3 + \frac{\beta(1-\gamma_{\pi,H})}{1-\beta\gamma_{\pi,H}}\Tilde{E}_2\pi_3.
\end{equation*}
Due to the positive prior expectations, $\Tilde{E}_2\pi_3 > 0$, the AS curve does not fully shift back to its initial position but remains above it. Due to the increased steepness of the AD curve, however, the shift in the AS curve leads to a stronger reduction in inflation compared to a flatter AD curve. 

\begin{figure}[!ht]
\caption{Illustrative example: after the shock has died out}
\label{fig:asad3}\vspace{0.15cm} \centering%
\begin{tabular}{c}
 \includegraphics[width=.45\textwidth]{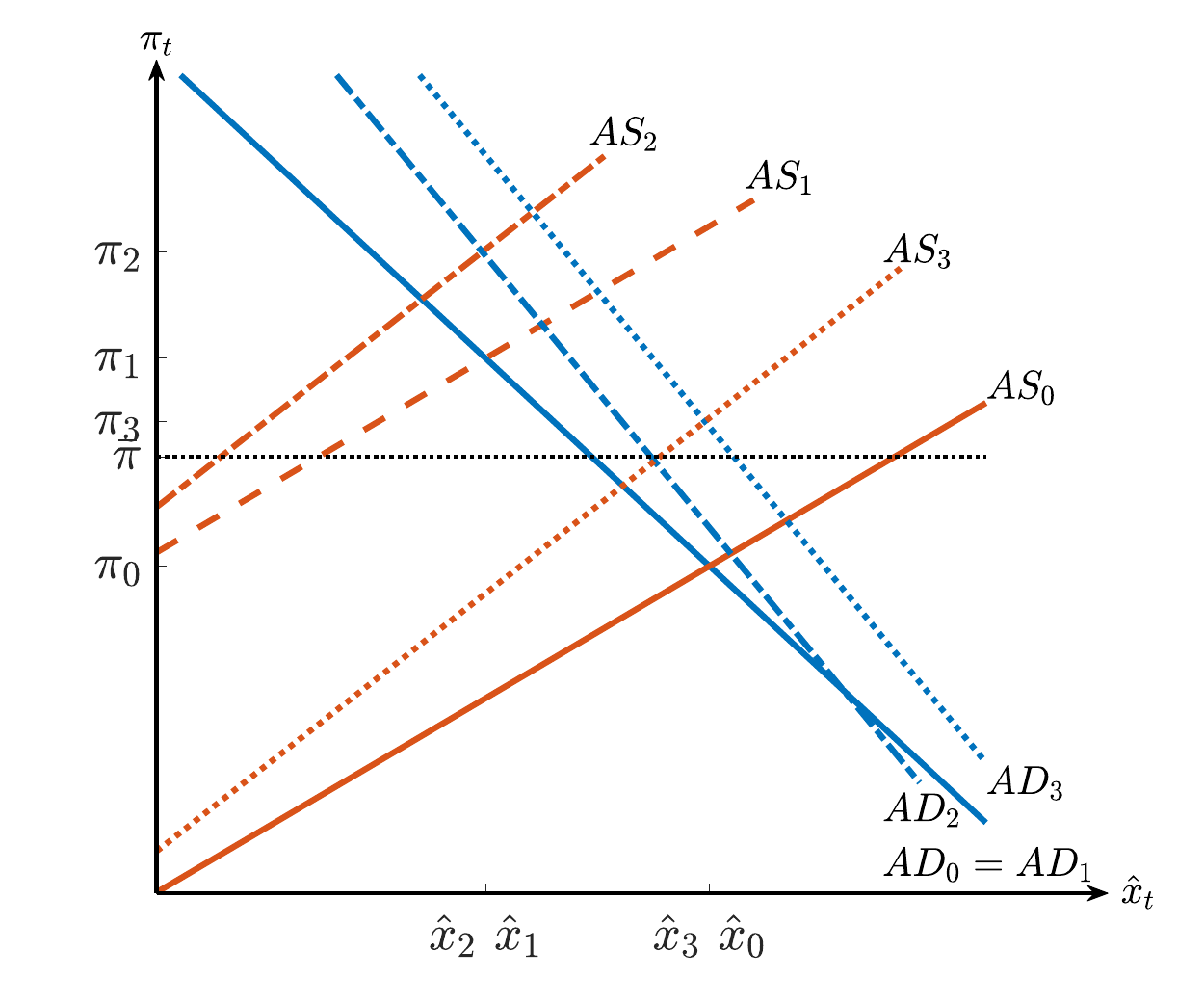} 
\end{tabular}%

\begin{minipage}{1\textwidth}
\footnotesize{\emph{Notes}: This figure depicts the third period in which the cost-push shock has died out (after being positive for two consecutive periods). The black-dotted line at $\Bar{\pi}$ depicts the inflation attention threshold.
}%
 \end{minipage}
\end{figure}

While the AS curve comes back down, the AD curve shifts further out due to the positive prior expectations that agents have when going into period 3. The AD curve is given by
\begin{equation*}
    AD_3: \quad \pi_3 = -\frac{1}{\phi_{\pi}-\gamma_{\pi,H}}\widehat{x}_3 + \frac{1-\gamma_{\pi,H}}{\phi_{\pi}-\gamma_{\pi,H}}\Tilde{E}_2\pi_3.
\end{equation*}
Thus, output recovers more strongly during this disinflationary period than what would be the case absent this shift in the AD curve, and disinflation occurs more gradually.
These results are shown graphically in Figure \ref{fig:asad3}.

\clearpage\newpage

\subsection{Additional figures to Section \ref{sec:irfs}}\label{app:additonal_model}
\begin{figure}[h]
\caption{Inflation dynamics under FIRE or absent the attention change}
\label{fig:pi_irfs_fire}\vspace{0.05cm} \centering%
\begin{tabular}{c}
\includegraphics[width=.8\textwidth]{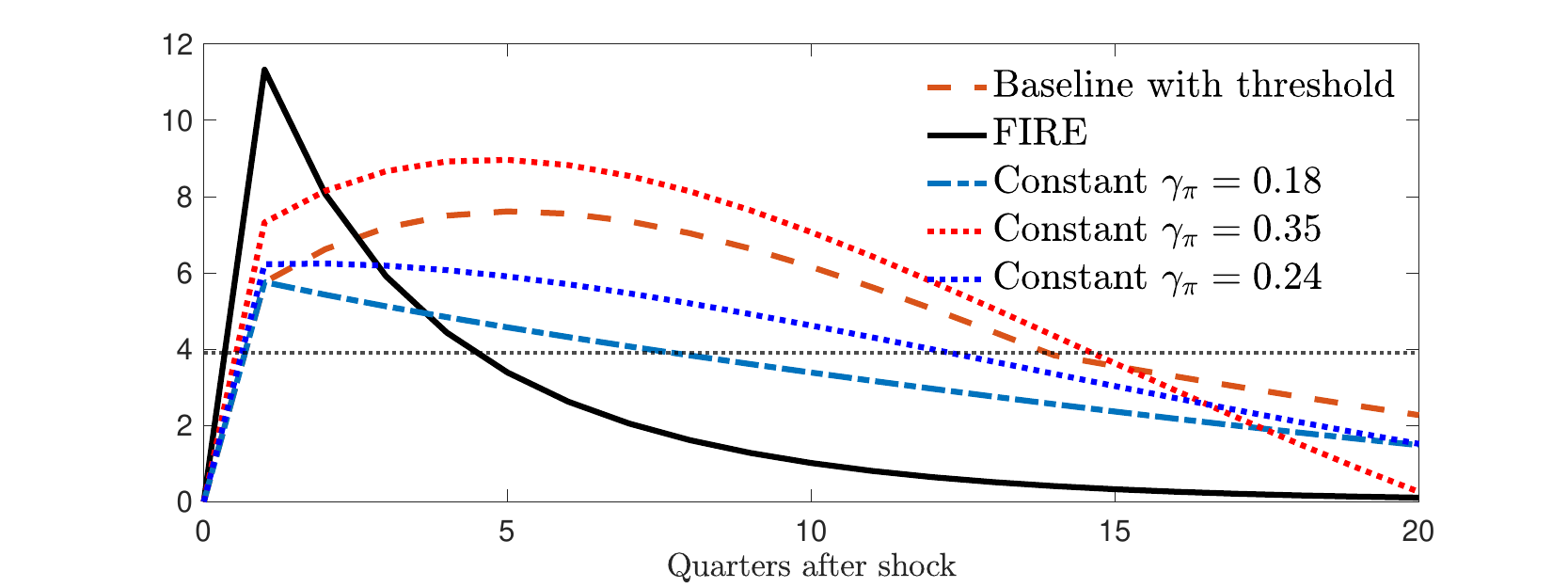}  
\end{tabular}%

\begin{minipage}{1\textwidth}
\footnotesize{\emph{Notes}: This figure shows the inflation response to a cost-push shock for different model specifications.
}%
 \end{minipage}
\end{figure}

\vspace{-0.2cm}
Figure \ref{fig:pi_irfs_fire} shows the inflation response to a cost-push shock for five different models. My model with the attention threshold (labelled "Baseline with threshold" and depicted by the red-dashed line) shows the discussed hump-shaped pattern of the inflation surge and the decrease in the speed of disinflation once inflation falls back below the threshold. As the other four illustrate, these inflation dynamics are unique to the model with the attention threshold. In particular, the models under FIRE (black-solid line) or in which attention remains low (blue-dashed-dotted line) predict that inflation peaks on impact and then comes down relatively quickly. The same is true for the model with a constant attention parameter of 0.24, which corresponds to the estimated attention level in the data if I do not impose a threshold (the inflation response is captured by the blue-dotted line). Additionally, models with FIRE could not account for the observed dynamics of forecast errors. A model variant in which attention to inflation is always high (red-dotted), on the other hand, can produce the initial hump-shaped response but then predicts a very fast decline back to target, i.e., this model would not explain why we see such a long last mile of disinflation. Furthermore, a model in which attention is always high would have predicted a very pronounced deflationary period after the Great Financial crisis. However, once thing that the model with high-attention illustrates, is that the inflation response is hump shaped when attention is high whereas it peaks on impact when the shock hits in times of low attenion, consistent with the empirical findings (panels (a) and (b) in Figure \ref{fig:emp_irfs_inflation_disentangle}).

\begin{figure}[h]
\caption{Inflation dynamics after a demand shock}
\label{fig:pi_irfs_r}\vspace{0.05cm} \centering%
\begin{tabular}{c}
\includegraphics[width=.8\textwidth]{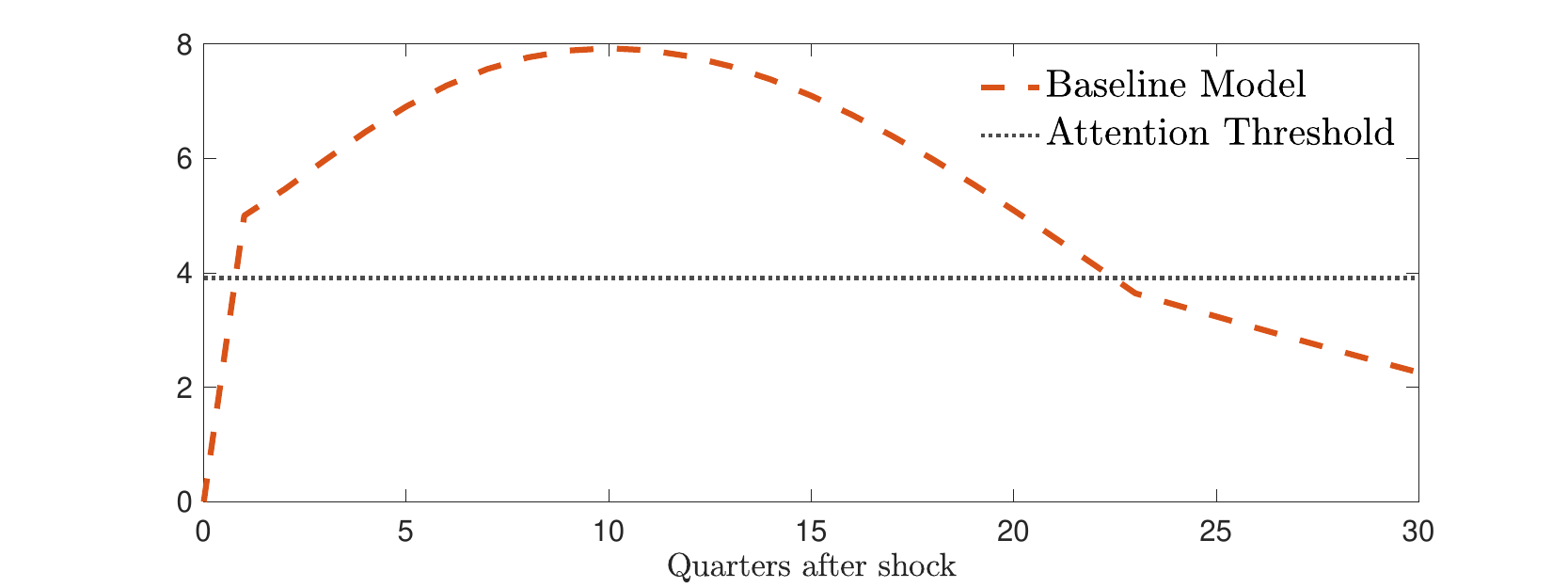}  
\end{tabular}%

\begin{minipage}{1\textwidth}
\footnotesize{\emph{Notes}: This figure shows the inflation response in the model with the attention threshold in response to a demand shock $r^*_t$ with persistence $\rho_r = 0.55$.
}%
 \end{minipage}
\end{figure}

Figure \ref{fig:pi_irfs_r} shows the inflation response to a positive demand shock. We see that the discussed inflation dynamics after a relatively large supply shock are similar to the ones after a demand shock: inflation keeps on increasing, fueled by the attention increase, and the speed of disinflation slows substantially once inflation falls back below the threshold and attention therefore decreases. Note, however, that the inflation surge is somewhat longer lived than in response to the supply shock.

\begin{figure}[h]
\caption{Inflation dynamics with $\gamma_{x,L} = 0.25$ and $\gamma_{x,H} = 0.1$}
\label{fig:pi_irfs_ychange}\vspace{0.05cm} \centering%
\begin{tabular}{c}
\includegraphics[width=.8\textwidth]{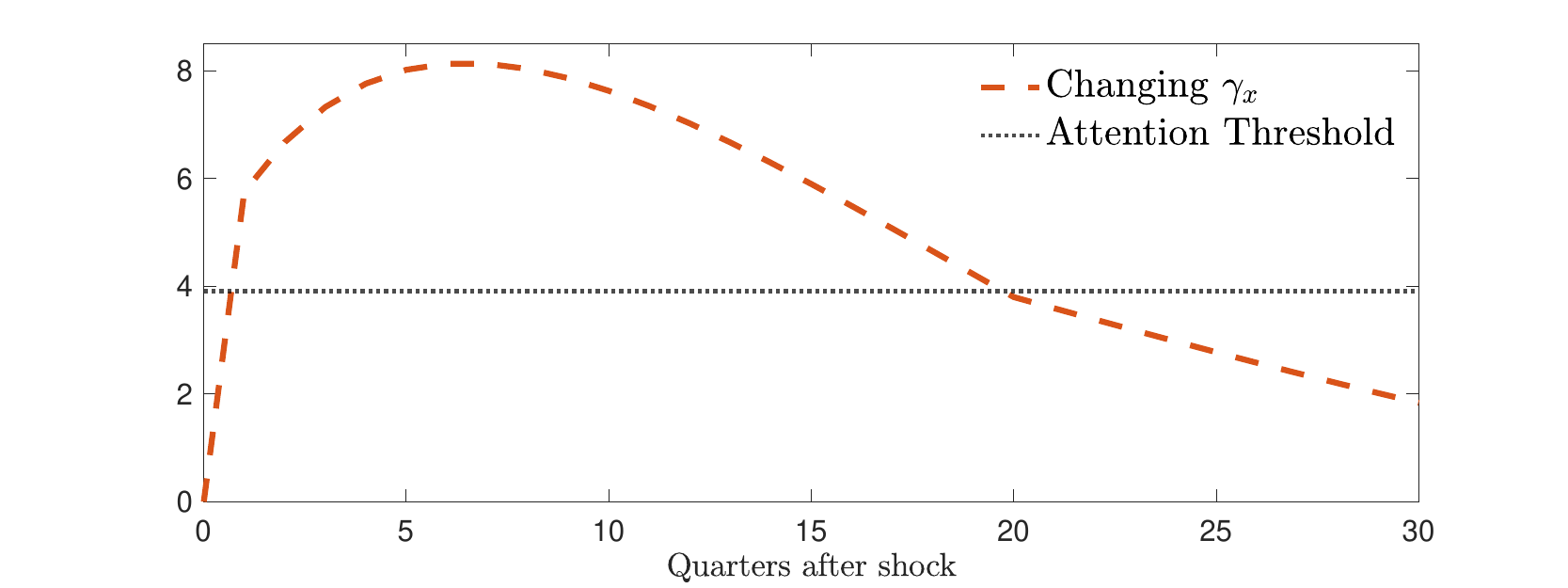}  
\end{tabular}%

\begin{minipage}{1\textwidth}
\footnotesize{\emph{Notes}: This figure shows the inflation response to a cost-push shock when attention to the output gap decreases in times of high attention to inflation.
}%
 \end{minipage}
\end{figure}

Figure \ref{fig:pi_irfs_ychange} shows the inflation response for the case in which attention to the output gap, $\gamma_{x}$, decreases when attention to inflation increases. This may reflect that agents shift their attention from the output gap to inflation when inflation is high. We see that the inflation surge becomes more persistent in that case (note that the x-axis now shows 30 periods rather than 20). The reason is that the negative output gap due to the adverse cost-push shock leads to a smaller downward revision of output gap expectations, thus, agents remain relatively more optimistic about the output gap, and hence, consume relatively more, leading to additional upward pressure on inflation. 

\begin{figure}[h]
\caption{Inflation dynamics with rational consumption expectations}
\label{fig:pi_irfs_fire_c}\vspace{0.05cm} \centering%
\begin{tabular}{c}
\includegraphics[width=.8\textwidth]{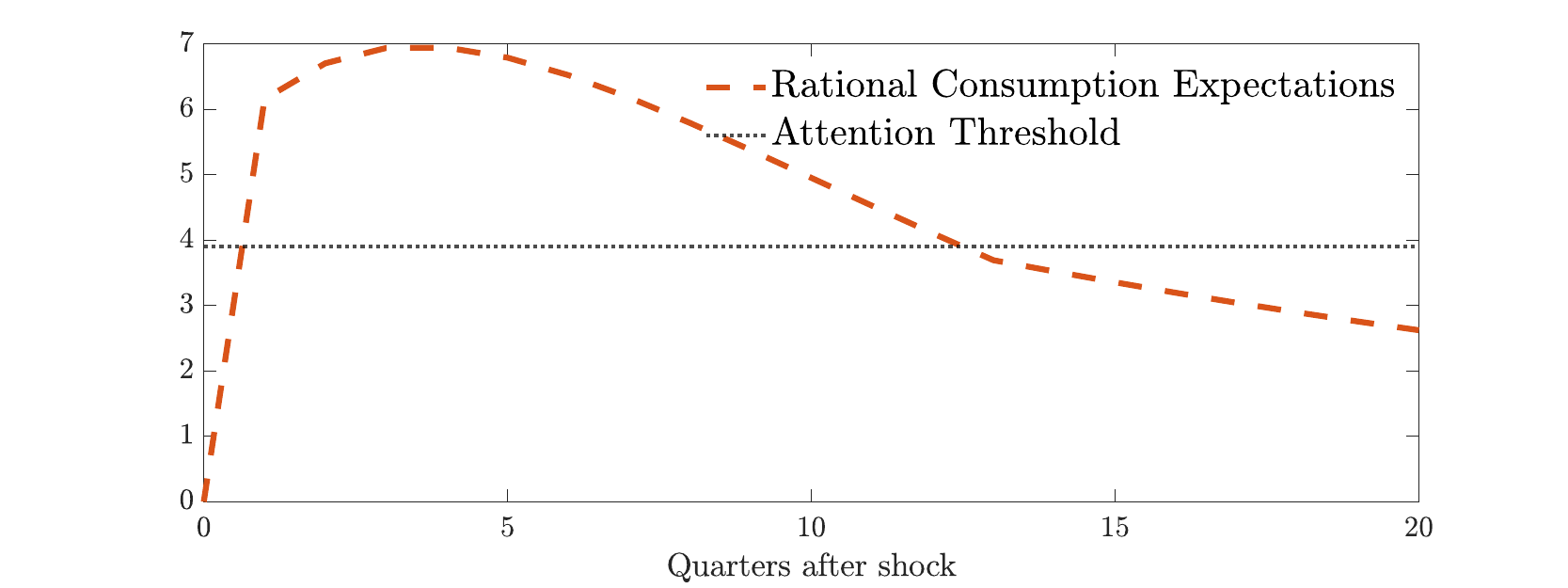}  
\end{tabular}%

\begin{minipage}{1\textwidth}
\footnotesize{\emph{Notes}: This figure shows the inflation response to a cost-push shock when consumption expectations are formed under full-information rational expectations.
}%
 \end{minipage}
\end{figure}

Figure \ref{fig:pi_irfs_fire_c} shows the inflation response to a cost-push shock when consumption expectations are formed under full-information rational expectations. We see that the dynamics are qualitatively similar to the baseline model. The endogenous inflation increase is somewhat muted here because the forward-looking consumption expectations do not induce endogenous persistence in the way they do in the baseline model. When increasing the exogenous persistence of the shock slightly, however, the models are very similar (not shown).

\end{document}